\documentclass[trackchanges,twocolumn]{aastex701}

\newcommand{\nuclei}[2]{\textbf{$^{#1}${#2}}}

\usepackage{float}

\graphicspath{{./}{figs/}}

\begin{document}

\title{Gamma-Ray Spectra of $R$-Process Nuclei}

\author[orcid=0000-0002-7893-4183]{Axel Gross}
\affiliation{Theoretical Division, Los Alamos National Laboratory, Los Alamos, NM 87545, USA}
\affiliation{ Center for Theoretical Astrophysics, Los Alamos National Laboratory, Los Alamos, NM 87545, USA}
\email[show]{agross@lanl.gov}  

\author[orcid=0000-0003-1758-8376]{Samuel Cupp}
\affiliation{Theoretical Division, Los Alamos National Laboratory, Los Alamos, NM 87545, USA}
\affiliation{ Center for Theoretical Astrophysics, Los Alamos National Laboratory, Los Alamos, NM 87545, USA}
\email{scupp@lanl.gov}  

\author[orcid=0000-0002-9950-9688]{Matthew R. Mumpower}
\affiliation{Obsidian Research, Fort Wayne, IN 46835, USA}
\affiliation{Department of Physics and Astronomy, University of Notre Dame, Notre Dame, IN, 46656, USA}
\affiliation{Computational and Artificial Intelligence Division, Los Alamos National Laboratory, Los Alamos, NM 87545, USA}
\affiliation{ Center for Theoretical Astrophysics, Los Alamos National Laboratory, Los Alamos, NM 87545, USA}
\email{matthew@mumpower.net}

\begin{abstract}
The radioactive decay of unstable nuclei created in the rapid neutron capture process releases a large amount of $\gamma$-rays. When the ejecta are optically thick, these $\gamma$-rays may contribute to an associated kilonova. Once transparent, prominent spectral features will be directly observable in current and future $\gamma$-ray detectors. In this work, we study and compare the $\gamma$-ray spectra of different representative $r$-process trajectories across a broad range of timescales, from hours to 50,000 years, identifying the nuclei which significantly contribute. We discuss these findings in the context of observability, noting that there are several practical challenges to connecting observed signatures to specific nuclei. However, if these challenges can be overcome, direct observation of $\gamma$-rays from $r$-process sites can provide insight into the fundamental physics underpinning the $r$-process.
\end{abstract}


% =============================================
\section{Introduction}
It has long been inferred from solar abundance data (see e.g. \cite{1957RvMP...29..547B}, \cite{2023A&ARv..31....1A}) that the majority of elements heavier than iron were created by a combination of two different mechanisms: the slow ($s$) and rapid ($r$) neutron capture processes. 
However, while it is clear that the $r$-process is a significant component of galactic chemical evolution, the site(s) where the $r$-process occurs and their relative significance are much less certain. 
The $r$-process occurs in extreme environments with large numbers of free neutrons and involves capture on neutron-rich isotopes far from stability. There are several proposed sites of the $r$-process, including core-collapse supernovae (\citealt{1994ApJ...433..229W,1996ApJ...471..331Q,2001ApJ...554..578W,2006ApJ...646L.131F,2012ApJ...750L..22W,2015ApJ...810..109N,2019Natur.569..241S,2021Natur.595..223Y}), compact object mergers (\citealt{1974ApJ...192L.145L,1989Natur.340..126E,1999A&A...341..499R,1999ApJ...525L.121F,2006ApJ...646L.131F,2011ApJ...738L..32G,2012MNRAS.426.1940K,PhysRevD.87.024001,2013ApJ...773...78B,2014ApJ...789L..39W}), collapsars (\citealt{2013ApJ...778....8B,2024ApJ...962...68A,2025ApJ...982...81M}), and magnetars (\citealt{2024MNRAS.528.5323C,2025ApJ...985..234P}). 

There have been observational hints of the $r$-process at some sites---notably, the multimessenger observations of GW170817 \citep{2017ApJ...848L..12A}, as well as the magnetar giant flare SGR 1806-20 \citep{2025ApJ...984L..29P} and the long-duration gamma-ray bursts GRB211211A and GRB230307A \citep{2025arXiv250903003R}. Due to the extreme conditions present in these environments, as well as limited experimental data for many neutron-rich isotopes far from stability where the \textit{r}-process occurs, it is difficult to directly link observational data to specific physics. For example, the observed kilonova associated with GW170817---in particular, its peak in the near-infrared and optical---was widely interpreted as evidence that NSMs undergo a strong $r$-process (\citealt{2017Sci...358.1570D,2017Natur.551...80K,2017Sci...358.1559K,2017ApJ...848L..34M,2017PASJ...69..102T,2018MNRAS.481.3423W,Brethauer:2024zxg}). This conclusion stems from the argument that the production of lanthanides, which have high opacity when ionized, (see e.g.,  \citealt{2022ApJ...939....8D,2025arXiv250707785F}), results in a kilonova which peaks in the infrared (e.g., \citep{2013ApJ...774...25K,2013ApJ...775..113T}), while a weaker $r$-process which does not produce lanthanides would produce a kilonova which peaks in the blue and optical bands (e.g., \citealt{2010MNRAS.406.2650M,2011ApJ...736L..21R,2014MNRAS.441.3444M}). While this interpretation has merit, the multi-physics required in the associated kilonova modeling is challenging, and a large number of poorly understood parameters can have significant impact on the resultant light curve (see e.g., \citealt{2016ApJ...829..110B,2021ApJ...906...94Z,2021ApJ...910..116K,Fryer:2023pew,2023ApJ...944..144L}), suggesting that there are other interpretations of the observational data. For example, \cite{2023ApJ...958..121T} and \cite{Fryer:2023osz} found that different ejecta velocity assumptions can lead to late-time features that mimic the suggested contributions from lanthanides. \cite{2025arXiv250903003R} modeled the kilonova of a long-duration GRB associated with a weak $r$-process (no lanthanides) and found consistency with the observed near-infrared and optical peaks observed in GRB211211A and GRB230307A.

While the challenges associated with the modeling of kilonova make it very difficult to disentangle and understand the detailed physics of the $r$-process from kilonova light curves, there may also be more direct observational signatures. The large number of $\beta$-decays associated with $r$-process nucleosynthesis generate an enormous amount of emitted particles, including $\gamma$-rays. These $\gamma$-rays will initially be opaque to observations, and are a significant contributor to the aforementioned kilonova. However, in the days and weeks after the event, as the opacity lowers, specific strong spectral emission lines may be directly observable. Because of the numerous long-lived nuclei which are synthesized, some emission lines are potentially observable for hundreds of thousands of years in remnants. The observation of prominent $\gamma$-rays have been of interest to the astrophysical community outside the $r$-process: the $\gamma$-rays from decays of \nuclei{44}{Ti} and \nuclei{56}{Ni} have been informative to the detonation mechanism of core-collapse supernovae \citep{1988Natur.332..516L,1994A&A...284L...1I,Mochizuki:1999yg,2014Natur.512..406C,2015Sci...348..670B,2020A&A...638A..83W,2015A&A...574A..72D}, and the observations of the 1.8 MeV line from \nuclei{26}{Al} have informed both star formation and nucleosynthetic activity in our galaxy \citep{1977ApJ...213L...5R,1982ApJ...262..742M,2023A&A...672A..53P,2025A&A...695A.190W}. The potential of $\gamma$-ray observations from $r$-process sites have also been explored. \cite{2016MNRAS.459...35H} calculated the $\gamma$-ray signal from kilonova ejecta and found it would be observable to $\sim$3-10 Mpc with current detectors. \cite{2020ApJ...889..168K} expanded on this work by modeling $\gamma$-ray transport, identifying specific spectral lines which may be observable, and also examined kilonova remnants in detail. \cite{1998ApJ...506..868Q} studied $\gamma$-rays from the $r$-process in supernova, identifying several potential spectral lines. \cite{2022ApJ...933..111T} studied neutron star merger remnants in detail, and proposed line diagnostics to infer the initial electron fraction of the ejecta. \cite{Vassh2024} highlighted the potential signal of \nuclei{208}{Tl} resulting from the decays of long-lived actinides at both prompt ($\sim$ days) and longer ($~\sim$ years) timescales, and \cite{2020ApJ...903L...3W} highlighted that the prompt emission spectrum from the fission of actinides could produce a significant amount of MeV $\gamma$-rays. In general, these studies have found that $\gamma$-rays from $r$-process nucleosynthesis may be observable in current and/or next generation detectors at galactic scales, highlighting the importance for these signatures to be studied in detail. 

While the physics associated with the direct observation of strong spectral lines is more straightforward than kilonova modeling, it is still nontrivial for a multitude of reasons. Spectral lines will broaden due to the expansion of the ejecta, softening the magnitude of the spectral peaks. Transport of the emitted $\gamma$-rays through the medium may cause significant redistribution of energy, especially at earlier times when the opacity is higher. There may be other significant sources of $\gamma$-rays at the site which must be compared to the potential $r$-process emission.  The consequences of this are that the specific spectral features which are observable are dependent on both the choice of astrophysical site and the details of the physics which comprise the model, and therefore, studies of $\gamma$-ray observability are very specific to the scenario which is considered.  

In this work, we take a different approach than the above works. Rather than considering the observability of $\gamma$-rays, we narrow our focus, characterizing the nuclei which significantly contribute to the $\gamma$-ray emission spectra across a representative set of $r$-process scenarios. Our goal is to be comprehensive, considering these contributions across a broad timescale. In this manner, our results are independent of the $r$-process site considered, and while not all of the spectral features we discuss may be observable at all sites, spectral features which are observed by current and future $\gamma$-ray detectors may be identified and matched with the relevant nucleus to infer the nature of the $r$-process which may have occurred. 

We organize our paper as follows: In $\S\ref{sec:method}$, we detail our methodology for calculating the $\gamma$-spectra.  In $\S\ref{sec:spectra}$, we present our main results and compare the spectra for the different types of $r$-process, highlighting the most significant spectral features. In $\S\ref{sec:discuss}$, we qualitatively discuss our results in the context of observability, highlighting several important lessons learned. We summarize and conclude in $\S\ref{sec:conclude}$.

\section{Methodology}
\label{sec:method}
To calculate the $\gamma$-ray spectra, we follow the approach of \cite{2025ApJ...995L..28G}, which combines detailed spectral calculations for the $\beta$-decay of individual nuclei with a nuclear reaction network which calculates the number of nuclei of each species which decay as a function of time. The total $\gamma$-ray emission spectrum $S^{\gamma}(E,t)$ can be factorized as:
\begin{equation}
\label{eq:spectra}
S^{\gamma}(E,t)= \sum_{^A_ZX} F_{^A_ZX}(t) S^{\gamma}_{^A_ZX}(E)
\end{equation}
where $S^{\gamma}_{^A_ZX}(E)$ is the $\gamma$-ray emission of the nucleus ${^A_ZX}$ as a function of energy  (number per unit energy per decay), and $F_{^A_ZX}(t)$ is the $\beta$-decay reaction flow as a function of time (decays per second per unit ejecta mass), and the sum is over every nucleus which decays in the network. We discuss the calculation of these two quantities individually below.
\subsection{Spectra of Individual Nuclei}
For each nucleus which $\beta$-decays, we use the $\gamma$-ray and x-ray spectra of ENDF/B-VIII.0 \citep{brown2018endfb} if available. If there is no available $\beta$-decay spectrum for a specific nucleus, we use the tabulated results of \cite{MUMPOWER2025101736}, which provides the theoretical $\beta$-decay emission spectra for neutron-rich nuclei, including $\gamma$-rays, electrons, neutrinos, and neutrons. We additionally include the $\gamma$-ray and x-ray spectra for $\alpha$-decay if available in ENDF. While we do not have theoretical $\alpha$-decay spectra, we do not anticipate that this will be a large contribution, as $\alpha$-decays primarily transition to the ground state in the daughter nucleus. Any nucleus which $\alpha$-decays but does not have a measured spectra is far from stability, and therefore short-lived enough that it should decay before the relevant timescales for potential direct observation ($\sim$ hours or longer).

\begin{figure}[ht!]
\plotone{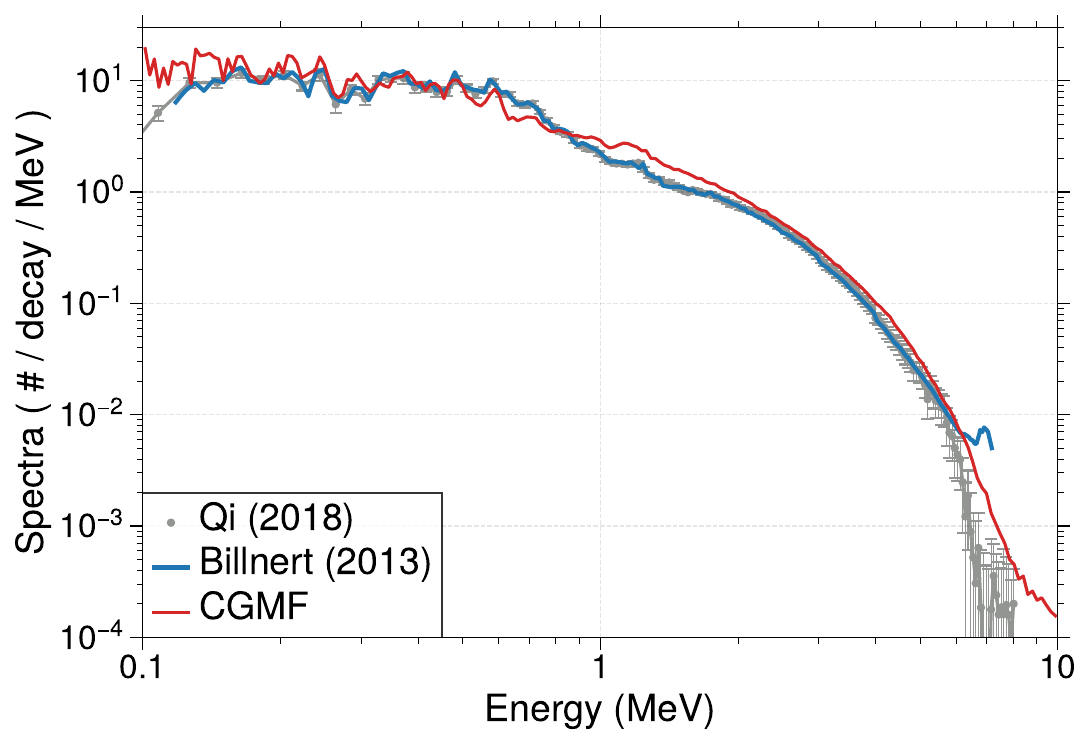}
\caption{Prompt fission spectrum of \nuclei{252}{Cf} from the CGMF code as compared to the experimental results of \cite{2013PhRvC..87b4601B} and \cite{2018PhRvC..98a4612Q}. In each case, the experimental spectrum shown is the obtained from the LaBr$_{3}$ detector. The energy resolution of these detectors varies as a function of energy, from $\sim$ 10 keV at 100 keV to $\sim$ 100 keV at 8 MeV.
\label{fig:promptfission}}
\end{figure}

The fission of nuclei will introduce an additional prompt source of $\gamma$-rays. It is generally challenging to compute the prompt fission emission spectrum for a nucleus. As a stand-in, for all fissioning nuclei, we will use the prompt fission emission spectrum calculated from CGMF \citep{2021CoPhC.26908087T} for \nuclei{252}{Cf}. This emission spectrum is shown in Fig. \ref{fig:promptfission} as compared to the experimental measurements of \cite{2013PhRvC..87b4601B} and \cite{2018PhRvC..98a4612Q}. For a detailed discussion of the impact of fission on the $\gamma$-ray spectrum, see \cite{2020ApJ...903L...3W}.

For the rest of the text, when we refer to the $\gamma$-ray spectra, we are referring to the combined photon spectra, including $\gamma$-rays (de-excitation of the nucleus), x-rays (de-excitation of the electrons), and the photons promptly emitted by fissioning nuclei. 

\subsection{Nuclear Reaction Flow}
We simulate nucleosynthesis with version 1.6.0 of the Portable Routines for Integrated nucleoSynthesis Modeling (PRISM) reaction network \citep{2021PhRvC.104a5803S}. 
The nuclear input to PRISM is based on the 2012 version of the Finite Range Droplet Model
\citep{2012PhRvL.108e2501M,2016ADNDT.109....1M}. Radiative capture rates are calculated with the CoH$_3$ statistical Hauser–Feshbach code \citep{2019arXiv190105641K,2021EPJA...57...16K}. 
$\beta$-decay rates, including delayed neutron emission, are calculated assuming statistical de-excitation from excited states \citep{2016PhRvC..94f4317M,2018ApJ...869...14M}. 
The remaining reaction rates (e.g. $\alpha$-decay and other less substantive reaction types for the $r$-process) are obtained from the REACLIB database \citep{2010ApJS..189..240C}. 
Nuclear fission is handled as in \cite{Vassh2019}. From PRISM, we extract the nuclear reaction flow, which is given in units of number per second per ejected nucleon. This can be scaled to more conventional units by multiplying by the number of nucleons in the ejecta, given by $M_{ejecta}/M_N$, where $M_N = 1.67 \times 10^{-24}$ g is the average nucleon mass.

We are interested in understanding the differences in $\gamma$-ray spectra which may result from different $r$-process scenarios, as well as the most significant spectral features of each scenario. The primary cause of these differences is the strength of the $r$-process which occurs. In a stronger $r$-process, higher mass nuclei are produced, which allows for the additional contributions from the unique signatures of the associated $\beta$-decays. In addition, there are relatively less lower mass nuclei which $\beta$-decay, which suppresses their associated signatures. We use the temporal evolution of temperature and density of trajectory (b) from \cite{2025ApJ...982...81M}. This trajectory is obtained through modeling the cocoon of a $\gamma$-ray burst, which has been suggested to be a site of the $r$-process due to photo-hadronic interactions in the jet head. The cocoon is modeled with the density profile of:
\begin{equation}
    \rho(t) = \rho_0 \left( 1 + \frac{t}{\tau_1} + \left(\frac{t}{\tau_2}\right)^\xi \right)^{-1} \ ,
\end{equation}
where $\xi$ = 2, $\tau_1$, $\tau_2$ are characteristic timescales, for which we use $\tau_1 = \tau_2 = 3.5 \times 10^{-2}$ s, and $\rho_0$ is the initial density, for which we use $3.2 \times 10^4$ g/cm$^3$. The temperature is assumed to evolve as an adiabatic gas:
\begin{equation}
    \label{eqn:tempfromrho}
    T(t) = T_0 \left( \frac{\rho(t)}{\rho_0} \right)^{\gamma-1} \ , 
\end{equation}
where $\gamma=4/3$ (radiation dominated), and we take $T_0 = 2$ GK. Our choice of trajectory, specifically the unique early-time behavior as compared to more conventional trajectories, does not have a significant impact on the observability of $\gamma$-rays, which are primarily determined by the final abundance pattern which is produced.

\begin{figure}[ht!]
\plotone{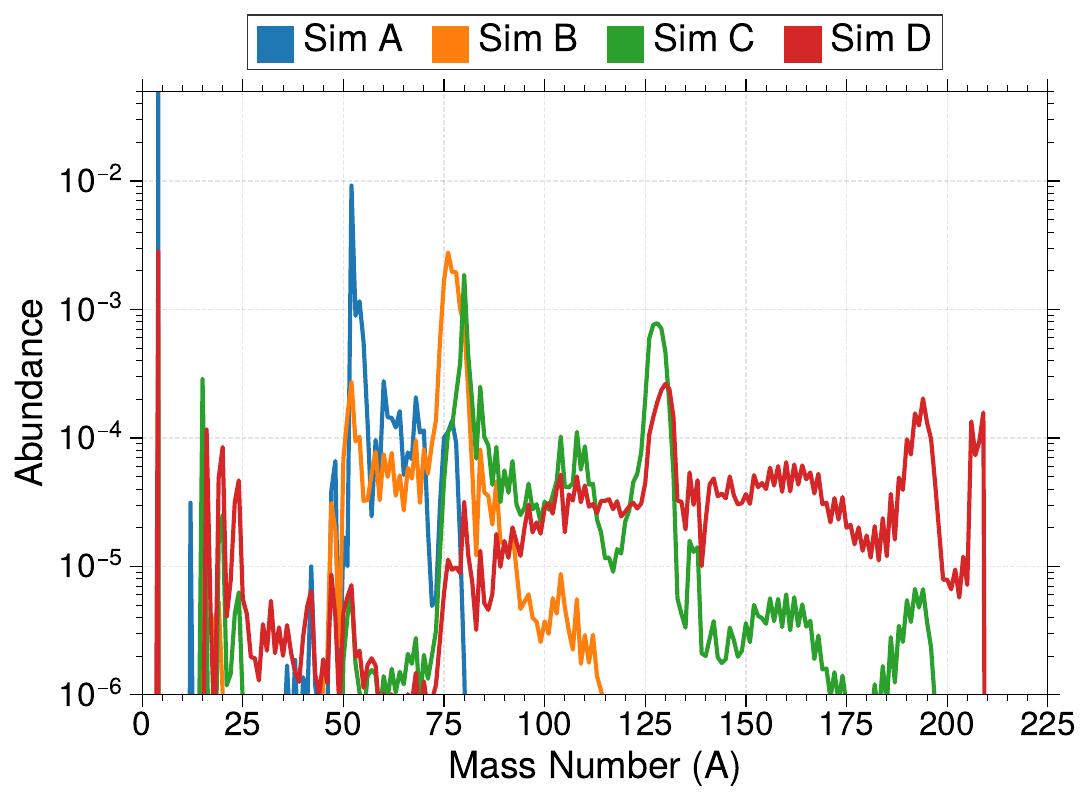}
\caption{Final abundance pattern (at 100 Myr) of each of our $r$-process scenarios: a limited $r$-process (Simulation A), a weak $r$-process, (Simulation B), a strong $r$-process (Simulation C), and an extended $r$-process (Simulation D). All four patterns are normalized to the abundance per nucleon. 
\label{fig:abund}}
\end{figure}

With this trajectory, we vary the initial electron fraction to control the strength of the $r$-process which occurs, adopting values of $Y_e$ = 0.4, 0.25, 0.175, and 0.034 for Simulations A, B, C, and D, respectively. These scenarios produce abundance patterns which are characteristic of: a limited $r$-process which does not produce the first abundance peak, a weak $r$-process which consists of primarily the first abundance peak, a strong $r$-process which produces the first and second abundance peaks, and an extended $r$-process which produces all three abundance peaks, including an extensive amount of actinides. Fig. \ref{fig:abund} shows the final abundance patterns of these scenarios.

\begin{figure*}[t]
\digitalasset
\plotone{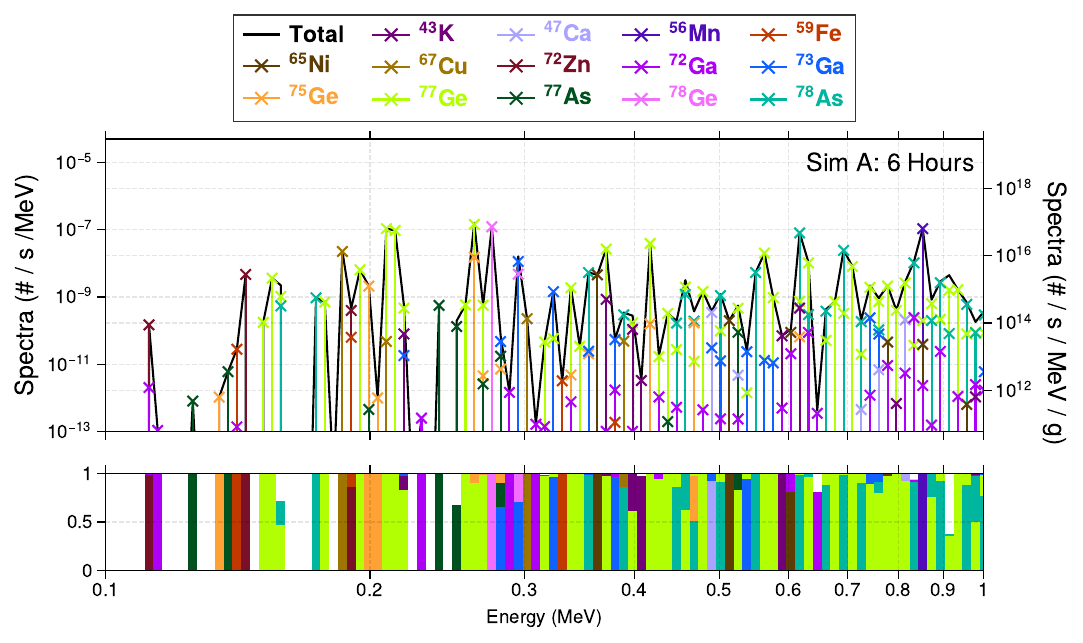}
\caption{Each panel gives the spectral decomposition for one of the four simulations at one of the eight timescales in the region 0.1 $<$ E $<$ 1 MeV, in units of (\#/s/MeV) (per nucleon) (left axis) and (\#/s/MeV/g) (right axis). The solid black line gives the total spectra, while the colored crosses denote the contributions to this spectra from an individual nucleus. Colors for nuclei are consistent across simulations and timescales, though some colors represent multiple nuclei (on different panels). The complete figure set (32 images) is available in the online journal and in the appendix of this pdf version.
\label{fig:xray}}
\end{figure*}

\begin{figure*}[t]
\digitalasset
\plotone{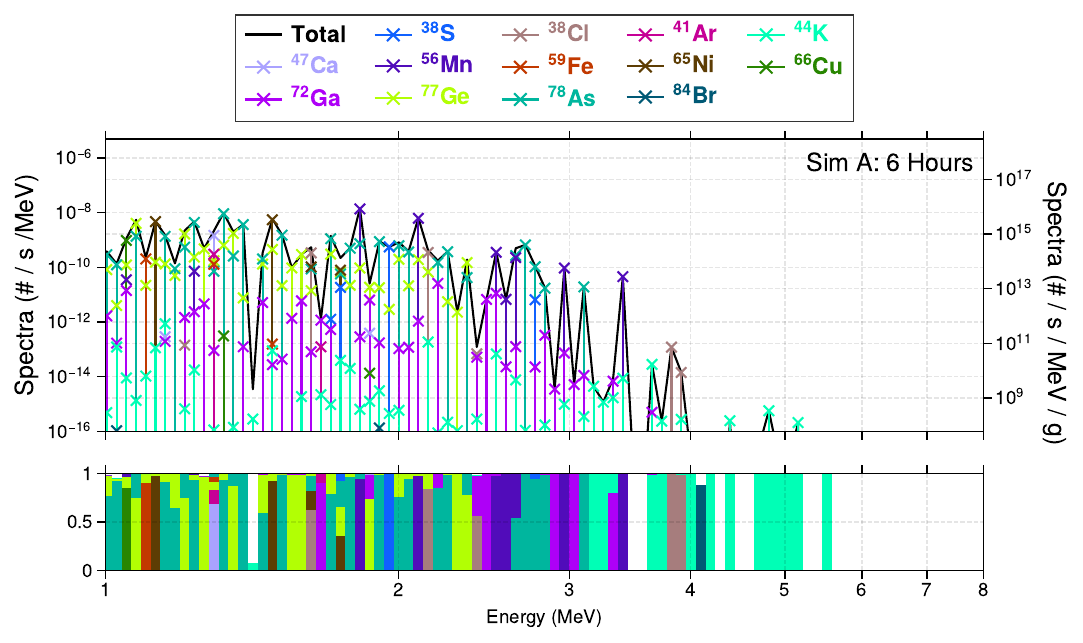}
\caption{Each panel gives the spectral decomposition for one of the four simulations at one of the eight timescales in the region E $>1$ MeV, in units of (\#/s/MeV) (per nucleon) (left axis) and (\#/s/MeV/g) (right axis). The solid black line gives the total spectra, while the colored crosses denote the contributions to this spectra from an individual nucleus. Colors for nuclei are consistent across simulations and timescales, though some colors represent multiple nuclei (on different panels). The complete figure set (32 images) is available in the online journal and in the appendix of this pdf version. \label{fig:gammaray}}

\end{figure*}

While these scenarios are representative of the different strengths of the $r$-process which can occur, the inherent uncertainties in modeling the $r$-process as well as the differences in trajectories from different $r$-process sites make it possible to generate generally similar abundance patterns to the ones presented here, but with significant differences in the abundances of individual mass numbers. The effects of these differences are that individual spectral features could be boosted or suppressed by a significant factor in a different scenario. Therefore, in an effort to make our results more generally applicable, we also discuss the nuclei which have significant but subdominant contributions, as in a different scenario, these contributions may be more dominant. 

\section{Gamma-Ray Spectra}
\label{sec:spectra}
We consider the $\gamma$-ray spectrum for each of our simulations at 8 representative timescales: 6 hours, 1 day, 1 week, 1 month, 1 year, 50 years, 1000 years, and 50,000 years. For each simulation at each of these times, we calculate the associated $\gamma$-ray spectrum and perform a spectral decomposition, identifying the nuclei which significantly contribute to the spectral features. We focus on the energy region E $>$ 0.1 MeV, as the lower energy region generally has contributions from a larger number of nuclei, less prominent spectral features, and an increased likelihood of background sources which lower the observability. Figs. \ref{fig:xray} and \ref{fig:gammaray} show these spectral decompositions at each of the 8 timescales for the regions 0.1 $<$ E $<$ 1 MeV and E $>$ 1 MeV, respectively.  

We have tabulated each nucleus shown in Figs. \ref{fig:xray} or \ref{fig:gammaray} in Table \ref{tab:gammaray}, along with the nuclei which drive the timescale at which they appear, the simulations they appear in, and the specific spectral lines which contribute significantly to the overall spectra. Bolded spectral lines indicates spectral features of notable prominence in at least one simulation. 

\subsection{Comparison of Spectral Profiles}
\begin{figure*}[t]
\plotone{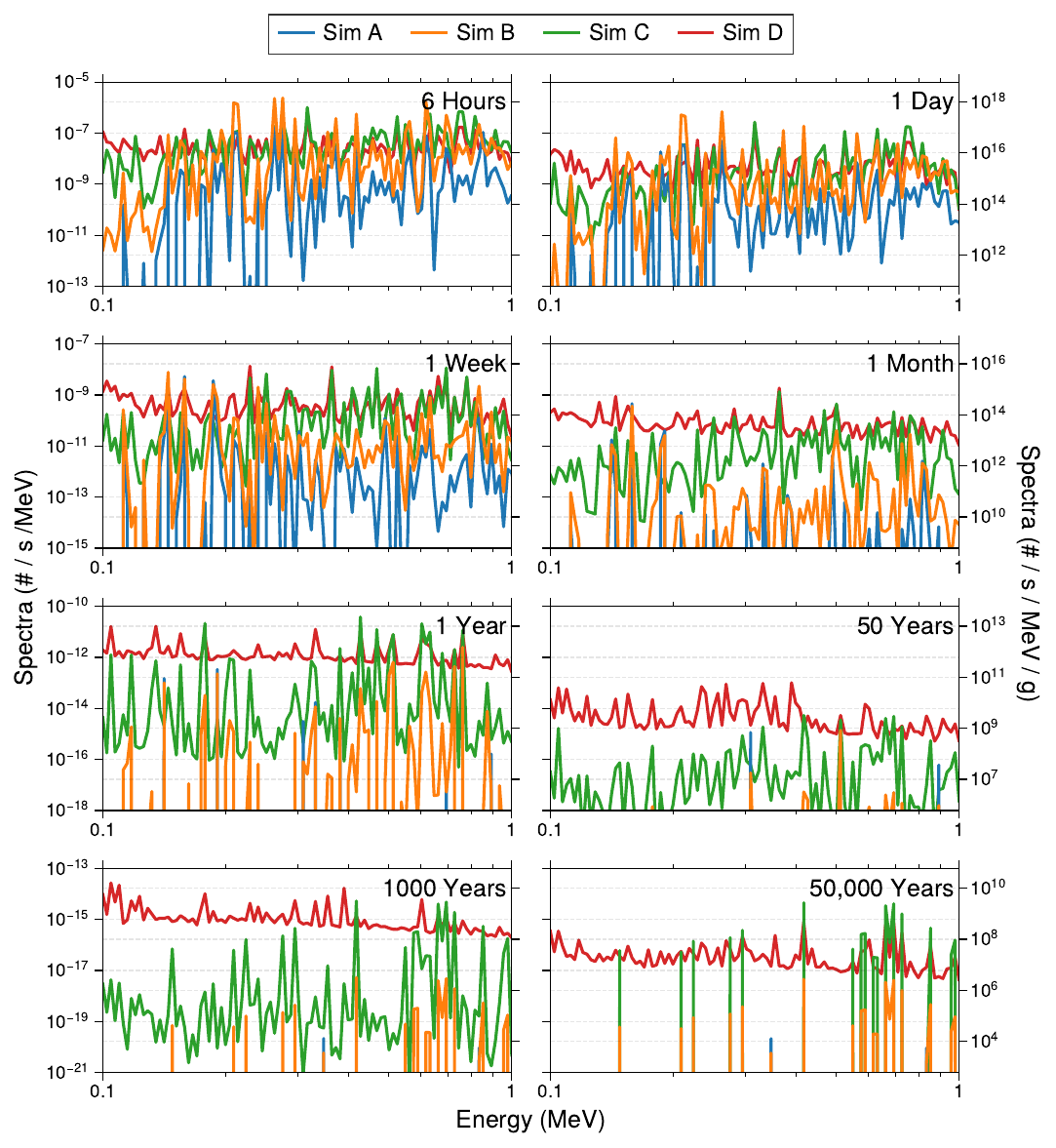}
\caption{Each panel gives the comparison of the spectra for the four representative simulations at one of the eight representative timescales for the energy region 0.1 $< E <$ 1 MeV, in units of (\#/s/MeV) (per nucleon) (left axis) and (\#/s/MeV/g) (right axis).
\label{fig:comparexray}}
\end{figure*}

\begin{figure*}[t]
\plotone{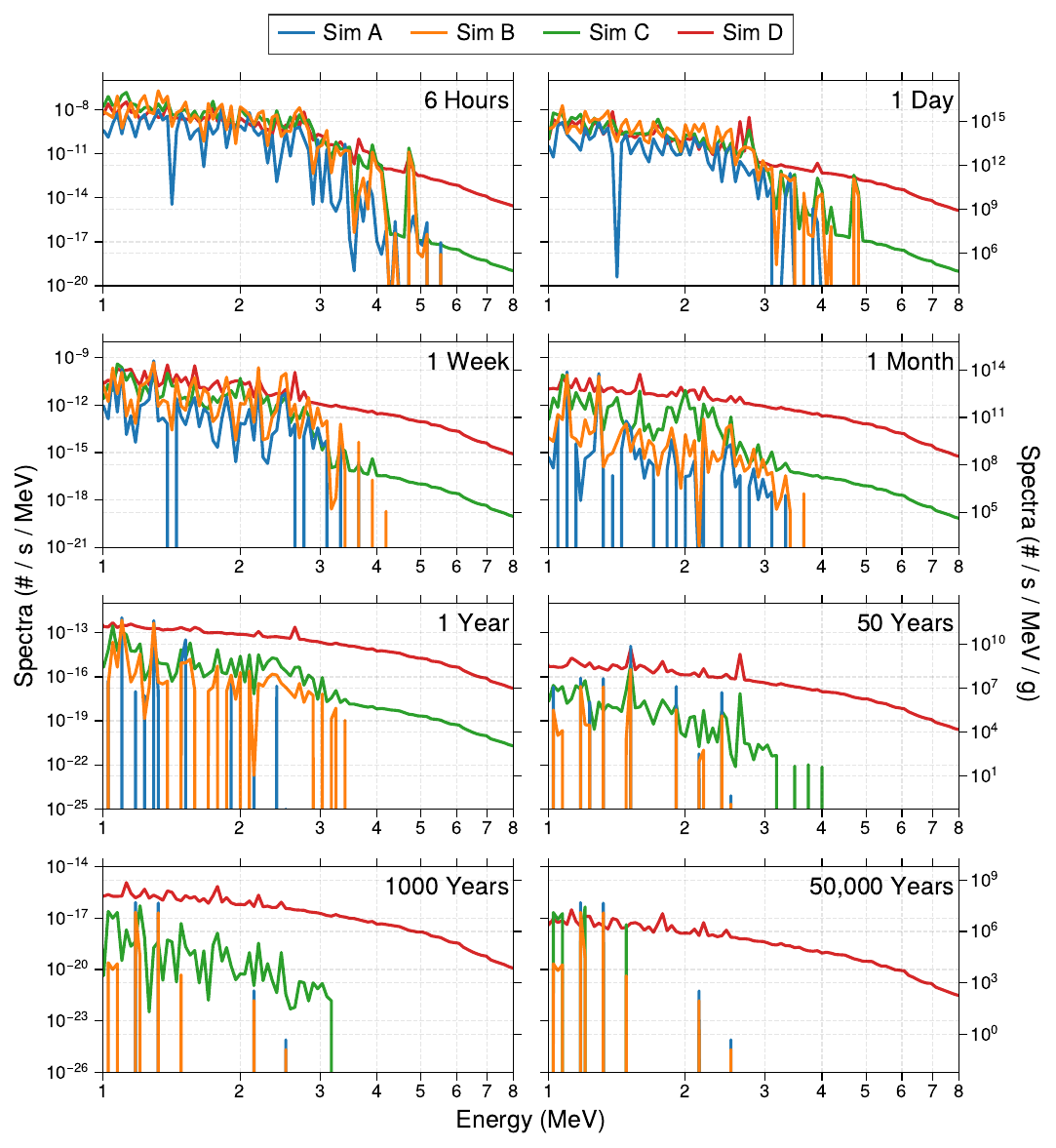}
\caption{Each panel gives the comparison of the spectra for the four representative simulations at one of the eight representative timescales for the energy region $E >$ 1 MeV, in units of (\#/s/MeV) (per nucleon) (left axis) and (\#/s/MeV/g) (right axis).
\label{fig:comparegamma}}
\end{figure*}

In Figs. \ref{fig:comparexray} and \ref{fig:comparegamma}, we compare the $\gamma$-ray emission profiles at each representative timescale for the energy for the regions 0.1 $<$ E $<$ 1 MeV and E $>$ 1 MeV, respectively. Below, we highlight the most significant features of each simulation at each timescale.


At 6 hours, there are a large number of nuclei contributing to the spectrum for all simulations, and all simulations are relatively comparable in the lower energy region ($\lessapprox$ 3 MeV). Due to the lower number of nuclei contributing, Sim A and B have the most prominent spectral lines, with strong contributions (particularly for Sim B) from \nuclei{73}{Ga}, \nuclei{77}{Ge}, \nuclei{78}{Ge}, and \nuclei{78}{As}. Sim C and D have several prominent lines from \nuclei{128}{Sb} and \nuclei{129}{Sb} at different characteristic energies; however, their overall magnitudes are somewhat less than the extremely strong features of Sim B. In Sim D, these lines are somewhat less prominent; however, there are a few additional spectral lines from \nuclei{184}{Hf} and \nuclei{117}{In}.

The higher energy region ($\gtrapprox$ 3 MeV) is characteristically very different from the lower energy region. There are only a few spectral features in Sim A, B, and C from \nuclei{56}{Mn}, \nuclei{84}{Br}, and \nuclei{88}{Rb}. Sim D also has a prominent contribution from \nuclei{142}{La}, as well as significant background from fissioning nuclei, predominantly \nuclei{254}{Cf}, \nuclei{267}{Rf}, and \nuclei{273}{Rf}, as well as several nuclei (\nuclei{86}{Br}, \nuclei{88}{Br} , \nuclei{92}{Rb}) which are fission byproducts. These fission byproducts have very short lifetimes ($\sim$ seconds to minutes) but can appear on very long timescales due to the long lifetime of the fissioning nucleus.

At one day, the lower energy region of Sim A and B is already mostly composed of only a few nuclei: \nuclei{72}{Zn}, \nuclei{72}{Ga}, and \nuclei{77}{Ge}, though the prominence of these features are significantly less for Sim A. The most prominent spectral features of Sim C: \nuclei{127}{Sb}, \nuclei{128}{Sb}, \nuclei{129}{Sb}, \nuclei{131}{I}, and \nuclei{132}{Te} are the most significant spectral features across all nuclei. Once again, Sim D has a much stronger background from the large number of decaying nuclei, and thus the spectral lines are much less prominent. Sim D additionally includes higher energy lines from \nuclei{24}{Na} and \nuclei{208}{Tl}. The higher energy region ($\gtrapprox$ 3 MeV) is similar to the 6 hour case, though the prominence of Sim A, B, and C are significantly lower relative to Sim D.

By one week, the prominence of Sim A and B are greatly diminished, with only a few strong spectral lines able to compete with the much more significant features of Sims C and D. The nuclei driving these features are \nuclei{47}{Sc}, \nuclei{47}{Sc}, \nuclei{59}{Fe}, \nuclei{72}{Zn}, and \nuclei{78}{As}. The higher energy region of both Sim A and B ($\gtrapprox$ 1.5 MeV) is completely dominated by \nuclei{72}{Ga}.

In contrast, the spectra of Sim C and D are still significant. The spectrum of Sim C is predominantly composed of \nuclei{125}{Sn}, \nuclei{127}{Sb}, and \nuclei{132}{I}, with strong spectral lines from \nuclei{131}{I} and \nuclei{132}{Te}. Although the spectrum of Sim D is similar in overall prominence, it is composed of many more decaying nuclei, including significant contributions from the heavy elements. The most prominent spectral features of Sim D are from \nuclei{131}{I}, \nuclei{132}{Te} ,\nuclei{140}{La}, and \nuclei{208}{Tl}. At this point, the dominant fission comes from \nuclei{254}{Cf}, whose importance was shown in \cite{2018ApJ...863L..23Z}.

At one month, the $\gamma$-ray spectrum from Sim D has surpassed all but the strongest features from Sims A, B, and C. These include \nuclei{47}{Sc}, \nuclei{59}{Fe}, \nuclei{103}{Rh}, \nuclei{125}{Sn}, and \nuclei{127}{Sb}. Sim C has additional contributions from \nuclei{140}{La} and \nuclei{156}{Eu} in the higher energy region ($\sim$ 1.5-3 MeV). The prompt emission spectrum from fissioning of \nuclei{254}{Cf} is so strong in Sim D that nearly every spectral feature is reduced in prominence, though the majority of the above mentioned spectral features are still present in Sim D.

By 1 year, there are only a few nuclei which contribute to the spectra of Sim A and B, and even their contributions are not as significant as those in Sim C and D. The most notable nuclei include  \nuclei{42}{K}, \nuclei{59}{Fe}, \nuclei{106}{Rh}, \nuclei{95}{Zr}, and \nuclei{95}{Nb}. Sim C still has several nuclei which produce strong spectral features, in particular, \nuclei{125}{Sb}, but also including \nuclei{123}{Sn} , \nuclei{144}{Ce} , \nuclei{144}{Pr}, \nuclei{155}{Eu}, and \nuclei{188}{Re}. Sim D is still washed out by the prompt fission spectra--the lower prominence of \nuclei{254}{Cf} (with a half-life of 60 days) is compensated not only by the lower overall spectra, but also by the increasing prominence of \nuclei{252}{Cf}. Many of the spectral features prominent in Sim C are still prominent in Sim D, though relatively less so. 

By 50 years, there are only a few remaining significant features for Sim A, B, and C: \nuclei{42}{K}, \nuclei{84}{Kr}, \nuclei{125}{Sb}, \nuclei{126}{Sb}, \nuclei{155}{Eu}, \nuclei{194}{Ir}, and \nuclei{208}{Tl}, which has re-emerged due to its production from the $\alpha$-decay of the long-lived \nuclei{228}{Ra}. Sim D is no longer as washed out by the prompt fission spectra, which is now dominantly produced from \nuclei{250}{Cm}, \nuclei{262}{Fm}, and \nuclei{265}{No}. There are a large number of spectral features in the lower energies ($\lessapprox$ 3 MeV), primarily generated either directly from the $\alpha$-decay of long-lived nuclei (e.g. \nuclei{251}{Cf}), or nuclei which are produced from the $\alpha$-decay chains of these long-lived nuclei (e.g. \nuclei{214}{Bi}).

By 1000 years, the only significant feature of Sim A and B is the two strong spectral lines of \nuclei{60}{Co}. Sim C has strong contributions from \nuclei{126}{Sb}, even surpassing Sim D for its most prominent spectral features due to the higher abundance of A=126. For Sim D, the high energy spectrum ($\gtrapprox$ 1 MeV) is dominantly produced by the prompt fission of \nuclei{250}{Cm}, as well as the contributions of \nuclei{214}{Bi}. However, the lower energy spectrum is still quite diverse, with contributions from many nuclei, most prominently \nuclei{214}{Pb}, \nuclei{239}{Np}, \nuclei{246}{Am}, and  \nuclei{249}{Cf}.

By 50,000 years, the long lifetime of \nuclei{60}{Fe} (2.62 Myr) has allowed the spectral lines of \nuclei{60}{Co} in Sim A and B to grow in prominence, surpassing the spectra of Sim C and D. Sim C is almost completely composed of the spectral emission of \nuclei{126}{Sb}. Sim D is similar to its emission profile at 1000 years, though its overall prominence relative to Sims A, B, and C is diminished. At this point, prompt fission emission is dominantly \nuclei{248}{Cm} and \nuclei{250}{Cm}, and the relative prominence of the other spectral features are diminished, though we still see contributions from \nuclei{126}{Sb}, \nuclei{209}{Tl}, \nuclei{214}{Pb}, \nuclei{214}{Bi}, \nuclei{246}{Pu}, \nuclei{246}{Am}, and \nuclei{250}{Bk}. 

At very late times, there are additional long-lived nuclei which have decay timescales hundreds of thousands to millions of years. Such nuclei are unlikely to be directly observable in a remnant object as the flux is too low. As an example of this, we consider the incident flux from a remnant object at a given energy:
\begin{equation}
F = \frac{S^\gamma(E_\gamma) E_\gamma w_{bin} M_{ejecta}}{4\pi D^2 M_{N}},
\end{equation}
where $S^\gamma(E_\gamma)$ is the spectral emission at energy $E_\gamma$, $w_{bin}$ is the bin width, $M_{ejecta}$ is the ejecta mass, D is the distance to source, and $M_N = 1.67 \times 10^{-24}$ g is the nucleon mass. Taking optimistic values of $S(E) =$ 10$^6$ s$^{-1}$ MeV$^{-1}$ g$^{-1}$, $M_{ejecta} = 0.1 M_\odot$, and $D = 8$ kpc at E = 1 MeV ($w_{bin} = 0.023$ MeV), we find $F = 6.0 \times 10^{-10} $ MeV s$^{-1}$ cm$^{-2}$, approximately three orders of magnitude below the sensitivity thresholds of current and future $\gamma$-ray observatories (e.g. \cite{2019BAAS...51g.123M}).\\
While such nuclei are unlikely to be directly observable from a remnant object, they may be a source of an observable diffuse background. The observability of such a background is dependent on the frequency and magnitude of nucleosynthetic events, and has been studied by others (e.g. \cite{1998ApJ...506..868Q,2019ApJ...880...23W,Terada:2022hut}).
\subsection{Integrated Spectra}
\begin{figure*}[t]
\plotone{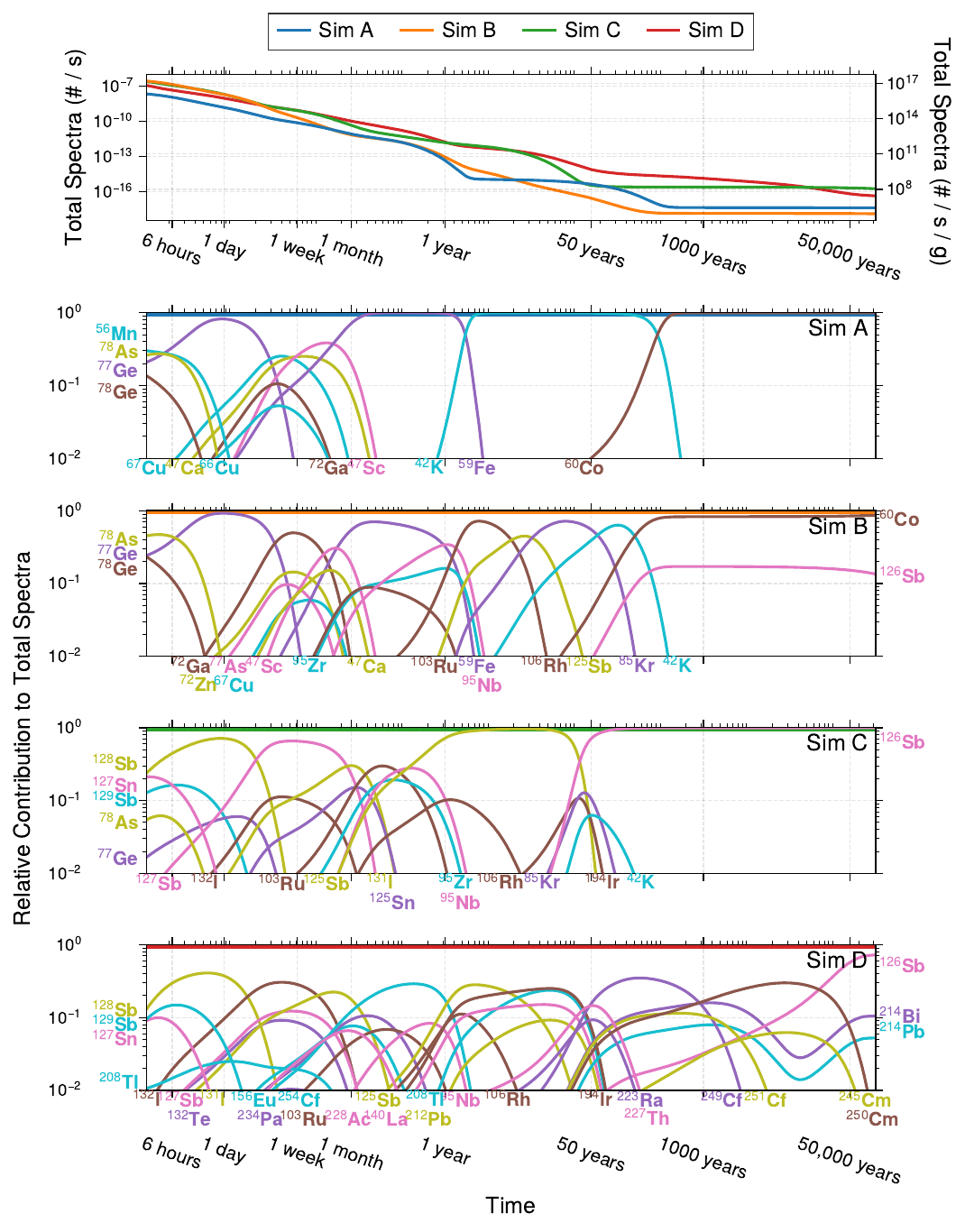}
\caption{The top panel compares the integrated spectra across the energy range E $>$ 0.1 MeV for each of our four simulations as a function of time, in units of (\#/s) (per nucleon) (left axis) and (\#/s/g) (right axis). In each of the bottom 4 panels, the fractional contributions to the corresponding spectrum from individual nuclei are shown. 
\label{fig:integrated}}
\end{figure*}

In Fig. \ref{fig:integrated}, we compare the integrated spectrum (E $>$ 0.1 MeV) for each of our simulations as a function of time, as well as show the spectral decomposition of each of these integrated spectra. We note that each simulation has a different characteristic shape, but the overall magnitude of the $\gamma$-ray spectra are comparable for the first $\sim$ year. After that, there is a general divergence, with Sim C and D having many more longer-lived nuclei and thus a larger spectra than Sim A and B. The characteristic bumps in each spectrum can be easily matched to the corresponding nuclei which dominates the spectra. For example, the bump around 50 years for Sim A which causes the spectrum of Sim A to be significantly larger than Sim B is due to the higher amount of $\nuclei{42}{K}$, which decays on a 32.9 year timescale due to the lifetime of its parent, \nuclei{42}{Ar}.

In contrast to the complexity of the energy-dependent spectra, where over a hundred nuclei emit a line strong enough to potentially be significant, the integrated spectra are dominated by a small number of nuclei. This suggests that one could perhaps search for the spectral patterns of these individual nuclei. For example, \nuclei{125}{Sb} is a dominant contributor to Sim C between 1 and 50 years, with several strong $\gamma$-ray lines between 100 keV and 1 MeV. If these $\gamma$-ray lines are observed in ratios consistent with experimental measurements, then identification of production of this nucleus can be made.

\subsection{Low Energy Spectra}
\begin{figure*}[t]
\plotone{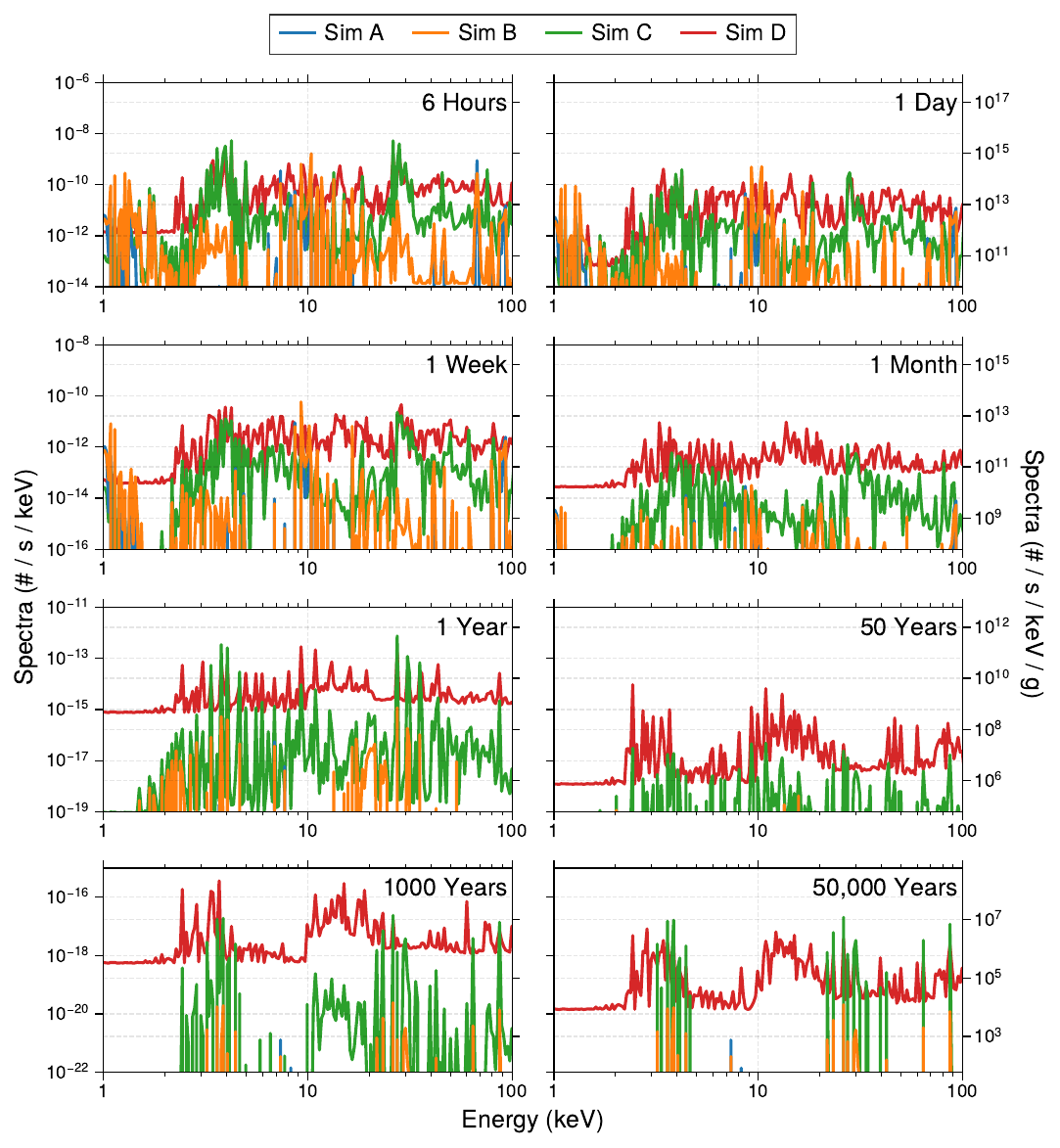}
\caption{Each panel gives the comparison of the spectra for the four representative simulations at one of the eight representative timescales for the energy region $E <$ 0.1 MeV, in units of (\#/s/MeV) (per nucleon) (left axis) and (\#/s/MeV/g) (right axis).
\label{fig:comparelowe}}
\end{figure*}

While we have focused on the higher energy regime (E $>$ 0.1 MeV), there are also potential observational signatures in the lower energy regimes, especially on longer timescales. In Fig. \ref{fig:comparelowe}, we compare the $\gamma$-ray emission profiles for 0.001 $<$ E $<$ 0.1 MeV at each representative timescale. It can be seen that there are significant differences between the $r$-process scenarios, and that each scenario has specific, prominent lines across broad timescales. These lines may be observable by current or future $x$-ray telescopes, providing a complimentary observable to potential $\gamma$-ray signatures, though a more detailed analysis of these signatures and the relevant backgrounds are required.

\section{Discussion}
\label{sec:discuss}
Our results above not only provide a detailed accounting of the spectral contributions in different $r$-process scenarios, but they also highlight several important details which are crucial for the process of connecting observational signals to $r$-process physics.

First, the calculations that we have performed give only the $\gamma$-rays resultant from r-process decays. In realistic astrophysical scenarios, there are potentially significant background sources which obscure the $r$-process signal, as well as alternative explanations to significant $\gamma$-ray signals such as cosmic ray excitation. Therefore, it may be challenging to identify with certainty that observations are a consequence of $r$-process nucleosynthesis. In addition, $\gamma$-rays will undergo doppler broadening due to the expansion of the ejecta. This effect can be significant at early times ( $v_{ejecta} \gtrapprox$ 0.1 c), and relevant even at much longer timescales ($v_{ejecta} \sim 0.001 - 0.01$ c at 10,000 years) \citep{2020ApJ...889..168K}. This makes it more difficult to identify spectral features, especially when combined with the large number of nuclei which contribute $\gamma$-ray signals.

Second, the most observable spectral contributions are not necessarily the dominant spectral contributions for a nucleus, because the strength of a spectral line must be considered relative to the $\gamma$-ray production from other nuclei. For example, the decay of \nuclei{144}{Pr} produces (among other lines) a 696.5 keV line 1.34\% of the time and a 2186 keV line 0.69\% of the time. The 2186 keV line is prominent in Sim C, while the 696.5 keV line is completely superseded by the much more prominent lines emitted by \nuclei{125}{Sb} and \nuclei{95}{Zr}.  

Third, prominent spectral lines which have been referenced in the literature are not necessarily the most prominent sources of $\gamma$-rays at that line energy. For example, the 1384 keV line produced by \nuclei{92}{Sr} has been noted as a potential $\gamma$-ray signature in multiple works \citep{2020ApJ...889..168K,2025ApJ...984L..29P}. In practice, this line is very close to the 1369 keV line of \nuclei{24}{Na} and the 1374 keV line of \nuclei{78}{As}. In Sims A and B, \nuclei{78}{As} is dominant, in Sim C, \nuclei{92}{Sr}, and in Sim D, \nuclei{24}{Na}. Taken with the possible spectral broadening, this makes the determination of the nuclei responsible for an observed $\sim$ 1370 keV peak a challenge. 

Fourth, the spectral lines which are prominent may change significantly based on the detailed physics of the $r$-process that occurs. The prominence of a particular spectral line is mostly tied to the relative production of the corresponding mass number (though things are much more involved for the actinide region). Different proposed $r$-process sites impart different conditions, and when combined with the inherent uncertainties, the detailed production patterns for different characteristic $r$-process types are not well-understood. This adds to the challenge of identifying nuclei responsible for significant spectral features, as these features could be the result of a nucleus which is dominant in the $r$-process which occurred observationally, but subdominant in the $r$-process model which we have used. However, if nuclei can be confidently identified through detailed observational modeling and spectral analysis, then much can be learned about the $r$-process.

Many of the nuclei with significant spectral features also have long-lived nuclear isomers. These isomers may be significantly populated, affecting the timescale of $\beta$-decay, and may also directly $\beta$-decay, producing a different $\gamma$-ray spectra.  Both of these will change the observability of spectral features. Understanding which isomers are populated requires detailed understanding of the overall nuclear structure, as transitions through all possible nuclear states (not just the isomers) must be considered (see e.g. \cite{2021ApJS..252....2M}). One nucleus significantly affected is \nuclei{128}{Sb}, which has an isomer with excitation energy 43.9 keV \citep{PhysRevLett.131.262701} with $\beta$-decay lifetime 10.8 minutes (ground state lifetime is 9 hours), but we also expect that there will be many other affected nuclei.

For very strong $r$-processes, the large amount of uncertainty in the nuclear data in the very high mass region creates additional uncertainty in the resultant spectra. The extended $r$-process (Sim D) produces a significant amount of high mass material (A$>$250). In this region it is generally possible for nuclei to decay via $\alpha$, $\beta^-$, $\beta^+$, and fission, with the exact branchings for an individual nucleus not well measured. The lifetimes of these nuclei (and thus the timescale which their children emit $\gamma$-rays) are also poorly understood. As a consequence, the prominence of spectral features from nuclei in the decay chain of these heavy mass nuclei is uncertain. As a prime example of this,  \nuclei{250}{Cm} has a lifetime of $\sim$ 8300 years. with uncertain decay branching, estimated at 8\% $\beta^{-}$, 18\% $\alpha$, and 74\% fission. All three decay modes contribute significant spectral features: the prompt fission spectrum of \nuclei{250}{Cm} dominates the fission spectra from $\sim$ 50 to $\sim$ 50,000 years, \nuclei{250}{Bk} has spectral lines at 989 and 1032 keV, and both \nuclei{246}{Am} and its child \nuclei{246}{Pu} have spectral lines between 100 and 200 keV. With a different branching ratio, the relative prominence of these spectral features could be significantly affected.

There is also much uncertainty surrounding the details of fissioning nuclei. For extended $r$-processes, the strength of the prompt fission spectrum is competitive with the spectral features of $\beta$-decay at higher energies. However, the prompt fission spectrum that is used for all nuclei is a reference curve based on the fission of \nuclei{252}{Cf}. In addition, fission produces short-lived nuclei with strong $\gamma$-ray lines, such as \nuclei{97}{Y}. The strength of these spectral lines is mediated by the probability that the given nucleus fissions into a fragment with the correct A value. Taken in concert with the uncertainties in the prompt fission spectra, it is unclear if the spectral features of these short-lived nuclei will be observable above the prompt fission background.

Given the aforementioned challenges, observational searches could instead consider searching for a template spectrum. However, such an approach has its own difficulties, notably that since $r$-process events are a mixing of many different nucleosynthetic conditions, the detailed spectral features of such a template are unclear. As mentioned above, it may be possible to identify the characteristic spectral features of the most dominant nuclei. In practice, a template will be muddied by significant lines of other nuclei as well as doppler broadening of ejecta. 
\section{Conclusion}
\label{sec:conclude}
We have analyzed the $\gamma$-ray emission from the decay of $r$-process nuclei across a broad timescale for 4 different characteristic $r$-process types: a limited, weak, strong, and extended $r$-process. For each of these profiles, we have identified the nuclei which are responsible for significant spectral features, which we have tabulated in Table \ref{tab:gammaray}. We find that a large number of nuclei are capable of producing prominent spectral lines. The relative prominence of these lines can be influenced by a variety of factors, including the strength of $r$-process, the site which the $r$-process occurs at (which affects the detailed abundance pattern), and uncertainties in both the nuclear data input and the $\gamma$-ray emission spectra, especially in the actinide and super-heavy region. In addition, spectral broadening and the contributions from other background sources of $\gamma$-rays at the site will make it challenging to identify with certainty the nuclei responsible for spectral features from the plethora of possible emission lines at a given energy. These factors emphasize the importance of detailed modeling of both the $\gamma$-ray emission spectra and the environment in which the $r$-process occurs. However, if these spectral features can be identified, much can be learned about nature of the $r$-process. With the increase in observations of $r$-process $\gamma$-rays from next-generation $\gamma$-ray observatories, the provided reference tables will aid in the identification of the nuclear species responsible for observed spectral features and enable us to learn about the nature of the $r$-process which has occurred.

\begin{acknowledgements}
 LANL is operated by Triad National Security, LLC, for the National Nuclear Security Administration of U.S. Department of Energy (Contract No. 89233218CNA000001). 
Research presented in this article was supported by the Laboratory Directed Research and Development program of Los Alamos National Laboratory under project numbers 20230052ER and 20240004DR.
\end{acknowledgements}

\bibliography{refs}{}

@ARTICLE{2023A&ARv..31....1A,
       author = {{Arcones}, Almudena and {Thielemann}, Friedrich-Karl},
        title = "{Origin of the elements}",
      journal = {\aapr},
         year = 2023,
        month = dec,
       volume = {31},
       number = {1},
          eid = {1},
        pages = {1},
          doi = {10.1007/s00159-022-00146-x},
       adsurl = {https://ui.adsabs.harvard.edu/abs/2023A&ARv..31....1A}
}

@ARTICLE{1957RvMP...29..547B,
       author = {{Burbidge}, E. Margaret and {Burbidge}, G.~R. and {Fowler}, William A. and {Hoyle}, F.},
        title = "{Synthesis of the Elements in Stars}",
      journal = {Reviews of Modern Physics},
         year = 1957,
        month = oct,
       volume = {29},
       number = {4},
        pages = {547-650},
          doi = {10.1103/RevModPhys.29.547},
       adsurl = {https://ui.adsabs.harvard.edu/abs/1957RvMP...29..547B}
}

@ARTICLE{1996ApJ...471..331Q,
       author = {{Qian}, Y. -Z. and {Woosley}, S.~E.},
        title = "{Nucleosynthesis in Neutrino-driven Winds. I. The Physical Conditions}",
      journal = {\apj},
         year = 1996,
        month = nov,
       volume = {471},
        pages = {331},
          doi = {10.1086/177973},
archivePrefix = {arXiv},
       eprint = {astro-ph/9611094},
 primaryClass = {astro-ph},
       adsurl = {https://ui.adsabs.harvard.edu/abs/1996ApJ...471..331Q}
}

@ARTICLE{2019Natur.569..241S,
       author = {{Siegel}, Daniel M. and {Barnes}, Jennifer and {Metzger}, Brian D.},
        title = "{Collapsars as a major source of r-process elements}",
      journal = {\nat},
         year = 2019,
        month = may,
       volume = {569},
       number = {7755},
        pages = {241-244},
          doi = {10.1038/s41586-019-1136-0},
archivePrefix = {arXiv},
       eprint = {1810.00098},
 primaryClass = {astro-ph.HE},
       adsurl = {https://ui.adsabs.harvard.edu/abs/2019Natur.569..241S}
}

@ARTICLE{2015ApJ...810..109N,
       author = {{Nishimura}, Nobuya and {Takiwaki}, Tomoya and {Thielemann}, Friedrich-Karl},
        title = "{The r-process Nucleosynthesis in the Various Jet-like Explosions of Magnetorotational Core-collapse Supernovae}",
      journal = {\apj},
         year = 2015,
        month = sep,
       volume = {810},
       number = {2},
          eid = {109},
        pages = {109},
          doi = {10.1088/0004-637X/810/2/109},
archivePrefix = {arXiv},
       eprint = {1501.06567},
 primaryClass = {astro-ph.SR},
       adsurl = {https://ui.adsabs.harvard.edu/abs/2015ApJ...810..109N}
}

@ARTICLE{2012ApJ...750L..22W,
       author = {{Winteler}, C. and {K{\"a}ppeli}, R. and {Perego}, A. and {Arcones}, A. and {Vasset}, N. and {Nishimura}, N. and {Liebend{\"o}rfer}, M. and {Thielemann}, F. -K.},
        title = "{Magnetorotationally Driven Supernovae as the Origin of Early Galaxy r-process Elements?}",
      journal = {\apjl},
         year = 2012,
        month = may,
       volume = {750},
       number = {1},
          eid = {L22},
        pages = {L22},
          doi = {10.1088/2041-8205/750/1/L22},
archivePrefix = {arXiv},
       eprint = {1203.0616},
 primaryClass = {astro-ph.SR},
       adsurl = {https://ui.adsabs.harvard.edu/abs/2012ApJ...750L..22W}
}

@ARTICLE{2021Natur.595..223Y,
       author = {{Yong}, D. and {Kobayashi}, C. and {Da Costa}, G.~S. and {Bessell}, M.~S. and {Chiti}, A. and {Frebel}, A. and {Lind}, K. and {Mackey}, A.~D. and {Nordlander}, T. and {Asplund}, M. and {Casey}, A.~R. and {Marino}, A.~F. and {Murphy}, S.~J. and {Schmidt}, B.~P.},
        title = "{r-Process elements from magnetorotational hypernovae}",
      journal = {\nat},
         year = 2021,
        month = jul,
       volume = {595},
       number = {7866},
        pages = {223-226},
          doi = {10.1038/s41586-021-03611-2},
archivePrefix = {arXiv},
       eprint = {2107.03010},
 primaryClass = {astro-ph.SR},
       adsurl = {https://ui.adsabs.harvard.edu/abs/2021Natur.595..223Y}
}

@ARTICLE{1994ApJ...433..229W,
       author = {{Woosley}, S.~E. and {Wilson}, J.~R. and {Mathews}, G.~J. and {Hoffman}, R.~D. and {Meyer}, B.~S.},
        title = "{The r-Process and Neutrino-heated Supernova Ejecta}",
      journal = {\apj},
         year = 1994,
        month = sep,
       volume = {433},
        pages = {229},
          doi = {10.1086/174638},
       adsurl = {https://ui.adsabs.harvard.edu/abs/1994ApJ...433..229W}
}

@ARTICLE{2001ApJ...554..578W,
       author = {{Wanajo}, Shinya and {Kajino}, Toshitaka and {Mathews}, Grant J. and {Otsuki}, Kaori},
        title = "{The r-Process in Neutrino-driven Winds from Nascent, ``Compact'' Neutron Stars of Core-Collapse Supernovae}",
      journal = {\apj},
         year = 2001,
        month = jun,
       volume = {554},
       number = {1},
        pages = {578-586},
          doi = {10.1086/321339},
archivePrefix = {arXiv},
       eprint = {astro-ph/0102261},
 primaryClass = {astro-ph},
       adsurl = {https://ui.adsabs.harvard.edu/abs/2001ApJ...554..578W}
}

@ARTICLE{2006ApJ...646L.131F,
       author = {{Fryer}, Christopher L. and {Herwig}, Falk and {Hungerford}, Aimee and {Timmes}, F.~X.},
        title = "{Supernova Fallback: A Possible Site for the r-Process}",
      journal = {\apjl},
         year = 2006,
        month = aug,
       volume = {646},
       number = {2},
        pages = {L131-L134},
          doi = {10.1086/507071},
archivePrefix = {arXiv},
       eprint = {astro-ph/0606450},
 primaryClass = {astro-ph},
       adsurl = {https://ui.adsabs.harvard.edu/abs/2006ApJ...646L.131F}
}

@article{PhysRevD.87.024001,
  title = {Mass ejection from the merger of binary neutron stars},
  author = {Hotokezaka, Kenta and Kiuchi, Kenta and Kyutoku, Koutarou and Okawa, Hirotada and Sekiguchi, Yu-ichiro and Shibata, Masaru and Taniguchi, Keisuke},
  journal = {Phys. Rev. D},
  volume = {87},
  issue = {2},
  pages = {024001},
  numpages = {27},
  year = {2013},
  month = {Jan},
  publisher = {American Physical Society},
  doi = {10.1103/PhysRevD.87.024001},
  url = {https://link.aps.org/doi/10.1103/PhysRevD.87.024001}
}

@ARTICLE{2011ApJ...738L..32G,
       author = {{Goriely}, Stephane and {Bauswein}, Andreas and {Janka}, Hans-Thomas},
        title = "{r-process Nucleosynthesis in Dynamically Ejected Matter of Neutron Star Mergers}",
      journal = {\apjl},
         year = 2011,
        month = sep,
       volume = {738},
       number = {2},
          eid = {L32},
        pages = {L32},
          doi = {10.1088/2041-8205/738/2/L32},
archivePrefix = {arXiv},
       eprint = {1107.0899},
 primaryClass = {astro-ph.SR},
       adsurl = {https://ui.adsabs.harvard.edu/abs/2011ApJ...738L..32G}
}

@ARTICLE{1989Natur.340..126E,
       author = {{Eichler}, David and {Livio}, Mario and {Piran}, Tsvi and {Schramm}, David N.},
        title = "{Nucleosynthesis, neutrino bursts and {\ensuremath{\gamma}}-rays from coalescing neutron stars}",
      journal = {\nat},
         year = 1989,
        month = jul,
       volume = {340},
       number = {6229},
        pages = {126-128},
          doi = {10.1038/340126a0},
       adsurl = {https://ui.adsabs.harvard.edu/abs/1989Natur.340..126E}
}

@ARTICLE{1999ApJ...525L.121F,
       author = {{Freiburghaus}, C. and {Rosswog}, S. and {Thielemann}, F. -K.},
        title = "{R-Process in Neutron Star Mergers}",
      journal = {\apjl},
         year = 1999,
        month = nov,
       volume = {525},
       number = {2},
        pages = {L121-L124},
          doi = {10.1086/312343},
       adsurl = {https://ui.adsabs.harvard.edu/abs/1999ApJ...525L.121F}
}

@ARTICLE{1974ApJ...192L.145L,
       author = {{Lattimer}, J.~M. and {Schramm}, D.~N.},
        title = "{Black-Hole-Neutron-Star Collisions}",
      journal = {\apjl},
         year = 1974,
        month = sep,
       volume = {192},
        pages = {L145},
          doi = {10.1086/181612},
       adsurl = {https://ui.adsabs.harvard.edu/abs/1974ApJ...192L.145L}
}

@ARTICLE{2012MNRAS.426.1940K,
       author = {{Korobkin}, O. and {Rosswog}, S. and {Arcones}, A. and {Winteler}, C.},
        title = "{On the astrophysical robustness of the neutron star merger r-process}",
      journal = {\mnras},
         year = 2012,
        month = nov,
       volume = {426},
       number = {3},
        pages = {1940-1949},
          doi = {10.1111/j.1365-2966.2012.21859.x},
archivePrefix = {arXiv},
       eprint = {1206.2379},
 primaryClass = {astro-ph.SR},
       adsurl = {https://ui.adsabs.harvard.edu/abs/2012MNRAS.426.1940K}
}

@ARTICLE{2013ApJ...773...78B,
       author = {{Bauswein}, A. and {Goriely}, S. and {Janka}, H. -T.},
        title = "{Systematics of Dynamical Mass Ejection, Nucleosynthesis, and Radioactively Powered Electromagnetic Signals from Neutron-star Mergers}",
      journal = {\apj},
         year = 2013,
        month = aug,
       volume = {773},
       number = {1},
          eid = {78},
        pages = {78},
          doi = {10.1088/0004-637X/773/1/78},
archivePrefix = {arXiv},
       eprint = {1302.6530},
 primaryClass = {astro-ph.SR},
       adsurl = {https://ui.adsabs.harvard.edu/abs/2013ApJ...773...78B}
}

@ARTICLE{2014ApJ...789L..39W,
       author = {{Wanajo}, Shinya and {Sekiguchi}, Yuichiro and {Nishimura}, Nobuya and {Kiuchi}, Kenta and {Kyutoku}, Koutarou and {Shibata}, Masaru},
        title = "{Production of All the r-process Nuclides in the Dynamical Ejecta of Neutron Star Mergers}",
      journal = {\apjl},
         year = 2014,
        month = jul,
       volume = {789},
       number = {2},
          eid = {L39},
        pages = {L39},
          doi = {10.1088/2041-8205/789/2/L39},
archivePrefix = {arXiv},
       eprint = {1402.7317},
 primaryClass = {astro-ph.SR},
       adsurl = {https://ui.adsabs.harvard.edu/abs/2014ApJ...789L..39W}
}

@ARTICLE{1999A&A...341..499R,
       author = {{Rosswog}, S. and {Liebend{\"o}rfer}, M. and {Thielemann}, F. -K. and {Davies}, M.~B. and {Benz}, W. and {Piran}, T.},
        title = "{Mass ejection in neutron star mergers}",
      journal = {\aap},
         year = 1999,
        month = jan,
       volume = {341},
        pages = {499-526},
          doi = {10.48550/arXiv.astro-ph/9811367},
archivePrefix = {arXiv},
       eprint = {astro-ph/9811367},
 primaryClass = {astro-ph},
       adsurl = {https://ui.adsabs.harvard.edu/abs/1999A&A...341..499R}
}

@ARTICLE{2017ApJ...848L..12A,
       author = {{Abbott}, B.~P. and {Abbott}, R. and {Abbott}, T.~D. and {Acernese}, F. and {Ackley}, K. and {Adams}, C. and {Adams}, T. and {Addesso}, P. and {Adhikari}, R.~X. and {Adya}, V.~B. and {Affeldt}, C. and {Afrough}, M. and {Agarwal}, B. and {Agathos}, M. and {Agatsuma}, K. and {Aggarwal}, N. and {Aguiar}, O.~D. and {Aiello}, L. and {Ain}, A. and {Ajith}, P. and {Allen}, B. and {Allen}, G. and {Allocca}, A. and {Altin}, P.~A. and {Amato}, A. and {Ananyeva}, A. and {Anderson}, S.~B. and {Anderson}, W.~G. and {Angelova}, S.~V. and {Antier}, S. and {Appert}, S. and {Arai}, K. and {Araya}, M.~C. and {Areeda}, J.~S. and {Arnaud}, N. and {Arun}, K.~G. and {Ascenzi}, S. and {Ashton}, G. and {Ast}, M. and {Aston}, S.~M. and {Astone}, P. and {Atallah}, D.~V. and {Aufmuth}, P. and {Aulbert}, C. and {AultONeal}, K. and {Austin}, C. and {Avila-Alvarez}, A. and {Babak}, S. and {Bacon}, P. and {Bader}, M.~K.~M. and {Bae}, S. and {Baker}, P.~T. and {Baldaccini}, F. and {Ballardin}, G. and {Ballmer}, S.~W. and {Banagiri}, S. and {Barayoga}, J.~C. and {Barclay}, S.~E. and {Barish}, B.~C. and {Barker}, D. and {Barkett}, K. and {Barone}, F. and {Barr}, B. and {Barsotti}, L. and {Barsuglia}, M. and {Barta}, D. and {Barthelmy}, S.~D. and {Bartlett}, J. and {Bartos}, I. and {Bassiri}, R. and {Basti}, A. and {Batch}, J.~C. and {Bawaj}, M. and {Bayley}, J.~C. and {Bazzan}, M. and {B{\'e}csy}, B. and {Beer}, C. and {Bejger}, M. and {Belahcene}, I. and {Bell}, A.~S. and {Berger}, B.~K. and {Bergmann}, G. and {Bero}, J.~J. and {Berry}, C.~P.~L. and {Bersanetti}, D. and {Bertolini}, A. and {Betzwieser}, J. and {Bhagwat}, S. and {Bhandare}, R. and {Bilenko}, I.~A. and {Billingsley}, G. and {Billman}, C.~R. and {Birch}, J. and {Birney}, R. and {Birnholtz}, O. and {Biscans}, S. and {Biscoveanu}, S. and {Bisht}, A. and {Bitossi}, M. and {Biwer}, C. and {Bizouard}, M.~A. and {Blackburn}, J.~K. and {Blackman}, J. and {Blair}, C.~D. and {Blair}, D.~G. and {Blair}, R.~M. and {Bloemen}, S. and {Bock}, O. and {Bode}, N. and {Boer}, M. and {Bogaert}, G. and {Bohe}, A. and {Bondu}, F. and {Bonilla}, E. and {Bonnand}, R. and {Boom}, B.~A. and {Bork}, R. and {Boschi}, V. and {Bose}, S. and {Bossie}, K. and {Bouffanais}, Y. and {Bozzi}, A. and {Bradaschia}, C. and {Brady}, P.~R. and {Branchesi}, M. and {Brau}, J.~E. and {Briant}, T. and {Brillet}, A. and {Brinkmann}, M. and {Brisson}, V. and {Brockill}, P. and {Broida}, J.~E. and {Brooks}, A.~F. and {Brown}, D.~A. and {Brown}, D.~D. and {Brunett}, S. and {Buchanan}, C.~C. and {Buikema}, A. and {Bulik}, T. and {Bulten}, H.~J. and {Buonanno}, A. and {Buskulic}, D. and {Buy}, C. and {Byer}, R.~L. and {Cabero}, M. and {Cadonati}, L. and {Cagnoli}, G. and {Cahillane}, C. and {Calder{\'o}n Bustillo}, J. and {Callister}, T.~A. and {Calloni}, E. and {Camp}, J.~B. and {Canepa}, M. and {Canizares}, P. and {Cannon}, K.~C. and {Cao}, H. and {Cao}, J. and {Capano}, C.~D. and {Capocasa}, E. and {Carbognani}, F. and {Caride}, S. and {Carney}, M.~F. and {Casanueva Diaz}, J. and {Casentini}, C. and {Caudill}, S. and {Cavagli{\`a}}, M. and {Cavalier}, F. and {Cavalieri}, R. and {Cella}, G. and {Cepeda}, C.~B. and {Cerd{\'a}-Dur{\'a}n}, P. and {Cerretani}, G. and {Cesarini}, E. and {Chamberlin}, S.~J. and {Chan}, M. and {Chao}, S. and {Charlton}, P. and {Chase}, E. and {Chassande-Mottin}, E. and {Chatterjee}, D. and {Chatziioannou}, K. and {Cheeseboro}, B.~D. and {Chen}, H.~Y. and {Chen}, X. and {Chen}, Y. and {Cheng}, H. -P. and {Chia}, H. and {Chincarini}, A. and {Chiummo}, A. and {Chmiel}, T. and {Cho}, H.~S. and {Cho}, M. and {Chow}, J.~H. and {Christensen}, N. and {Chu}, Q. and {Chua}, A.~J.~K. and {Chua}, S. and {Chung}, A.~K.~W. and {Chung}, S. and {Ciani}, G.},
        title = "{Multi-messenger Observations of a Binary Neutron Star Merger}",
      journal = {\apjl},
         year = 2017,
        month = oct,
       volume = {848},
       number = {2},
          eid = {L12},
        pages = {L12},
          doi = {10.3847/2041-8213/aa91c9},
archivePrefix = {arXiv},
       eprint = {1710.05833},
 primaryClass = {astro-ph.HE},
       adsurl = {https://ui.adsabs.harvard.edu/abs/2017ApJ...848L..12A}
}

@ARTICLE{2016ApJ...829..110B,
       author = {{Barnes}, Jennifer and {Kasen}, Daniel and {Wu}, Meng-Ru and {Mart{\'\i}nez-Pinedo}, Gabriel},
        title = "{Radioactivity and Thermalization in the Ejecta of Compact Object Mergers and Their Impact on Kilonova Light Curves}",
      journal = {\apj},
         year = 2016,
        month = oct,
       volume = {829},
       number = {2},
          eid = {110},
        pages = {110},
          doi = {10.3847/0004-637X/829/2/110},
archivePrefix = {arXiv},
       eprint = {1605.07218},
 primaryClass = {astro-ph.HE},
       adsurl = {https://ui.adsabs.harvard.edu/abs/2016ApJ...829..110B}
}

@ARTICLE{2021ApJ...906...94Z,
       author = {{Zhu}, Y.~L. and {Lund}, K.~A. and {Barnes}, J. and {Sprouse}, T.~M. and {Vassh}, N. and {McLaughlin}, G.~C. and {Mumpower}, M.~R. and {Surman}, R.},
        title = "{Modeling Kilonova Light Curves: Dependence on Nuclear Inputs}",
      journal = {\apj},
         year = 2021,
        month = jan,
       volume = {906},
       number = {2},
          eid = {94},
        pages = {94},
          doi = {10.3847/1538-4357/abc69e},
archivePrefix = {arXiv},
       eprint = {2010.03668},
 primaryClass = {astro-ph.HE},
       adsurl = {https://ui.adsabs.harvard.edu/abs/2021ApJ...906...94Z}
}

@ARTICLE{2021ApJ...910..116K,
       author = {{Korobkin}, Oleg and {Wollaeger}, Ryan T. and {Fryer}, Christopher L. and {Hungerford}, Aimee L. and {Rosswog}, Stephan and {Fontes}, Christopher J. and {Mumpower}, Matthew R. and {Chase}, Eve A. and {Even}, Wesley P. and {Miller}, Jonah and {Misch}, G. Wendell and {Lippuner}, Jonas},
        title = "{Axisymmetric Radiative Transfer Models of Kilonovae}",
      journal = {\apj},
         year = 2021,
        month = apr,
       volume = {910},
       number = {2},
          eid = {116},
        pages = {116},
          doi = {10.3847/1538-4357/abe1b5},
archivePrefix = {arXiv},
       eprint = {2004.00102},
 primaryClass = {astro-ph.HE},
       adsurl = {https://ui.adsabs.harvard.edu/abs/2021ApJ...910..116K}
}

@article{MUMPOWER2025101736,
title = {Nuclear β−-decay with statistical de-excitation},
journal = {Atomic Data and Nuclear Data Tables},
pages = {101736},
year = {2025b},
issn = {0092-640X},
doi = {https://doi.org/10.1016/j.adt.2025.101736},
url = {https://www.sciencedirect.com/science/article/pii/S0092640X25000294},
author = {M.R. Mumpower and T. Kawano and O. Korobkin and G.W. Misch and T.M. Sprouse}
}

@ARTICLE{2021PhRvC.104a5803S,
       author = {{Sprouse}, T.~M. and {Mumpower}, M.~R. and {Surman}, R.},
        title = "{Following nuclei through nucleosynthesis: A novel tracing technique}",
      journal = {\prc},
         year = 2021,
        month = jul,
       volume = {104},
       number = {1},
          eid = {015803},
        pages = {015803},
          doi = {10.1103/PhysRevC.104.015803},
archivePrefix = {arXiv},
       eprint = {2008.06075},
 primaryClass = {nucl-th},
       adsurl = {https://ui.adsabs.harvard.edu/abs/2021PhRvC.104a5803S}
}

@ARTICLE{2025ApJ...982...81M,
       author = {{Mumpower}, Matthew R. and {Lee}, Tsung-Shung H. and {Lloyd-Ronning}, Nicole and {Barker}, Brandon L. and {Gross}, Axel and {Cupp}, Samuel and {Miller}, Jonah M.},
        title = "{Let There Be Neutrons! Hadronic Photoproduction from a Large Flux of High-energy Photons}",
      journal = {\apj},
         year = {2025a},
        month = apr,
       volume = {982},
       number = {2},
          eid = {81},
        pages = {81},
          doi = {10.3847/1538-4357/adb1e3},
archivePrefix = {arXiv},
       eprint = {2411.11831},
 primaryClass = {astro-ph.HE},
       adsurl = {https://ui.adsabs.harvard.edu/abs/2025ApJ...982...81M}
}

@ARTICLE{2012PhRvL.108e2501M,
       author = {{M{\"o}ller}, Peter and {Myers}, William D. and {Sagawa}, Hiroyuki and {Yoshida}, Satoshi},
        title = "{New Finite-Range Droplet Mass Model and Equation-of-State Parameters}",
      journal = {\prl},
         year = 2012,
        month = feb,
       volume = {108},
       number = {5},
          eid = {052501},
        pages = {052501},
          doi = {10.1103/PhysRevLett.108.052501},
       adsurl = {https://ui.adsabs.harvard.edu/abs/2012PhRvL.108e2501M}
}

@ARTICLE{2016ADNDT.109....1M,
       author = {{M{\"o}ller}, P. and {Sierk}, A.~J. and {Ichikawa}, T. and {Sagawa}, H.},
        title = "{Nuclear ground-state masses and deformations: FRDM(2012)}",
      journal = {Atomic Data and Nuclear Data Tables},
         year = 2016,
        month = may,
       volume = {109},
        pages = {1-204},
          doi = {10.1016/j.adt.2015.10.002},
archivePrefix = {arXiv},
       eprint = {1508.06294},
 primaryClass = {nucl-th},
       adsurl = {https://ui.adsabs.harvard.edu/abs/2016ADNDT.109....1M}
}

@ARTICLE{2019arXiv190105641K,
       author = {{Kawano}, Toshihiko},
        title = "{Unified Coupled-Channels and Hauser-Feshbach Model Calculation for Nuclear Data Evaluation}",
      journal = {arXiv e-prints},
         year = 2019,
        month = jan,
          eid = {arXiv:1901.05641},
        pages = {arXiv:1901.05641},
          doi = {10.48550/arXiv.1901.05641},
archivePrefix = {arXiv},
       eprint = {1901.05641},
 primaryClass = {nucl-th},
       adsurl = {https://ui.adsabs.harvard.edu/abs/2019arXiv190105641K}
}

@ARTICLE{2021EPJA...57...16K,
       author = {{Kawano}, Toshihiko},
        title = "{Unified description of the coupled-channels and statistical Hauser-Feshbach nuclear reaction theories for low energy neutron incident reactions}",
      journal = {European Physical Journal A},
         year = 2021,
        month = jan,
       volume = {57},
       number = {1},
          eid = {16},
        pages = {16},
          doi = {10.1140/epja/s10050-020-00311-9},
archivePrefix = {arXiv},
       eprint = {2009.12736},
 primaryClass = {nucl-th},
       adsurl = {https://ui.adsabs.harvard.edu/abs/2021EPJA...57...16K}
}

@ARTICLE{2016PhRvC..94f4317M,
       author = {{Mumpower}, M.~R. and {Kawano}, T. and {M{\"o}ller}, P.},
        title = "{Neutron-{\ensuremath{\gamma}} competition for {\ensuremath{\beta}} -delayed neutron emission}",
      journal = {\prc},
         year = 2016,
        month = dec,
       volume = {94},
       number = {6},
          eid = {064317},
        pages = {064317},
          doi = {10.1103/PhysRevC.94.064317},
archivePrefix = {arXiv},
       eprint = {1608.01956},
 primaryClass = {nucl-th},
       adsurl = {https://ui.adsabs.harvard.edu/abs/2016PhRvC..94f4317M}
}

@ARTICLE{2018ApJ...869...14M,
       author = {{Mumpower}, M.~R. and {Kawano}, T. and {Sprouse}, T.~M. and {Vassh}, N. and {Holmbeck}, E.~M. and {Surman}, R. and {M{\"o}ller}, P.},
        title = "{{\ensuremath{\beta}}-delayed Fission in r-process Nucleosynthesis}",
      journal = {\apj},
         year = 2018,
        month = dec,
       volume = {869},
       number = {1},
          eid = {14},
        pages = {14},
          doi = {10.3847/1538-4357/aaeaca},
archivePrefix = {arXiv},
       eprint = {1802.04398},
 primaryClass = {nucl-th},
       adsurl = {https://ui.adsabs.harvard.edu/abs/2018ApJ...869...14M}
}

@ARTICLE{2010ApJS..189..240C,
       author = {{Cyburt}, Richard H. and {Amthor}, A. Matthew and {Ferguson}, Ryan and {Meisel}, Zach and {Smith}, Karl and {Warren}, Scott and {Heger}, Alexander and {Hoffman}, R.~D. and {Rauscher}, Thomas and {Sakharuk}, Alexander and {Schatz}, Hendrik and {Thielemann}, F.~K. and {Wiescher}, Michael},
        title = "{The JINA REACLIB Database: Its Recent Updates and Impact on Type-I X-ray Bursts}",
      journal = {\apjs},
         year = 2010,
        month = jul,
       volume = {189},
       number = {1},
        pages = {240-252},
          doi = {10.1088/0067-0049/189/1/240},
       adsurl = {https://ui.adsabs.harvard.edu/abs/2010ApJS..189..240C}
}

@ARTICLE{2017Sci...358.1570D,
       author = {{Drout}, M.~R. and {Piro}, A.~L. and {Shappee}, B.~J. and {Kilpatrick}, C.~D. and {Simon}, J.~D. and {Contreras}, C. and {Coulter}, D.~A. and {Foley}, R.~J. and {Siebert}, M.~R. and {Morrell}, N. and {Boutsia}, K. and {Di Mille}, F. and {Holoien}, T.~W. -S. and {Kasen}, D. and {Kollmeier}, J.~A. and {Madore}, B.~F. and {Monson}, A.~J. and {Murguia-Berthier}, A. and {Pan}, Y. -C. and {Prochaska}, J.~X. and {Ramirez-Ruiz}, E. and {Rest}, A. and {Adams}, C. and {Alatalo}, K. and {Ba{\~n}ados}, E. and {Baughman}, J. and {Beers}, T.~C. and {Bernstein}, R.~A. and {Bitsakis}, T. and {Campillay}, A. and {Hansen}, T.~T. and {Higgs}, C.~R. and {Ji}, A.~P. and {Maravelias}, G. and {Marshall}, J.~L. and {Moni Bidin}, C. and {Prieto}, J.~L. and {Rasmussen}, K.~C. and {Rojas-Bravo}, C. and {Strom}, A.~L. and {Ulloa}, N. and {Vargas-Gonz{\'a}lez}, J. and {Wan}, Z. and {Whitten}, D.~D.},
        title = "{Light curves of the neutron star merger GW170817/SSS17a: Implications for r-process nucleosynthesis}",
      journal = {Science},
         year = 2017,
        month = dec,
       volume = {358},
       number = {6370},
        pages = {1570-1574},
          doi = {10.1126/science.aaq0049},
archivePrefix = {arXiv},
       eprint = {1710.05443},
 primaryClass = {astro-ph.HE},
       adsurl = {https://ui.adsabs.harvard.edu/abs/2017Sci...358.1570D}
}

@ARTICLE{2017Natur.551...80K,
       author = {{Kasen}, Daniel and {Metzger}, Brian and {Barnes}, Jennifer and {Quataert}, Eliot and {Ramirez-Ruiz}, Enrico},
        title = "{Origin of the heavy elements in binary neutron-star mergers from a gravitational-wave event}",
      journal = {\nat},
         year = 2017,
        month = nov,
       volume = {551},
       number = {7678},
        pages = {80-84},
          doi = {10.1038/nature24453},
archivePrefix = {arXiv},
       eprint = {1710.05463},
 primaryClass = {astro-ph.HE},
       adsurl = {https://ui.adsabs.harvard.edu/abs/2017Natur.551...80K}
}

@ARTICLE{2017Sci...358.1559K,
       author = {{Kasliwal}, M.~M. and {Nakar}, E. and {Singer}, L.~P. and {Kaplan}, D.~L. and {Cook}, D.~O. and {Van Sistine}, A. and {Lau}, R.~M. and {Fremling}, C. and {Gottlieb}, O. and {Jencson}, J.~E. and {Adams}, S.~M. and {Feindt}, U. and {Hotokezaka}, K. and {Ghosh}, S. and {Perley}, D.~A. and {Yu}, P. -C. and {Piran}, T. and {Allison}, J.~R. and {Anupama}, G.~C. and {Balasubramanian}, A. and {Bannister}, K.~W. and {Bally}, J. and {Barnes}, J. and {Barway}, S. and {Bellm}, E. and {Bhalerao}, V. and {Bhattacharya}, D. and {Blagorodnova}, N. and {Bloom}, J.~S. and {Brady}, P.~R. and {Cannella}, C. and {Chatterjee}, D. and {Cenko}, S.~B. and {Cobb}, B.~E. and {Copperwheat}, C. and {Corsi}, A. and {De}, K. and {Dobie}, D. and {Emery}, S.~W.~K. and {Evans}, P.~A. and {Fox}, O.~D. and {Frail}, D.~A. and {Frohmaier}, C. and {Goobar}, A. and {Hallinan}, G. and {Harrison}, F. and {Helou}, G. and {Hinderer}, T. and {Ho}, A.~Y.~Q. and {Horesh}, A. and {Ip}, W. -H. and {Itoh}, R. and {Kasen}, D. and {Kim}, H. and {Kuin}, N.~P.~M. and {Kupfer}, T. and {Lynch}, C. and {Madsen}, K. and {Mazzali}, P.~A. and {Miller}, A.~A. and {Mooley}, K. and {Murphy}, T. and {Ngeow}, C. -C. and {Nichols}, D. and {Nissanke}, S. and {Nugent}, P. and {Ofek}, E.~O. and {Qi}, H. and {Quimby}, R.~M. and {Rosswog}, S. and {Rusu}, F. and {Sadler}, E.~M. and {Schmidt}, P. and {Sollerman}, J. and {Steele}, I. and {Williamson}, A.~R. and {Xu}, Y. and {Yan}, L. and {Yatsu}, Y. and {Zhang}, C. and {Zhao}, W.},
        title = "{Illuminating gravitational waves: A concordant picture of photons from a neutron star merger}",
      journal = {Science},
         year = 2017,
        month = dec,
       volume = {358},
       number = {6370},
        pages = {1559-1565},
          doi = {10.1126/science.aap9455},
archivePrefix = {arXiv},
       eprint = {1710.05436},
 primaryClass = {astro-ph.HE},
       adsurl = {https://ui.adsabs.harvard.edu/abs/2017Sci...358.1559K}
}

@ARTICLE{2017ApJ...848L..34M,
       author = {{Murguia-Berthier}, A. and {Ramirez-Ruiz}, E. and {Kilpatrick}, C.~D. and {Foley}, R.~J. and {Kasen}, D. and {Lee}, W.~H. and {Piro}, A.~L. and {Coulter}, D.~A. and {Drout}, M.~R. and {Madore}, B.~F. and {Shappee}, B.~J. and {Pan}, Y. -C. and {Prochaska}, J.~X. and {Rest}, A. and {Rojas-Bravo}, C. and {Siebert}, M.~R. and {Simon}, J.~D.},
        title = "{A Neutron Star Binary Merger Model for GW170817/GRB 170817A/SSS17a}",
      journal = {\apjl},
         year = 2017,
        month = oct,
       volume = {848},
       number = {2},
          eid = {L34},
        pages = {L34},
          doi = {10.3847/2041-8213/aa91b3},
archivePrefix = {arXiv},
       eprint = {1710.05453},
 primaryClass = {astro-ph.HE},
       adsurl = {https://ui.adsabs.harvard.edu/abs/2017ApJ...848L..34M}
}

@ARTICLE{2017PASJ...69..102T,
       author = {{Tanaka}, Masaomi and {Utsumi}, Yousuke and {Mazzali}, Paolo A. and {Tominaga}, Nozomu and {Yoshida}, Michitoshi and {Sekiguchi}, Yuichiro and {Morokuma}, Tomoki and {Motohara}, Kentaro and {Ohta}, Kouji and {Kawabata}, Koji S. and {Abe}, Fumio and {Aoki}, Kentaro and {Asakura}, Yuichiro and {Baar}, Stefan and {Barway}, Sudhanshu and {Bond}, Ian A. and {Doi}, Mamoru and {Fujiyoshi}, Takuya and {Furusawa}, Hisanori and {Honda}, Satoshi and {Itoh}, Yoichi and {Kawabata}, Miho and {Kawai}, Nobuyuki and {Kim}, Ji Hoon and {Lee}, Chien-Hsiu and {Miyazaki}, Shota and {Morihana}, Kumiko and {Nagashima}, Hiroki and {Nagayama}, Takahiro and {Nakaoka}, Tatsuya and {Nakata}, Fumiaki and {Ohsawa}, Ryou and {Ohshima}, Tomohito and {Okita}, Hirofumi and {Saito}, Tomoki and {Sumi}, Takahiro and {Tajitsu}, Akito and {Takahashi}, Jun and {Takayama}, Masaki and {Tamura}, Yoichi and {Tanaka}, Ichi and {Terai}, Tsuyoshi and {Tristram}, Paul J. and {Yasuda}, Naoki and {Zenko}, Tetsuya},
        title = "{Kilonova from post-merger ejecta as an optical and near-Infrared counterpart of GW170817}",
      journal = {\pasj},
         year = 2017,
        month = dec,
       volume = {69},
       number = {6},
          eid = {102},
        pages = {102},
          doi = {10.1093/pasj/psx121},
archivePrefix = {arXiv},
       eprint = {1710.05850},
 primaryClass = {astro-ph.HE},
       adsurl = {https://ui.adsabs.harvard.edu/abs/2017PASJ...69..102T}
}

@ARTICLE{2018MNRAS.481.3423W,
       author = {{Waxman}, Eli and {Ofek}, Eran O. and {Kushnir}, Doron and {Gal-Yam}, Avishay},
        title = "{Constraints on the ejecta of the GW170817 neutron star merger from its electromagnetic emission}",
      journal = {\mnras},
         year = 2018,
        month = dec,
       volume = {481},
       number = {3},
        pages = {3423-3441},
          doi = {10.1093/mnras/sty2441},
archivePrefix = {arXiv},
       eprint = {1711.09638},
 primaryClass = {astro-ph.HE},
       adsurl = {https://ui.adsabs.harvard.edu/abs/2018MNRAS.481.3423W}
}

@ARTICLE{2013ApJ...774...25K,
       author = {{Kasen}, Daniel and {Badnell}, N.~R. and {Barnes}, Jennifer},
        title = "{Opacities and Spectra of the r-process Ejecta from Neutron Star Mergers}",
      journal = {\apj},
         year = 2013,
        month = sep,
       volume = {774},
       number = {1},
          eid = {25},
        pages = {25},
          doi = {10.1088/0004-637X/774/1/25},
archivePrefix = {arXiv},
       eprint = {1303.5788},
 primaryClass = {astro-ph.HE},
       adsurl = {https://ui.adsabs.harvard.edu/abs/2013ApJ...774...25K}
}

@ARTICLE{2013ApJ...775..113T,
       author = {{Tanaka}, Masaomi and {Hotokezaka}, Kenta},
        title = "{Radiative Transfer Simulations of Neutron Star Merger Ejecta}",
      journal = {\apj},
         year = 2013,
        month = oct,
       volume = {775},
       number = {2},
          eid = {113},
        pages = {113},
          doi = {10.1088/0004-637X/775/2/113},
archivePrefix = {arXiv},
       eprint = {1306.3742},
 primaryClass = {astro-ph.HE},
       adsurl = {https://ui.adsabs.harvard.edu/abs/2013ApJ...775..113T}
}

@ARTICLE{2010MNRAS.406.2650M,
       author = {{Metzger}, B.~D. and {Mart{\'\i}nez-Pinedo}, G. and {Darbha}, S. and {Quataert}, E. and {Arcones}, A. and {Kasen}, D. and {Thomas}, R. and {Nugent}, P. and {Panov}, I.~V. and {Zinner}, N.~T.},
        title = "{Electromagnetic counterparts of compact object mergers powered by the radioactive decay of r-process nuclei}",
      journal = {\mnras},
         year = 2010,
        month = aug,
       volume = {406},
       number = {4},
        pages = {2650-2662},
          doi = {10.1111/j.1365-2966.2010.16864.x},
archivePrefix = {arXiv},
       eprint = {1001.5029},
 primaryClass = {astro-ph.HE},
       adsurl = {https://ui.adsabs.harvard.edu/abs/2010MNRAS.406.2650M}
}

@ARTICLE{2011ApJ...736L..21R,
       author = {{Roberts}, L.~F. and {Kasen}, D. and {Lee}, W.~H. and {Ramirez-Ruiz}, E.},
        title = "{Electromagnetic Transients Powered by Nuclear Decay in the Tidal Tails of Coalescing Compact Binaries}",
      journal = {\apjl},
         year = 2011,
        month = jul,
       volume = {736},
       number = {1},
          eid = {L21},
        pages = {L21},
          doi = {10.1088/2041-8205/736/1/L21},
archivePrefix = {arXiv},
       eprint = {1104.5504},
 primaryClass = {astro-ph.HE},
       adsurl = {https://ui.adsabs.harvard.edu/abs/2011ApJ...736L..21R}
}

@ARTICLE{2014MNRAS.441.3444M,
       author = {{Metzger}, Brian D. and {Fern{\'a}ndez}, Rodrigo},
        title = "{Red or blue? A potential kilonova imprint of the delay until black hole formation following a neutron star merger}",
      journal = {\mnras},
         year = 2014,
        month = jul,
       volume = {441},
       number = {4},
        pages = {3444-3453},
          doi = {10.1093/mnras/stu802},
archivePrefix = {arXiv},
       eprint = {1402.4803},
 primaryClass = {astro-ph.HE},
       adsurl = {https://ui.adsabs.harvard.edu/abs/2014MNRAS.441.3444M}
}

@ARTICLE{Vassh2019,
       author = {{Vassh}, N. and {Vogt}, R. and {Surman}, R. and {Randrup}, J. and {Sprouse}, T.~M. and {Mumpower}, M.~R. and {Jaffke}, P. and {Shaw}, D. and {Holmbeck}, E.~M. and {Zhu}, Y. and {McLaughlin}, G.~C.},
        title = "{Using excitation-energy dependent fission yields to identify key fissioning nuclei in r-process nucleosynthesis}",
      journal = {Journal of Physics G Nuclear Physics},
         year = 2019,
        month = jun,
       volume = {46},
       number = {6},
        pages = {065202},
          doi = {10.1088/1361-6471/ab0bea},
archivePrefix = {arXiv},
       eprint = {1810.08133},
 primaryClass = {nucl-th},
       adsurl = {https://ui.adsabs.harvard.edu/abs/2019JPhG...46f5202V}
}

@ARTICLE{Vassh2024,
       author = {{Vassh}, Nicole and {Wang}, Xilu and {Larivi{\`e}re}, Maude and {Sprouse}, Trevor and {Mumpower}, Matthew R. and {Surman}, Rebecca and {Liu}, Zhenghai and {McLaughlin}, Gail C. and {Denissenkov}, Pavel and {Herwig}, Falk},
        title = "{Thallium-208: A Beacon of In Situ Neutron Capture Nucleosynthesis}",
      journal = {\prl},
         year = 2024,
        month = jan,
       volume = {132},
       number = {5},
          eid = {052701},
        pages = {052701},
          doi = {10.1103/PhysRevLett.132.052701},
archivePrefix = {arXiv},
       eprint = {2311.10895},
 primaryClass = {nucl-th},
       adsurl = {https://ui.adsabs.harvard.edu/abs/2024PhRvL.132e2701V}
}

@inproceedings{Fryer:2023pew,
    author = "Fryer, C. L. and Fontes, C. J. and Korobkin, O. and Mumpower, M. and Wollaeger, R. and Holmbeck, E. M. and O{\textquoteright}Shaugnessy, R.",
    title = "{Uncertainties in kilonova modeling}",
    booktitle = "{16th Marcel Grossmann Meeting on~Recent Developments in Theoretical and Experimental General Relativity, Astrophysics and Relativistic Field Theories}",
    doi = "10.1142/9789811269776_0112",
    year = "2023"
}

@article{Fryer:2023osz,
    author = "Fryer, Chris L. and Hungerford, Aimee L. and Wollaeger, Ryan T. and Miller, Jonah M. and De, Soumi and Fontes, Christopher J. and Korobkin, Oleg and Kedia, Atul and Ristic, Marko and O{\textquoteright}Shaughnessy, Richard",
    title = "{The Effect of the Velocity Distribution on Kilonova Emission}",
    eprint = "2311.05005",
    archivePrefix = "arXiv",
    primaryClass = "astro-ph.HE",
    doi = "10.3847/1538-4357/ad1036",
    journal = "Astrophys. J.",
    volume = "961",
    number = "1",
    pages = "9",
    year = "2024"
}

@article{Brethauer:2024zxg,
    author = "Brethauer, Daniel and Kasen, Daniel and Margutti, Raffaella and Chornock, Ryan",
    title = "{Impact of Systematic Modeling Uncertainties on Kilonova Property Estimation}",
    eprint = "2408.02731",
    archivePrefix = "arXiv",
    primaryClass = "astro-ph.HE",
    doi = "10.3847/1538-4357/ad7d83",
    journal = "Astrophys. J.",
    volume = "975",
    number = "2",
    pages = "213",
    year = "2024"
}

@ARTICLE{2020ApJ...903L...3W,
       author = {{Wang}, Xilu and {N3AS Collaboration} and {Vassh}, Nicole and {FIRE Collaboration} and {Sprouse}, Trevor and {Mumpower}, Matthew and {Vogt}, Ramona and {Randrup}, Jorgen and {Surman}, Rebecca},
        title = "{MeV Gamma Rays from Fission: A Distinct Signature of Actinide Production in Neutron Star Mergers}",
      journal = {\apjl},
         year = 2020,
        month = nov,
       volume = {903},
       number = {1},
          eid = {L3},
        pages = {L3},
          doi = {10.3847/2041-8213/abbe18},
archivePrefix = {arXiv},
       eprint = {2008.03335},
 primaryClass = {astro-ph.HE},
       adsurl = {https://ui.adsabs.harvard.edu/abs/2020ApJ...903L...3W}
}

@ARTICLE{2025ApJ...995L..28G,
       author = {{Gross}, Axel and {Cupp}, Samuel and {Mumpower}, Matthew R.},
        title = "{Multimessengers from the Radioactive Decay of r-process Nuclei}",
      journal = {\apjl},
         year = 2025,
        month = dec,
       volume = {995},
       number = {1},
          eid = {L28},
        pages = {L28},
          doi = {10.3847/2041-8213/ae2465},
archivePrefix = {arXiv},
       eprint = {2509.00267},
 primaryClass = {astro-ph.HE},
       adsurl = {https://ui.adsabs.harvard.edu/abs/2025ApJ...995L..28G}
}

@techreport{brown2018endfb,
  title        = {{ENDF/B-VIII.0: The 8th Major Release of the Nuclear Reaction Data Library with CIELO-project Cross Sections, New Standards and Thermal Scattering Data}},
  author       = {D. A. Brown and M. B. Chadwick and R. Capote and A. D. Carlson and Y. Danon and A. C. Kahler and D. M. Smith and B. Vogt and H. Wienke and W. B. Wilson and R. D. McKnight and et al.},
  institution  = {Nuclear Data Sheets},
  year         = {2018},
  volume       = {148},
  pages        = {1--142},
  doi          = {10.1016/j.nds.2018.02.001},
  url          = {https://www.sciencedirect.com/science/article/pii/S0090375218300206},
}

@article{Gray:2016vsn,
    author = "Gray, Timothy J. and Reed, M. W. and Lane, G. J. and Akber, A. and Litvinov, Yu. A. and Walker, P. M.",
    editor = "Reed, M. W. and Simpson, E. C. and Mitchell, A. J.",
    title = "{Nuclear lifetime measurements from data with independently varying observation times}",
    doi = "10.1051/epjconf/201612304004",
    journal = "EPJ Web Conf.",
    volume = "123",
    pages = "04004",
    year = "2016"
}

@ARTICLE{2024ApJ...962...68A,
       author = {{Anand}, Shreya and {Barnes}, Jennifer and {Yang}, Sheng and {Kasliwal}, Mansi M. and {Coughlin}, Michael W. and {Sollerman}, Jesper and {De}, Kishalay and {Fremling}, Christoffer and {Corsi}, Alessandra and {Ho}, Anna Y.~Q. and {Balasubramanian}, Arvind and {Omand}, Conor and {Srinivasaragavan}, Gokul P. and {Cenko}, S. Bradley and {Ahumada}, Tom{\'a}s and {Andreoni}, Igor and {Dahiwale}, Aishwarya and {Das}, Kaustav Kashyap and {Jencson}, Jacob and {Karambelkar}, Viraj and {Kumar}, Harsh and {Metzger}, Brian D. and {Perley}, Daniel and {Sarin}, Nikhil and {Schweyer}, Tassilo and {Schulze}, Steve and {Sharma}, Yashvi and {Sit}, Tawny and {Stein}, Robert and {Tartaglia}, Leonardo and {Tinyanont}, Samaporn and {Tzanidakis}, Anastasios and {van Roestel}, Jan and {Yao}, Yuhan and {Bloom}, Joshua S. and {Cook}, David O. and {Dekany}, Richard and {Graham}, Matthew J. and {Groom}, Steven L. and {Kaplan}, David L. and {Masci}, Frank J. and {Medford}, Michael S. and {Riddle}, Reed and {Zhang}, Chaoran},
        title = "{Collapsars as Sites of r-process Nucleosynthesis: Systematic Photometric Near-infrared Follow-up of Type Ic-BL Supernovae}",
      journal = {\apj},
         year = 2024,
        month = feb,
       volume = {962},
       number = {1},
          eid = {68},
        pages = {68},
          doi = {10.3847/1538-4357/ad11df},
archivePrefix = {arXiv},
       eprint = {2302.09226},
 primaryClass = {astro-ph.HE},
       adsurl = {https://ui.adsabs.harvard.edu/abs/2024ApJ...962...68A}
}

@ARTICLE{2013ApJ...778....8B,
       author = {{Banerjee}, Indrani and {Mukhopadhyay}, Banibrata},
        title = "{Nucleosynthesis in the Outflows Associated with Accretion Disks of Type II Collapsars}",
      journal = {\apj},
         year = 2013,
        month = nov,
       volume = {778},
       number = {1},
          eid = {8},
        pages = {8},
          doi = {10.1088/0004-637X/778/1/8},
archivePrefix = {arXiv},
       eprint = {1309.0954},
 primaryClass = {astro-ph.HE},
       adsurl = {https://ui.adsabs.harvard.edu/abs/2013ApJ...778....8B}
}

@ARTICLE{2025ApJ...985..234P,
       author = {{Patel}, Anirudh and {Metzger}, Brian D. and {Goldberg}, Jared A. and {Cehula}, Jakub and {Thompson}, Todd A. and {Renzo}, Mathieu},
        title = "{r-process Nucleosynthesis and Radioactively Powered Transients from Magnetar Giant Flares}",
      journal = {\apj},
         year = 2025,
        month = jun,
       volume = {985},
       number = {2},
          eid = {234},
        pages = {234},
          doi = {10.3847/1538-4357/adceb7},
archivePrefix = {arXiv},
       eprint = {2501.17253},
 primaryClass = {astro-ph.HE},
       adsurl = {https://ui.adsabs.harvard.edu/abs/2025ApJ...985..234P}
}

@ARTICLE{2024MNRAS.528.5323C,
       author = {{Cehula}, Jakub and {Thompson}, Todd A. and {Metzger}, Brian D.},
        title = "{Dynamics of baryon ejection in magnetar giant flares: implications for radio afterglows, r-process nucleosynthesis, and fast radio bursts}",
      journal = {\mnras},
         year = 2024,
        month = mar,
       volume = {528},
       number = {3},
        pages = {5323-5345},
          doi = {10.1093/mnras/stae358},
archivePrefix = {arXiv},
       eprint = {2311.05681},
 primaryClass = {astro-ph.HE},
       adsurl = {https://ui.adsabs.harvard.edu/abs/2024MNRAS.528.5323C}
}

@ARTICLE{2025ApJ...984L..29P,
       author = {{Patel}, Anirudh and {Metzger}, Brian D. and {Cehula}, Jakub and {Burns}, Eric and {Goldberg}, Jared A. and {Thompson}, Todd A.},
        title = "{Direct Evidence for r-process Nucleosynthesis in Delayed MeV Emission from the SGR 1806{\textendash}20 Magnetar Giant Flare}",
      journal = {\apjl},
         year = 2025,
        month = may,
       volume = {984},
       number = {1},
          eid = {L29},
        pages = {L29},
          doi = {10.3847/2041-8213/adc9b0},
archivePrefix = {arXiv},
       eprint = {2501.09181},
 primaryClass = {astro-ph.HE},
       adsurl = {https://ui.adsabs.harvard.edu/abs/2025ApJ...984L..29P}
}

@ARTICLE{2025arXiv250903003R,
       author = {{Risti{\'c}}, Marko and {Barker}, Brandon L. and {Cupp}, Samuel and {Gross}, Axel and {Lloyd-Ronning}, Nicole and {Korobkin}, Oleg and {Miller}, Jonah M. and {Mumpower}, Matthew R.},
        title = "{Kilonovae and Long-duration Gamma-ray Bursts}",
      journal = {arXiv e-prints},
         year = 2025,
        month = sep,
          eid = {arXiv:2509.03003},
        pages = {arXiv:2509.03003},
          doi = {10.48550/arXiv.2509.03003},
archivePrefix = {arXiv},
       eprint = {2509.03003},
 primaryClass = {astro-ph.HE},
       adsurl = {https://ui.adsabs.harvard.edu/abs/2025arXiv250903003R}
}

@ARTICLE{2023ApJ...958..121T,
       author = {{Tak}, Donggeun and {Uhm}, Z. Lucas and {Gillanders}, James H.},
        title = "{Exploring the Impact of the Ejecta Velocity Profile on the Evolution of Kilonova: Diversity of the Kilonova Lightcurves}",
      journal = {\apj},
         year = 2023,
        month = dec,
       volume = {958},
       number = {2},
          eid = {121},
        pages = {121},
          doi = {10.3847/1538-4357/ad06b0},
archivePrefix = {arXiv},
       eprint = {2310.15608},
 primaryClass = {astro-ph.HE},
       adsurl = {https://ui.adsabs.harvard.edu/abs/2023ApJ...958..121T}
}

@ARTICLE{2016MNRAS.459...35H,
       author = {{Hotokezaka}, K. and {Wanajo}, S. and {Tanaka}, M. and {Bamba}, A. and {Terada}, Y. and {Piran}, T.},
        title = "{Radioactive decay products in neutron star merger ejecta: heating efficiency and {\ensuremath{\gamma}}-ray emission}",
      journal = {\mnras},
         year = 2016,
        month = jun,
       volume = {459},
       number = {1},
        pages = {35-43},
          doi = {10.1093/mnras/stw404},
archivePrefix = {arXiv},
       eprint = {1511.05580},
 primaryClass = {astro-ph.HE},
       adsurl = {https://ui.adsabs.harvard.edu/abs/2016MNRAS.459...35H}
}

@ARTICLE{2020ApJ...889..168K,
       author = {{Korobkin}, Oleg and {Hungerford}, Aimee M. and {Fryer}, Christopher L. and {Mumpower}, Matthew R. and {Misch}, G. Wendell and {Sprouse}, Trevor M. and {Lippuner}, Jonas and {Surman}, Rebecca and {Couture}, Aaron J. and {Bloser}, Peter F. and {Shirazi}, Farzane and {Even}, Wesley P. and {Vestrand}, W. Thomas and {Miller}, Richard S.},
        title = "{Gamma Rays from Kilonova: A Potential Probe of r-process Nucleosynthesis}",
      journal = {\apj},
         year = 2020,
        month = feb,
       volume = {889},
       number = {2},
          eid = {168},
        pages = {168},
          doi = {10.3847/1538-4357/ab64d8},
archivePrefix = {arXiv},
       eprint = {1905.05089},
 primaryClass = {astro-ph.HE},
       adsurl = {https://ui.adsabs.harvard.edu/abs/2020ApJ...889..168K}
}

@ARTICLE{1998ApJ...506..868Q,
       author = {{Qian}, Y. -Z. and {Vogel}, P. and {Wasserburg}, G.~J.},
        title = "{Supernovae as the Site of the r-Process: Implications for Gamma-Ray Astronomy}",
      journal = {\apj},
         year = 1998,
        month = oct,
       volume = {506},
       number = {2},
        pages = {868-873},
          doi = {10.1086/306285},
archivePrefix = {arXiv},
       eprint = {astro-ph/9803300},
 primaryClass = {astro-ph},
       adsurl = {https://ui.adsabs.harvard.edu/abs/1998ApJ...506..868Q}
}

@ARTICLE{2022ApJ...933..111T,
       author = {{Terada}, Yukikatsu and {Miwa}, Yuya and {Ohsumi}, Hayato and {Fujimoto}, Shin-ichiro and {Katsuda}, Satoru and {Bamba}, Aya and {Yamazaki}, Ryo},
        title = "{Gamma-Ray Diagnostics of r-process Nucleosynthesis in the Remnants of Galactic Binary Neutron-star Mergers}",
      journal = {\apj},
         year = 2022,
        month = jul,
       volume = {933},
       number = {1},
          eid = {111},
        pages = {111},
          doi = {10.3847/1538-4357/ac721f},
archivePrefix = {arXiv},
       eprint = {2205.05407},
 primaryClass = {astro-ph.HE},
       adsurl = {https://ui.adsabs.harvard.edu/abs/2022ApJ...933..111T}
}

@ARTICLE{2013PhRvC..87b4601B,
       author = {{Billnert}, R. and {Hambsch}, F. -J. and {Oberstedt}, A. and {Oberstedt}, S.},
        title = "{New prompt spectral {\ensuremath{\gamma}}-ray data from the reaction $^{252}$Cf(sf) and its implication on present evaluated nuclear data files}",
      journal = {\prc},
         year = 2013,
        month = feb,
       volume = {87},
       number = {2},
          eid = {024601},
        pages = {024601},
          doi = {10.1103/PhysRevC.87.024601},
       adsurl = {https://ui.adsabs.harvard.edu/abs/2013PhRvC..87b4601B}
}

@ARTICLE{2021CoPhC.26908087T,
       author = {{Talou}, P. and {Stetcu}, I. and {Jaffke}, P. and {Rising}, M.~E. and {Lovell}, A.~E. and {Kawano}, T.},
        title = "{Fission fragment decay simulations with the CGMF code}",
      journal = {Computer Physics Communications},
         year = 2021,
        month = dec,
       volume = {269},
          eid = {108087},
        pages = {108087},
          doi = {10.1016/j.cpc.2021.108087},
archivePrefix = {arXiv},
       eprint = {2011.10444},
 primaryClass = {nucl-th},
       adsurl = {https://ui.adsabs.harvard.edu/abs/2021CoPhC.26908087T}
}

@ARTICLE{2018PhRvC..98a4612Q,
       author = {{Qi}, L. and {Lebois}, M. and {Wilson}, J.~N. and {Chatillon}, A. and {Courtin}, S. and {Fruet}, G. and {Georgiev}, G. and {Jenkins}, D.~G. and {Laurent}, B. and {Le Meur}, L. and {Maj}, A. and {Marini}, P. and {Matea}, I. and {Morris}, L. and {Nanal}, V. and {Napiorkowski}, P. and {Oberstedt}, A. and {Oberstedt}, S. and {Schmitt}, C. and {Serot}, O. and {Stanoiu}, M. and {Wasilewska}, B.},
        title = "{Statistical study of the prompt-fission {\ensuremath{\gamma}} -ray spectrum for $^{238}$U(n , f ) in the fast-neutron region}",
      journal = {\prc},
         year = 2018,
        month = jul,
       volume = {98},
       number = {1},
          eid = {014612},
        pages = {014612},
          doi = {10.1103/PhysRevC.98.014612},
       adsurl = {https://ui.adsabs.harvard.edu/abs/2018PhRvC..98a4612Q}
}

@ARTICLE{2020A&A...638A..83W,
       author = {{Weinberger}, Christoph and {Diehl}, Roland and {Pleintinger}, Moritz M.~M. and {Siegert}, Thomas and {Greiner}, Jochen},
        title = "{$^{44}$Ti ejecta in young supernova remnants}",
      journal = {\aap},
         year = 2020,
        month = jun,
       volume = {638},
          eid = {A83},
        pages = {A83},
          doi = {10.1051/0004-6361/202037536},
archivePrefix = {arXiv},
       eprint = {2004.12688},
 primaryClass = {astro-ph.HE},
       adsurl = {https://ui.adsabs.harvard.edu/abs/2020A&A...638A..83W}
}

@article{Mochizuki:1999yg,
    author = "Mochizuki, Y. and Takahashi, K. and Janka, H. Th. and Hillebrandt, W. and Diehl, R.",
    title = "{Titanium-44: its effective decay rate in young supernova remnants, and its abundance in cas a}",
    eprint = "astro-ph/9904378",
    archivePrefix = "arXiv",
    reportNumber = "MPA-1161",
    journal = "Astron. Astrophys.",
    volume = "346",
    pages = "831",
    year = "1999"
}

@ARTICLE{2015Sci...348..670B,
       author = {{Boggs}, S.~E. and {Harrison}, F.~A. and {Miyasaka}, H. and {Grefenstette}, B.~W. and {Zoglauer}, A. and {Fryer}, C.~L. and {Reynolds}, S.~P. and {Alexander}, D.~M. and {An}, H. and {Barret}, D. and {Christensen}, F.~E. and {Craig}, W.~W. and {Forster}, K. and {Giommi}, P. and {Hailey}, C.~J. and {Hornstrup}, A. and {Kitaguchi}, T. and {Koglin}, J.~E. and {Madsen}, K.~K. and {Mao}, P.~H. and {Mori}, K. and {Perri}, M. and {Pivovaroff}, M.~J. and {Puccetti}, S. and {Rana}, V. and {Stern}, D. and {Westergaard}, N.~J. and {Zhang}, W.~W.},
        title = "{$^{44}$Ti gamma-ray emission lines from SN1987A reveal an asymmetric explosion}",
      journal = {Science},
         year = 2015,
        month = may,
       volume = {348},
       number = {6235},
        pages = {670-671},
          doi = {10.1126/science.aaa2259},
       adsurl = {https://ui.adsabs.harvard.edu/abs/2015Sci...348..670B}
}

@ARTICLE{1977ApJ...213L...5R,
       author = {{Ramaty}, R. and {Lingenfelter}, R.~E.},
        title = "{$^{26}$Al: a galactic source of gamma-ray line emission.}",
      journal = {\apjl},
         year = 1977,
        month = apr,
       volume = {213},
        pages = {L5-L7},
          doi = {10.1086/182397},
       adsurl = {https://ui.adsabs.harvard.edu/abs/1977ApJ...213L...5R}
}

@ARTICLE{2025A&A...695A.190W,
       author = {{Wehmeyer}, B. and {Kobayashi}, C. and {Yag{\"u}e L{\'o}pez}, A. and {Lugaro}, M.},
        title = "{The aluminium-26 distribution in a cosmological simulation of a Milky Way-type Galaxy}",
      journal = {\aap},
         year = 2025,
        month = mar,
       volume = {695},
          eid = {A190},
        pages = {A190},
          doi = {10.1051/0004-6361/202451915},
archivePrefix = {arXiv},
       eprint = {2504.17915},
 primaryClass = {astro-ph.GA},
       adsurl = {https://ui.adsabs.harvard.edu/abs/2025A&A...695A.190W}
}

@ARTICLE{2023A&A...672A..53P,
       author = {{Pleintinger}, Moritz M.~M. and {Diehl}, Roland and {Siegert}, Thomas and {Greiner}, Jochen and {Krause}, Martin G.~H.},
        title = "{$^{26}$Al gamma rays from the Galaxy with INTEGRAL/SPI}",
      journal = {\aap},
         year = 2023,
        month = apr,
       volume = {672},
          eid = {A53},
        pages = {A53},
          doi = {10.1051/0004-6361/202245069},
archivePrefix = {arXiv},
       eprint = {2212.11228},
 primaryClass = {astro-ph.HE},
       adsurl = {https://ui.adsabs.harvard.edu/abs/2023A&A...672A..53P}
}

@ARTICLE{2023ApJ...944..144L,
       author = {{Lund}, Kelsey A. and {Engel}, J. and {McLaughlin}, G.~C. and {Mumpower}, M.~R. and {Ney}, E.~M. and {Surman}, R.},
        title = "{The Influence of {\ensuremath{\beta}}-decay Rates on r-process Observables}",
      journal = {\apj},
         year = 2023,
        month = feb,
       volume = {944},
       number = {2},
          eid = {144},
        pages = {144},
          doi = {10.3847/1538-4357/acaf56},
archivePrefix = {arXiv},
       eprint = {2208.06373},
 primaryClass = {astro-ph.HE},
       adsurl = {https://ui.adsabs.harvard.edu/abs/2023ApJ...944..144L}
}

@ARTICLE{2018ApJ...863L..23Z,
       author = {{Zhu}, Y. and {Wollaeger}, R.~T. and {Vassh}, N. and {Surman}, R. and {Sprouse}, T.~M. and {Mumpower}, M.~R. and {M{\"o}ller}, P. and {McLaughlin}, G.~C. and {Korobkin}, O. and {Kawano}, T. and {Jaffke}, P.~J. and {Holmbeck}, E.~M. and {Fryer}, C.~L. and {Even}, W.~P. and {Couture}, A.~J. and {Barnes}, J.},
        title = "{Californium-254 and Kilonova Light Curves}",
      journal = {\apjl},
         year = 2018,
        month = aug,
       volume = {863},
       number = {2},
          eid = {L23},
        pages = {L23},
          doi = {10.3847/2041-8213/aad5de},
archivePrefix = {arXiv},
       eprint = {1806.09724},
 primaryClass = {astro-ph.HE},
       adsurl = {https://ui.adsabs.harvard.edu/abs/2018ApJ...863L..23Z}
}

@article{PhysRevLett.131.262701,
  title = {Direct Mass Measurements to Inform the Behavior of $^{128\mathrm{m}}\mathrm{Sb}$ in Nucleosynthetic Environments},
  author = {Hoff, D. E. M. and Kolos, K. and Misch, G. W. and Ray, D. and Liu, B. and Valverde, A. A. and Brodeur, M. and Burdette, D. P. and Callahan, N. and Clark, J. A. and Gallant, A. T. and Kondev, F. G. and Morgan, G. E. and Mumpower, M. R. and Orford, R. and Porter, W. S. and Rivero, F. and Savard, G. and Scielzo, N. D. and Sharma, K. S. and Sieja, K. and Sprouse, T. M. and Varriano, L.},
  journal = {Phys. Rev. Lett.},
  volume = {131},
  issue = {26},
  pages = {262701},
  numpages = {6},
  year = {2023},
  month = {Dec},
  publisher = {American Physical Society},
  doi = {10.1103/PhysRevLett.131.262701},
  url = {https://link.aps.org/doi/10.1103/PhysRevLett.131.262701}
}

@ARTICLE{2021ApJS..252....2M,
       author = {{Misch}, G. Wendell and {Ghorui}, Surja K. and {Banerjee}, Projjwal and {Sun}, Yang and {Mumpower}, Matthew R.},
        title = "{Astromers: Nuclear Isomers in Astrophysics}",
      journal = {\apjs},
         year = 2021,
        month = jan,
       volume = {252},
       number = {1},
          eid = {2},
        pages = {2},
          doi = {10.3847/1538-4365/abc41d},
archivePrefix = {arXiv},
       eprint = {2010.15238},
 primaryClass = {astro-ph.HE},
       adsurl = {https://ui.adsabs.harvard.edu/abs/2021ApJS..252....2M}
}

@ARTICLE{1988Natur.332..516L,
       author = {{Leising}, Mark D.},
        title = "{Gamma-rays and X-rays from SN1987A}",
      journal = {\nat},
         year = 1988,
        month = apr,
       volume = {332},
       number = {6164},
        pages = {516-518},
          doi = {10.1038/332516a0},
       adsurl = {https://ui.adsabs.harvard.edu/abs/1988Natur.332..516L}
}

@ARTICLE{2015A&A...574A..72D,
       author = {{Diehl}, Roland and {Siegert}, Thomas and {Hillebrandt}, Wolfgang and {Krause}, Martin and {Greiner}, Jochen and {Maeda}, Keiichi and {R{\"o}pke}, Friedrich K. and {Sim}, Stuart A. and {Wang}, Wei and {Zhang}, Xiaoling},
        title = "{SN2014J gamma rays from the $^{56}$Ni decay chain}",
      journal = {\aap},
         year = 2015,
        month = feb,
       volume = {574},
          eid = {A72},
        pages = {A72},
          doi = {10.1051/0004-6361/201424991},
archivePrefix = {arXiv},
       eprint = {1409.5477},
 primaryClass = {astro-ph.HE},
       adsurl = {https://ui.adsabs.harvard.edu/abs/2015A&A...574A..72D}
}

@ARTICLE{1994A&A...284L...1I,
       author = {{Iyudin}, A.~F. and {Diehl}, R. and {Bloemen}, H. and {Hermsen}, W. and {Lichti}, G.~G. and {Morris}, D. and {Ryan}, J. and {Schoenfelder}, V. and {Steinle}, H. and {Varendorff}, M. and {de Vries}, C. and {Winkler}, C.},
        title = "{COMPTEL observations of 44Ti gamma-ray line emission form CAS A.}",
      journal = {\aap},
         year = 1994,
        month = apr,
       volume = {284},
        pages = {L1-L4},
       adsurl = {https://ui.adsabs.harvard.edu/abs/1994A&A...284L...1I}
}

@ARTICLE{2014Natur.512..406C,
       author = {{Churazov}, E. and {Sunyaev}, R. and {Isern}, J. and {Kn{\"o}dlseder}, J. and {Jean}, P. and {Lebrun}, F. and {Chugai}, N. and {Grebenev}, S. and {Bravo}, E. and {Sazonov}, S. and {Renaud}, M.},
        title = "{Cobalt-56 {\ensuremath{\gamma}}-ray emission lines from the type Ia supernova 2014J}",
      journal = {\nat},
         year = 2014,
        month = aug,
       volume = {512},
       number = {7515},
        pages = {406-408},
          doi = {10.1038/nature13672},
archivePrefix = {arXiv},
       eprint = {1405.3332},
 primaryClass = {astro-ph.HE},
       adsurl = {https://ui.adsabs.harvard.edu/abs/2014Natur.512..406C}
}

@ARTICLE{1982ApJ...262..742M,
       author = {{Mahoney}, W.~A. and {Ling}, J.~C. and {Jacobson}, A.~S. and {Lingenfelter}, R.~E.},
        title = "{Diffuse galactic gamma-ray line emission from nucleosynthetic Fe-60, Al-26, and Na-22 - Preliminary limits from HEAO 3.}",
      journal = {\apj},
         year = 1982,
        month = nov,
       volume = {262},
        pages = {742-748},
          doi = {10.1086/160469},
       adsurl = {https://ui.adsabs.harvard.edu/abs/1982ApJ...262..742M}
}

@ARTICLE{2019ApJ...880...23W,
       author = {{Wu}, Meng-Ru and {Banerjee}, Projjwal and {Metzger}, Brian D. and {Mart{\'\i}nez-Pinedo}, Gabriel and {Aramaki}, Tsuguo and {Burns}, Eric and {Hailey}, Charles J. and {Barnes}, Jennifer and {Karagiorgi}, Georgia},
        title = "{Finding the Remnants of the Milky Way's Last Neutron Star Mergers}",
      journal = {\apj},
         year = 2019,
        month = jul,
       volume = {880},
       number = {1},
          eid = {23},
        pages = {23},
          doi = {10.3847/1538-4357/ab2593},
archivePrefix = {arXiv},
       eprint = {1905.03793},
 primaryClass = {astro-ph.HE},
       adsurl = {https://ui.adsabs.harvard.edu/abs/2019ApJ...880...23W}
}

@article{Terada:2022hut,
    author = "Terada, Yukikatsu and Miwa, Yuya and Ohsumi, Hayato and Fujimoto, Shin-ichiro and Katsuda, Satoru and Bamba, Aya and Yamazaki, Ryo",
    title = "{Gamma-ray Diagnostics of r-process Nucleosynthesis in the Remnants of Galactic Binary Neutron-Star Mergers}",
    eprint = "2205.05407",
    archivePrefix = "arXiv",
    primaryClass = "astro-ph.HE",
    doi = "10.3847/1538-4357/ac721f",
    month = "5",
    year = "2022"
}

@ARTICLE{2025arXiv250707785F,
       author = {{Fl{\"o}rs}, Andreas and {Ferreira da Silva}, Ricardo and {Marques}, Jos{\'e} P. and {Sampaio}, Jorge M. and {Mart{\'\i}nez-Pinedo}, Gabriel},
        title = "{Calibrated Lanthanide Atomic Data for Kilonova Radiative Transfer. I. Atomic Structure and Opacities}",
      journal = {arXiv e-prints},
         year = 2025,
        month = jul,
          eid = {arXiv:2507.07785},
        pages = {arXiv:2507.07785},
          doi = {10.48550/arXiv.2507.07785},
archivePrefix = {arXiv},
       eprint = {2507.07785},
 primaryClass = {astro-ph.HE},
       adsurl = {https://ui.adsabs.harvard.edu/abs/2025arXiv250707785F}
}

@ARTICLE{2022ApJ...939....8D,
       author = {{Domoto}, Nanae and {Tanaka}, Masaomi and {Kato}, Daiji and {Kawaguchi}, Kyohei and {Hotokezaka}, Kenta and {Wanajo}, Shinya},
        title = "{Lanthanide Features in Near-infrared Spectra of Kilonovae}",
      journal = {\apj},
         year = 2022,
        month = nov,
       volume = {939},
       number = {1},
          eid = {8},
        pages = {8},
          doi = {10.3847/1538-4357/ac8c36},
archivePrefix = {arXiv},
       eprint = {2206.04232},
 primaryClass = {astro-ph.HE},
       adsurl = {https://ui.adsabs.harvard.edu/abs/2022ApJ...939....8D}
}

@INPROCEEDINGS{2019BAAS...51g.123M,
       author = {{Miller}, Richard and {Ajello}, M. and {Beacom}, J.~F. and {Bloser}, P.~F. and {Burrows}, A. and {Fryer}, C.~L. and {Goldsten}, J.~O. and {Hartmann}, D.~H. and {Hoeflich}, P. and {Hungerford}, A. and {Lawrence}, D.~J. and {Leising}, M.~D. and {Milne}, P. and {Peplowski}, P.~N. and {Shirazi}, F. and {Sukhbold}, T. and {The}, L.-S. and {Yokley}, Z. and {Young}, C.~A.},
        title = "{Ex Luna Scientia: The Lunar Occultation eXplorer (LOX)}",
    booktitle = {Bulletin of the American Astronomical Society},
         year = 2019,
       volume = {51},
        month = sep,
          eid = {123},
        pages = {123},
          doi = {10.48550/arXiv.1907.07005},
archivePrefix = {arXiv},
       eprint = {1907.07005},
 primaryClass = {astro-ph.IM},
       adsurl = {https://ui.adsabs.harvard.edu/abs/2019BAAS...51g.123M}
}
\bibliographystyle{aasjournalv7}
\appendix
\startlongtable
\begin{deluxetable*}{LcLccc}
\tablewidth{0pt}
\tablecaption{Nuclei which contribute to the $\gamma$-ray spectra \label{tab:gammaray}\\}
\tablehead{
\colhead{Nuclei} & \colhead{$T_{1/2}$} & \colhead{Ancestor(s)} & \colhead{$T_{1/2}$}  & \colhead{Simulations Present} & \colhead{Spectral Lines [keV]$^{[1]}$ } 
}
\startdata
\nuclei{24}{Na} & 14.96 hr &  &   & A , B , C , D & 1369 , \textbf{2754 , 3866}  \\
\nuclei{28}{Al} & 2.245 min & \nuclei{28}{Mg} & 20.92 hr   & D & 1779  \\
\nuclei{38}{S} & 170.3 min &  &    & A , D & 1941  \\
\nuclei{38}{Cl} & 37.23 min &  \nuclei{38}{S} & 170.3 min    & A , B & 3810 , 3936  \\
\nuclei{41}{Ar} & 109.6 min &  &    & A , B & 1294  \\
\nuclei{42}{K} & 12.36 hr & \nuclei{42}{Ar} & 32.9 yr   & A , B , C , D & \textbf{312.6} , 692.0 , 899.7 , 1021 , \textbf{1525} , 1921 , 2424\\
\nuclei{43}{K} & 22.3 hr & &   & A & 372.8 , 396.9 , 593.4 , 617.5 , 1022\\
\nuclei{44}{K} & 22.13 min & &   & A , B & 3661 , 4409 , 4866 , 5162\\
\nuclei{47}{Ca} & 4.536 day & &   & A , B & \textbf{489.2} , 530.6 , 767.1 , 807.9 ,\textbf{1297}   \\
\nuclei{47}{Sc} & 3.349 day & \nuclei{47}{Ca} & 4.536 day & A , B , C , D & \textbf{159.4} \\
\nuclei{56}{Mn} & 2.579 hr & & & A , B & \textbf{846.8 , 1811 , 2113 , 2523 , 2658 , 2960 , 3370}\\
\nuclei{59}{Fe} & 44.49 day & &   & A , B & 142.7 , 192.3 , 334.8 , 382.0 , \textbf{1099 , 1292} , 1482 \\
\nuclei{60}{Co} & 1925 day & \nuclei{60}{Fe} & 2.62 Myr & A , B , C & 347.1 , 826.1 ,\textbf{1173 , 1332} , 2159\\
\nuclei{65}{Ni} & 2.518 hr & & & A , B & 366.3 ,   1116 , 1482\\
\nuclei{66}{Cu} & 5.120 min & \nuclei{66}{Ni} & 54.6 hr   & A , B & \textbf{1039} \\
\nuclei{67}{Cu} & 61.83 hr & & & A , B , C & \textbf{184.6} , 300.2 , 393.5\\
\nuclei{72}{Zn} & 46.5 hr & & & A , B , C & 102.8 , 112.1 , \textbf{144.7} , 191.5 \\ 
\nuclei{72}{Ga} & 14.10 hr & \nuclei{72}{Zn}& 46.5 hr& A , B , C & \textbf{336.7 , 381.7 , 600.9 , 630.0 , 786.5 , 810.3 ,}  \\ 
& & & & & \textbf{834.1 , 894.3 , 970.8, 1051 , 1231 , 1260 ,}\\
& & & & & \textbf{1277 , 1464 , 1597 , 1862 , 2109 , 2202 ,}\\
& & & & & \textbf{2491 , 2508 , 2844 , 2981 , 3094}\\
\nuclei{73}{Ga} & 4.86 hr & & & A , B & \textbf{297.3} , 325.7 \\
\nuclei{75}{Ge} & 82.78 min & & & A , B & 198.6 , 264.6 \\ 
\nuclei{77}{Ge} & 11.21 hr & & & A , B , C , D & \textbf{156.4 , 177.3 , 194.7 , 211.0 , 215.5 , 264.5 ,}\\
& & & & & \textbf{338.6 , 367.5 , 416.4 , 419.7 , 461.4 , 475.5 ,}\\
& & & & & \textbf{520.66 , 557.9 , 582.6 , 631.9 , 634.4 , 673.1 ,}\\
& & & & & \textbf{714.4 , 745.8 , 749.9 ,  766.8 , 781.3 , 784.8 ,}\\
& & & & & \textbf{810.4 , 875.2 , 907.0 , 923.1 , 925.5 , 928.9 ,}\\
& & & & & \textbf{1085 , 1193 , 1242 , 1264 , 1309 , 1313 ,}\\
& & & & & \textbf{1320 , 1368 , 1453 , 1477 , 1479 , 1496 ,}\\
& & & & & \textbf{1539 , 1574 , 1710 , 1720 , 1727 , 1847 ,}\\
& & & & & \textbf{2000 , 2077 , 2089 , 2126 , 2342}\\
\nuclei{77}{As} & 38.79 hr & & & A , B , C & \textbf{239.0 , 249.8} , 281.6 ,\textbf{520.7} \\ 
\nuclei{78}{Ge}&88.0 min & & & A , B , C , D & \textbf{277.3} , 293.9 \\
\nuclei{78}{As} &90.7 min & \nuclei{78}{Ge} & 88.0 min & A , B , C , D &  \textbf{174.2 , 354.3 , 462.2 , 497.0 , 503.7 , 545.3 ,}\\
& & & & & \textbf{551.8 , 613.8 , 637.1 , 657.9 , 686.3 , 687.5 ,}\\
& & & & & \textbf{694.9 , 722.4 , 828.1 , 841.5 , 842.6 , 882.0 ,}\\
& & & & & \textbf{884.9 , 959.0 , 968.2 , 988.2 , 1005 ,  1019 ,}\\
& & & & & \textbf{1080 , 1145 , 1240 , 1309 , 1374 , 1381 ,}\\
& & & & & \textbf{1530 , 1713 , 1792 , 1836 , 1921 , 1934 ,}\\
& & & & & \textbf{1996 , 2068 , 2188 , 2225 , 2616 , 2681 ,}\\
& & & & & \textbf{2759 , 2798 , 2839 , 3098}\\
\nuclei{84}{Br} & 31.76 min & &   & B , C & 3235 , 3266 , \textbf{3928} , 4085 \\
\nuclei{85}{Kr}& 10.74 yr & & & B , C , D & \textbf{514.0} \\
\nuclei{86}{Br} & 55.1 s & \text{Fission} &   & C , D & 5406 , 5518 , 6212 \\
\nuclei{87}{Kr} & 76.3 min & & & B , C & 402.6 ,  2555 , 2558 , 3235 , 3266 ,  3309 , 3705  \\ 
\nuclei{88}{Br} & 16.34 s & \text{Fission} &  & C , D &  3932, 4148 , 4563 , 6999 \\
\nuclei{88}{Kr} & 2.825 hr & & & B , C , D & 122.3 ,166.0 , 196.3 , 2196 , 2392\\
\nuclei{88}{Rb} & 17.77 min & \nuclei{88}{Kr} & 2.825 hr   &B , C , D & 1836  , 3010 , 3218 , 3486 , \textbf{4742}\\
\nuclei{90}{Rb} & 158 s & \text{Fission} &    & C , D & 4136 , 4366 \\
\nuclei{90}{Y} & 64.05 hr & \nuclei{90}{Sr} &  28.91 yr   & B , C & 1761\\
\nuclei{91}{Sr} & 9.65 hr & & & B , C , D & 118.5 , 652.3 , 652.9 , 749.8 ,1024\\ 
\nuclei{91}{Y} & 58.51 day & &   & B , C , D & 1205\\
\nuclei{92}{Rb} & 4.48 s & \text{Fission} &    & C , D & 4638 , 4809 , 4836 , 4923 , 5188, \\
& & & & & 5215 , 5249 , 5302 , 5377 , 5498 , 5574 , 5584,\\
& & & & & 5632 , 5739 , 5900 , 6004 , 6030 , 6115\\
\nuclei{92}{Sr} & 2.61 hr & &   & C , D & 1384\\
\nuclei{92}{Y} & 3.54 hr & \nuclei{92}{Sr} &  2.61 hr   & B , C & 1405 , 3264 , 3371\\
\nuclei{93}{Y} & 10.18 hr & & & C &266.9 , 1918 , 2191 \\
\nuclei{95}{Zr} & 64.03 day & & & B , C , D& \textbf{724.2} , 756.7\\
\nuclei{95}{Nb} & 34.99 day & \nuclei{95}{Zr} & 64.03 day & B , C , D &  204.1 , \textbf{765.8} \\
\nuclei{97}{Y} & 3.75 s & \text{Fission} &   & C , D & 3288 , 3401\\
\nuclei{97}{Nb} & 72.1 min & \nuclei{97}{Zr} & 16.75 hr & B & 657.9 \\
\nuclei{98}{Sr} & 0.653 s & \text{Fission} & & D & 119.4 \\
\nuclei{98}{Y} & 0.548 s  & \text{Fission} &    & C , D & 4399 , 4452 , 4492\\
\nuclei{99}{Mo} & 65.92 hr & & & B , C , D &142.7 , 181.1 , 739.5\\
\nuclei{103}{Ru} & 39.25 day & & & B , C , D & \textbf{497.1} \\ 
\nuclei{105}{Ru} & 4.439 hr & & & C & 724.2 \\
\nuclei{105}{Rh} & 35.34 hr & & & B , C & 306.3 , 319.2 \\
\nuclei{106}{Rh} & 30.07 s & \nuclei{106}{Ru} & 371.8 day & B , C , D & \textbf{511.9 , 621.9 , 873.5 , 1050 , 1128 , 1195 ,}\\
& & & & & \textbf{1496 , 1562 , 1766 , 1797 , 1927 , 1988 ,}\\
& & & & & \textbf{2112 , 2366 , 2406 , 2543 , 2705 , 2710 ,}\\
& & & & & \textbf{2821 , 3037 , 3273}\\
\nuclei{111}{Ag} &7.45 day & & & B , C , D & 245.4 , 342.1 \\
\nuclei{112}{Ag} & 3.130 hr & \nuclei{112}{Pd} & 21.04 hr & B , C , D & 617.5 ,1614 , 2106 , 2507\\
\nuclei{113}{Ag} &5.37 hr & & & C & 298.6 \\
\nuclei{116}{Pd} & 11.8 s & \text{Fission} & & D & 114.7 \\
\nuclei{117}{Cd} & 2.49 hr & & & D & 273.3 ,1577 \\
\nuclei{117}{In} & 43.2 min & \nuclei{117}{Cd} & 2.49 hr & C , D & \textbf{158.6} , 552.9 \\
\nuclei{123}{Sn} & 129.2 day & & & B , C , D& 160.3 , 1030 , \textbf{1089}\\ 
\nuclei{125}{Sn} & 9.64 day & & & B , C , D&  332.1 , 822.5 , 915.6 , \\
 &  & & & &\textbf{1067 , 1089 , 1420 , 1806 , 2002 , 2276}\\
\nuclei{125}{Sb} & 2.759 yr & & & B , C , D & \textbf{117.0 , 172.7 , 176.3 , 204.1 , 208.1 , 227.9 ,}\\
& & & & & \textbf{321.0 , 380.5 , 408.1 , 427.9 , 443.6 , 463.4 ,}\\
& & & & & \textbf{600.6 , 606.7 , 636.0 , 671.4}\\
\nuclei{126}{Sb} & 12.35 day & \nuclei{126}{Sn} & 218,000 yr & B , C , D & \textbf{414.7 , 556.3 , 573.9 , 593.2 , 656.3 , 666.5 ,}\\
& & & & & \textbf{674.8 , 695.0 , 697.0 , 720.7 , 856.8 , 953.7}\\
& & & & & \textbf{989.6 , 1036 , 1213}\\
 & &  &     &  & \textbf{1064 , 1191 , 1213 , 1290 , 1477 , 1589}\\
\nuclei{127}{Sn} & 2.10 hr & & & B , C , D & 104.1 , 110.1 , 119.7 , 124.0 , 141.9 , 143.7 ,\\
& & & & & 169.2 , 184.7 , 190.1 , 202.8 , 232.2 , 262.5,\\
& & & & & 266.2 , 284.3 , 292.9 , 390.5 , 407.1 , 438.2 ,\\
& & & & & 490.9 , 493.2 , 500.7 , 509.0 , 509.7 , 545.4 , \\
& & & & & 583.3 , 592.3 , 805.9 , 823.1 , 824.7 , 859.5 , \\
& & & & & 916.5 , 979.2 , 997.9 , 1003 , 1036 , 1093 ,\\
& & & & & 1096 , 1114 , 1159 , 1160 , 1179 , 1221 ,\\
& & & & & 1292 , 1368 , 1434 , 1471 , 1473 , 1584 ,\\
& & & & & 1648 , 1667 , 2003 , 2102 , 2317 , 2448 ,\\
& & & & & 2585 , 2696 , 2806 , 2846 , 2881 \\
\nuclei{127}{Sb} & 3.85 day & & & B , C , D & \textbf{154.3 , 252.4 , 280.4 , 290.8 , 293.3 , 310.0 ,}\\
& & & & & \textbf{391.8 , 412.1 , 441.0 , 445.1 , 451.0 , 456.0 ,}\\
& & & & & \textbf{473.0 , 502.9 , 543.3 , 584.2 , 603.5 , 637.8 ,}\\
& & & & & \textbf{652.3 , 667.5 , 682.3 , 685.7 , 698.5 , 722.2 ,}\\
& & & & & \textbf{745.9 , 783.7 , 817.0 , 820.6 , 924.4 , 1142 ,}\\
& & & & & \textbf{1290}\\
\nuclei{127}{Te} & 9.35 hr & \nuclei{127}{Sb} & 3.85 day & B , C & 417.9 \\
\nuclei{128}{In} & 0.84 s & \text{Fission}& & D & 1169 , 3520 , 4298\\
\nuclei{128}{Sn} & 59.07 min & & & C , D & 152.7 , 482.3\\
\nuclei{128}{Sb} &  9.05 hr & & & B , C , D & 118.4 , 152.6 , 204.4 , 214.8 , 227.3 , 249.7 ,\\
& & & & & \textbf{314.1 , 317.7 , 322.3 , 357.0 , 366.1 , 445.7 ,}\\
& & & & & \textbf{454.5 , 459.5 , 526.5 , 582.9 , 594.3 , 603.0 ,}\\
& & & & & \textbf{628.7 , 636.2 , 654.2 , 667.1 , 683.9 , 692.9 ,}\\
& & & & & \textbf{727.6 , 743,3 , 754.0 , 773.7 , 802.7 , 813.6 ,}\\
& & & & & \textbf{835.8 , 845.8 , 878.0 , 908.8 , 972.3 , 1048 ,}\\
& & & & & \textbf{1079 , 1113 , 1130 , 1158 , 1182 , 1251 ,}\\
& & & & & \textbf{1260 , 1340 , 1378 , 1593 , 1686 , 1708 ,}\\
& & & & & \textbf{1786}\\
\nuclei{129}{Sb} & 4.366 hr & & & C , D & \textbf{359.2 , 544.6 , 813.0 , 915.0 , 966.8 , 1031 ,}\\
& & & & & \textbf{1209 , 1263 , 1570 , 1656 , 1738 , 1772 ,}\\
& & & & & \textbf{2071 , 2115}\\
\nuclei{129}{Te} & 69.6 min & \nuclei{129}{Sb} & 4.366 hr & C , D & 459.6 \\
\nuclei{130}{Sb} & 39.5 min & & & C , D & \textbf{182.3} , 330.9 , 793.4\\
\nuclei{131}{I} & 8.025 day & & & B , C , D & 284.3 , \textbf{364.5} , 637.0 , 722.9\\
\nuclei{132}{Te} & 3.204 day & & & C , D & 111.8 , 116.3 , \textbf{228.2} \\
\nuclei{132}{I} & 2.295 hr & \nuclei{132}{Te} & 3.204 day & C , D& 522.7 , 630.2 , \textbf{667.7 , 772.6} , 954.6 , 1035 ,\\
& & & & & 1136 , 1143 , 1173 , 1291 , 1295 , 1298 ,\\
& & & & & 1372 , 1399 , 1443 , 1757 , 1921 , 2002 ,\\
& & & & & 2087 , 2173 , 2223 , 2390 , 2525\\
\nuclei{134}{I} & 52.5 min & \nuclei{134}{Te} & 41.8 min & D & 847 , 884 , 1073 , 1136 , 1455 , 1614 , 1741 , 1807\\
\nuclei{135}{I} & 6.58 hr & &   & C , D & 1132 , 1260 , 1458 , 1678 , 1791\\
\nuclei{138}{Cs} & 32.5 min & \text{Fission} &     & D & 1436\\
\nuclei{140}{Ba}&  12.75 day & & & C , D & 162.7 , 304.8 , 537.3 \\
\nuclei{140}{La}& 1.679 day & \nuclei{140}{Ba} &12.75 day & C , D & 328.8 , 487.0 , 815.8 , \textbf{1596} , 2521\\
\nuclei{141}{Ce}& 32.50 day & & & C , D & \textbf{145.4} \\
\nuclei{142}{La}& 91.1 min & & & C , D & \textbf{641.3 , 2971 , 3012 , 3034 , 3047 , 3076 ,}\\
& & & & & \textbf{3102 , 3122 , 3154 , 3180 , 3273 , 3314 ,}\\
& & & & & \textbf{3402 , 3459 , 3612 , 3633 , 3719 , 3850}\\
\nuclei{143}{Ce}& 33.04 hr & & & D & 293.3\\
\nuclei{144}{Ce}& 284.9 day & & & C , D & \textbf{133.5}\\
\nuclei{144}{Pr}& 17.28 min & \nuclei{144}{Ce} &  284.9 day & C , D & 696.5 , 1489 , \textbf{2186}\\
\nuclei{147}{Nd} & 11.03 day & & & D & 531.0\\
\nuclei{149}{Nd} & 1.726 hr & & & C & 114.3\\
\nuclei{151}{Pm} & 28.40 hr & & & C & 104.8\\
\nuclei{153}{Sm} & 46.28 hr & & & C , D & \textbf{103.2} \\
\nuclei{155}{Eu} & 4.753 yr & & & C , D & 105.3\\
\nuclei{156}{Sm} & 9.4 hr & & & C , D & 165.8 , 204.0\\
\nuclei{156}{Eu} & 15.19 day  & & & C , D & \textbf{1040 , 1065 , 1079 , 1154 , 1231 , 1242 ,}\\
& & & & & \textbf{1277 , 1366 , 1877 , 1938 , 1966 , 2027 ,}\\
& & & & & \textbf{2098 , 2181 , 2187 , 2205 , 2270}\\
\nuclei{157}{Eu} & 15.18 hr & & & D & 370.5 , \textbf{410.7}\\
\nuclei{167}{Ho} & 2.98 hr & & & D & \textbf{346.5}\\
\nuclei{171}{Er} & 7.516 hr & & & D & 295.9 , 308.3\\
\nuclei{172}{Er} & 49.3 hr & & & C , D & 407.3 , 610.1\\
\nuclei{172}{Tm} & 63.6 hr & \nuclei{172}{Er} &  49.3 hr   & C , D & 1094 , 1387 , 1466 , 1530 , 1608\\
\nuclei{173}{Tm} & 8.24 hr & & & D & 398.9\\
\nuclei{175}{Yb} & 4.185 day & & & C , D & 396.3\\
\nuclei{181}{Hf} & 42.39 day & & & C , D & \textbf{133.0} , 345.9 , 482.2\\
\nuclei{182}{Ta} & 114.74 day & \nuclei{182}{Hf} &  8.90 Myr   & D & 1121 , 1189 , 1221\\
\nuclei{183}{Ta} &5.1 day & & & C , D&133.0 , 346.0 , 482.2\\
\nuclei{184}{Hf} & 4.12 hr & & & C , D & \textbf{139.1} , 344.9\\
\nuclei{184}{Ta} & 8.7 hr & & & C , D & 111.2 , 215.3 , 252.9 , 414.0 , 1110 , 1174\\
\nuclei{187}{W} & 23.80 hr & & & C , D & 134.2\\
\nuclei{188}{W} & 69.78 day & & & C , D & 290.7\\
\nuclei{188}{Re} & 17.006 hr & \nuclei{188}{W} & 69.78 day & C , D & 155.0 , 1308 , 1457 , 1610\\
\nuclei{189}{Re} & 24.3 hr & & & D & 216.7 , 219.4\\
\nuclei{193}{Os} & 29.73 hr & $^{[2]}$ & & C , D & 138.9\\
\nuclei{194}{Ir} & 19.18 hr & \nuclei{194}{Os} & 6.0 yr & C , D & \textbf{293.5 , 328.5 , 622.7 , 645.2 , 938.7 , 1151 ,}\\
& & & & & \textbf{1184 , 1219 , 1294 , 1342 , 1469 , 1622 ,}\\
& & & & & \textbf{1797 , 1806 , 2044 , 2114}\\
\nuclei{195}{Ir} & 2.29 hr & & & C , D  & 129.7 \\
\nuclei{197}{Pt} & 19.89 hr & & & D & 197.6\\
\nuclei{199}{Au} & 3.139 day & & & D & 158.4\\
\nuclei{200}{Au} & 48.4 min & \nuclei{200}{Pt} &  12.6 hr   & D &1225 , 1263\\
\nuclei{203}{Hg} & 46.61 day & & & C , D & 279.2\\
\nuclei{207}{Tl} & 4.77 min & \nuclei{231}{Pa} & 32,570 yr & C , D & 897.8\\
\nuclei{208}{Tl}$^{[3]}$ & 3.053 min & \nuclei{212}{Pb} & 10.62 hr & C , D & 277.4 ,\textbf{ 510.8 , 583.2 , 860.6, 2615}\\
& & \nuclei{224}{Ra} & 3.6316 day & &  \\
& & \nuclei{228}{Th} & 1.912 yr & &  \\
& & \nuclei{228}{Ra} & 5.75 yr & &\\
\nuclei{209}{Tl}$^{[4]}$ & 2.162 min & \nuclei{225}{Ra} & 14.9 day  & D & \textbf{117.2 , 465.1 , 1567}\\
& & \nuclei{229}{Th} & 7880 yr & &   \\
\nuclei{211}{Pb} & 36.1 min & \nuclei{223}{Ra} & 11.43 day& C , D & \textbf{404.9 , 427.1}\\
& & \nuclei{227}{Th} & 18.70 day & &  \\
& & \nuclei{227}{Ac} & 21.77 yr& &  \\
& & \nuclei{231}{Pa} & 32,570 yr& &  \\
\nuclei{212}{Pb} & 10.62 hr & \nuclei{224}{Ra} & 3.6316 day  & C , D & \textbf{238.6} , 300.1 \\
& & \nuclei{228}{Th} & 1.912 yr & & \\
& & \nuclei{228}{Ra} & 5.75 yr &  &\\
\nuclei{212}{Bi} &60.55 min & \nuclei{232}{U}& 68.9 yr& D & 727.3 \\ 
\nuclei{213}{Bi} & 45.59 min & \nuclei{225}{Ra} & 14.9 day  & C , D & \textbf{440.5}\\
& & \nuclei{229}{Th} & 7880 yr & &   \\
& & \nuclei{233}{U} & 159,190 yr & &  \\
\nuclei{214}{Pb} & 27.06 min & \nuclei{222}{Rn} & 3.822 day & C , D & \textbf{242.0 , 295.2 , 351.9}\\
& & \nuclei{226}{Ra} & 1600 yr & &  \\
& & \nuclei{230}{Th} & 75,584 yr & &  \\
\nuclei{214}{Bi} & 19.71 min & \nuclei{226}{Ra} & 1600 yr & C , D & \textbf{ 386.8 , 388.9 , 609.3 , 768.4 , 934.1 , 1120 ,}\\
& &\nuclei{230}{Th} & 75,584 yr &  & \textbf{1155 , 1238 , 1281 , 1378 , 1402 , 1408 ,}\\
& & & & & \textbf{1509 , 1661 , 1730 , 1764 , 1847 , 2119 ,}\\
& & & & & \textbf{2204 , 2448}\\
\nuclei{219}{Rn}$^{[5]}$ & 3.96 s & \nuclei{227}{Ac} & 21.77 yr & C , D & 271.2 , 401.8 \\
& & \nuclei{231}{Pa} & 32,570 yr & &   \\
\nuclei{221}{Fr}$^{[6]}$ & 22.00 min & \nuclei{225}{Ac} & 9.920 day &  C , D & 134.6 , 204.9 , 234.8\\
& & \nuclei{225}{Ra} & 14.9 day & &  \\
& & \nuclei{229}{Th} & 7880 yr & &   \\
\nuclei{222}{Rn}$^{[6]}$ & 3.822 day & \nuclei{226}{Ra} & 1,600 yr & C , D & 510.0 \\
\nuclei{223}{Ra}$^{[6]}$ & 11.43 day &  \nuclei{227}{Ac} & 21.77 yr & C , D & 144.2 , 154.2\\
& & \nuclei{231}{Pa} & 32,570 yr & &   \\
\nuclei{224}{Fr} & 3.33 min & \nuclei{224}{Rn} & 107 min & D & 131.6\\
\nuclei{226}{Ra}$^{[6]}$ & 1600 yr & \nuclei{230}{Th} & 75,584 yr & C , D & 186.2 \\
\nuclei{227}{Th}$^{[6]}$ & 18.68 day &  \nuclei{227}{Ac} & 21.77 yr  & C , D & 210.6 , \textbf{236.0 , 256.2} , 286.1 , 289.6 , 300.0 , 304.5 \\
& & \nuclei{231}{Pa} & 32,570 yr & &  \\
\nuclei{228}{Ac} & 6.15 hr & \nuclei{228}{Ra} & 5.75 yr& C , D & 129.1 , 154.0 , 209.3 , 270.2 , 328.0 ,\\
& & & & & 338.3 , 409.4 , 463.0 , 794.9 , 835.7 , 911.2\\
& & & & & 964.8 , 969.0 , 1588 , 1631\\
\nuclei{229}{Th}$^{[6]}$ & 7880 yr & &  &  C , D & 137.0 , 193.5\\
\nuclei{233}{Pa} & 26.975 day & \nuclei{237}{Np} & 2.14 Myr  & C , D & \textbf{311.9} , 340.5\\
\nuclei{234}{Pa} & 6.70 hr & \nuclei{234}{Th} & 24.10 day & D & \textbf{131.3 , 152.7 , 226.5 , 227.3 , 293.8 , 369.5} \\
& & & & & \textbf{568.9 , 569.5 , 666.5 , 669.5 , 669.7 , 692.6 ,}  \\
& & & & & \textbf{699.0 , 705.9 , 733.4 , 738.0 , 742.8 , 755.0 ,}\\
& & & & & \textbf{786.3 , 796.1 , 805.8 , 819.2 , 824.2 , 825.1 ,}\\
& & & & & \textbf{831.5 , 876.0 , 880.6 , 883.2 , 898.7 , 925.0 ,}\\
& & & & & \textbf{926.0 , 926.7 , 946.0 , 947.7 , 980.3 , 984.2 ,}\\
& & & & & \textbf{1353 , 1394}\\
\nuclei{237}{U} & 6.752 day & & & D & \textbf{101.1}$^{[7]}$ , [113.3 - 117.9] $^{[7]}$ , \textbf{208.0}\\
\nuclei{239}{Np} &2.356 day & \nuclei{243}{Am} & 7364 yr & D & 106.1 , 209.8 , 228.2 , 277.6 \\
\nuclei{240}{Np} & 61.9 min & \nuclei{240}{U} & 14.1 hr & D & 103.7 $^{[7]}$ , [116.2-120.7]$^{[7]}$ , 152.7 ,\\
& & & & & 175.4 , 193.3 , 566.3 , 973.9\\
\nuclei{245}{Cm}$^{[6]}$ & 8243 yr & & & C , D &\textbf{103.7 , [116.2-121.0]$^{[7]}$ , 133.1 , 175.0}\\
\nuclei{246}{Pu} & 10.84 day & \nuclei{250}{Cm}$^{[8]}$& 8300 yr & D & \textbf{102.0$^{[7]}$ , 106.5$^{[7]}$ , [119.2 - 123.8]$^{[7]}$ , 180.0 , 223.8}\\
\nuclei{246}{Am} & 39 min & \nuclei{250}{Cm}$^{[8]}$ & 8300 yr & D & \textbf{127.4 , 153.5 , 205 , 679.2 , 756.0}\\
\nuclei{247}{Am} & 23.0 min & \nuclei{247}{Pu} & 2.27 day & D & 104.6$^{[7]}$ , 109.3$^{[7]}$ , 285.0\\
\nuclei{248}{Cm} & 348,000 yr & & & D & Prompt Fission \\
\nuclei{249}{Cf}$^{[6]}$ & 351 yr & & & C , D & 252.8 , \textbf{333.3 , 388.2} \\
\nuclei{250}{Bk} & 3.212 hr & \nuclei{250}{Cm}$^{[8]}$ & 8300 yr & D & 890.0 , \textbf{989.1} , \textbf{1032}\\
\nuclei{250}{Cm}$^{[8]}$ & 8300 yr & & & D & Prompt Fission\\
\nuclei{251}{Cf}$^{[6]}$ & 898 yr & & & C , D & \textbf{104.6$^{[7]}$ , 109.3 $^{[7]}$ , [122.3-127.4]$^{[7]}$ , 177.5 , 227.4}\\
\nuclei{252}{Cf} & 2.647 yr & & & D & Prompt Fission \\
\nuclei{254}{Cf} & 60.5 day & & & D & Prompt Fission \\
\nuclei{257}{Fm}$^{[6]}$ & 100.5 day & & & D & 109.9$^{[7]}$ , 115.1$^{[7]}$ , [128.6-134.0]$^{[7]}$\\
\nuclei{258}{Cf} & 2145 yr $^{[9]}$ & & & D & Prompt Fission\\
\nuclei{262}{Fm} & 39.85 yr$^{[9]}$ & & & D & Prompt Fission\\
\nuclei{265}{No} & 118.4 yr$^{[9]}$ & & & D & Prompt Fission\\
\nuclei{267}{Rf} & 60.5 day & \nuclei{267}{Lr}& 1.9 day$^{[9]}$ & D & Prompt Fission \\
\nuclei{273}{Rf} & 2.269 hr$^{[9]}$ & & & D & Prompt Fission\\
\enddata
\tablecomments{$^{[1]}$ Bolded spectral lines are more significant than unbolded lines.\\
$^{[2]}$ The lifetime of \nuclei{193}{Re} is not known, but it is estimated to be $\sim$ 40 s \citep{Gray:2016vsn}, and therefore does not affect the decay timescale of \nuclei{193}{Os}.\\
$^{[3]}$ \nuclei{212}{Bi} only $\alpha$ decays to \nuclei{208}{Tl} 35.94\% of the time, otherwise $\beta$-decaying to \nuclei{212}{Po} which directly $\alpha$ decays to \nuclei{208}{Pb}.\\
$^{[4]}$ \nuclei{213}{Bi} only $\alpha$ decays to \nuclei{209}{Tl} 2.14\% of the time, otherwise $\beta$-decaying to \nuclei{213}{Po} which directly $\alpha$ decays to \nuclei{209}{Pb}.\\
$^{[5]}$ \nuclei{214}{Bi} only $\alpha$ decays to \nuclei{210}{Tl} 0.02\% of the time, otherwise $\beta$-decaying to \nuclei{214}{Po} which directly $\alpha$ decays to \nuclei{210}{Pb}.\\
$^{[6]}$ These nuclei decay via $\alpha$ decay.\\
$^{[7]}$ These spectral lines come from x-ray transitions.\\
$^{[8]}$ The branching ratio of \nuclei{250}{Cm} is uncertain, estimated as 74\% fission, 18\% of the time $\alpha$ to \nuclei{246}{Pu}, and 8\% of the time $\beta-$ to \nuclei{250}{Bk}.\\
$^{[9]}$  The lifetime of these nuclei are unknown, and the values used here are estimated from systematics.\\
}
\end{deluxetable*}
\section{Full Version of Figures 3 and 4}
\includegraphics{LowETime1SimA.pdf}
\includegraphics{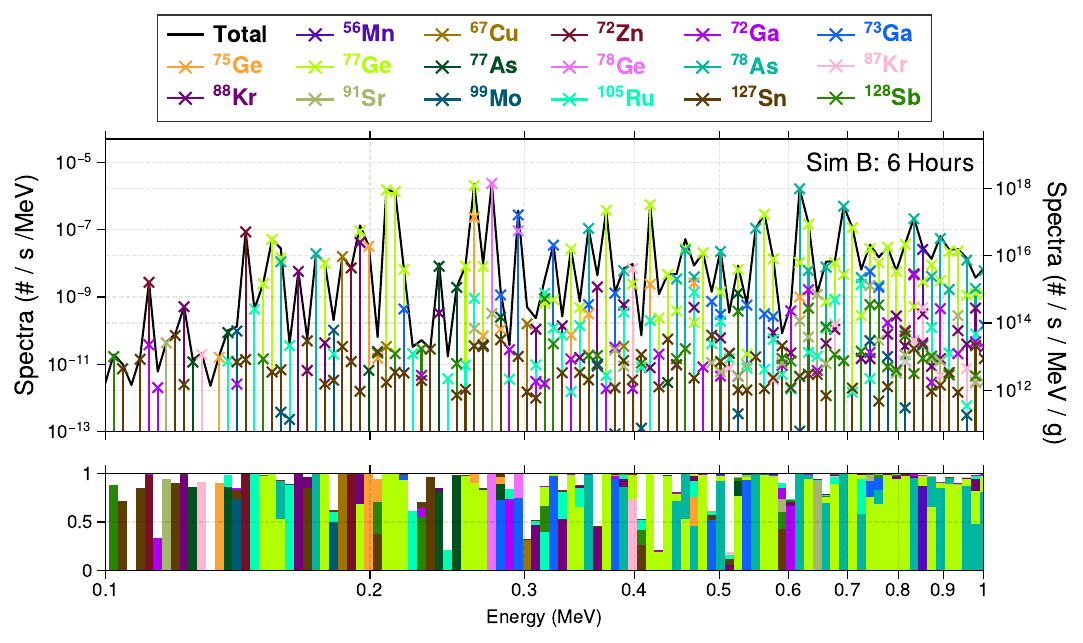}
\includegraphics{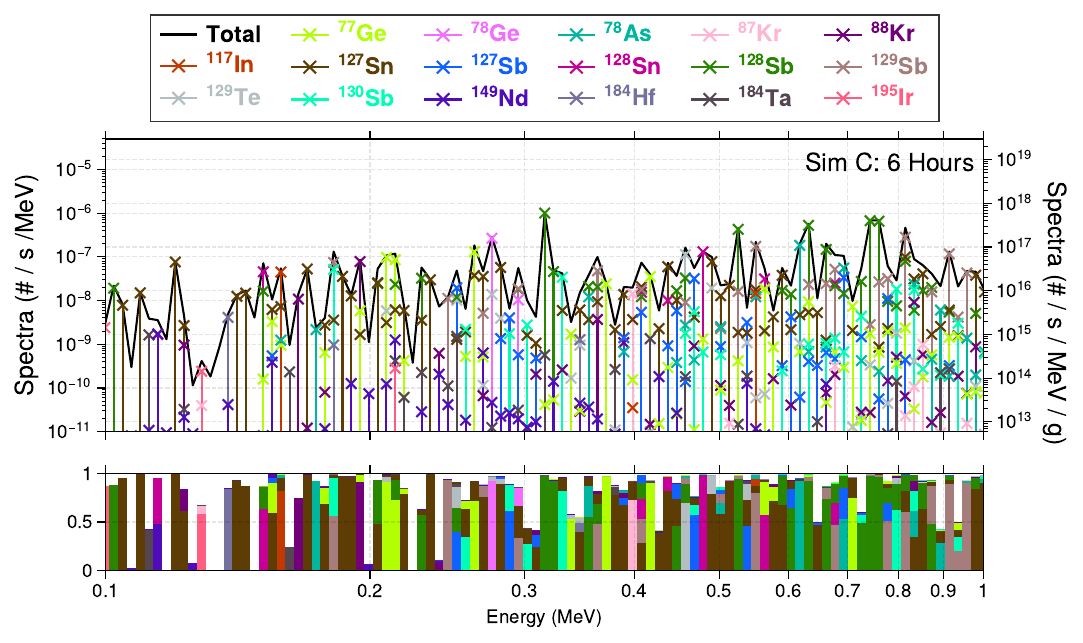}
\includegraphics{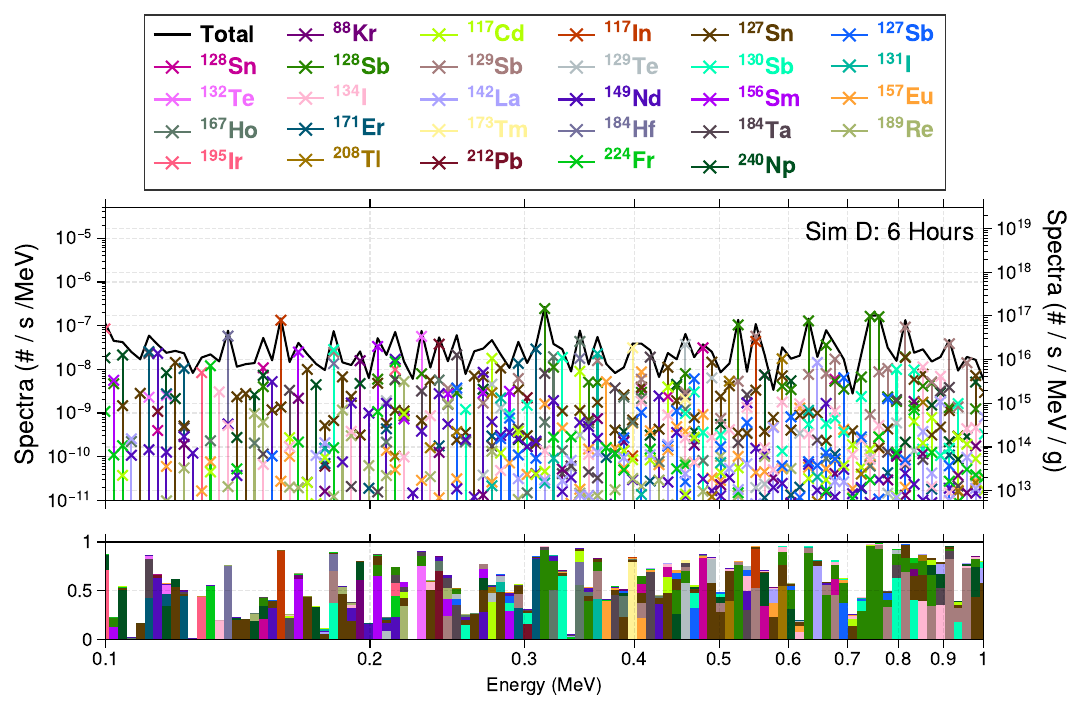}

\includegraphics{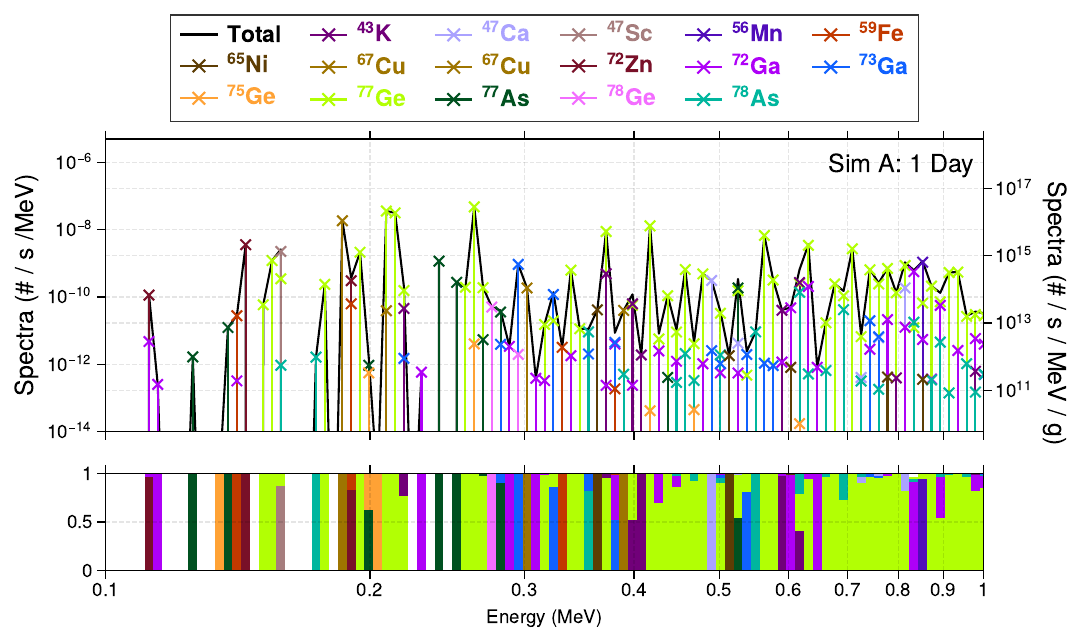}
\includegraphics{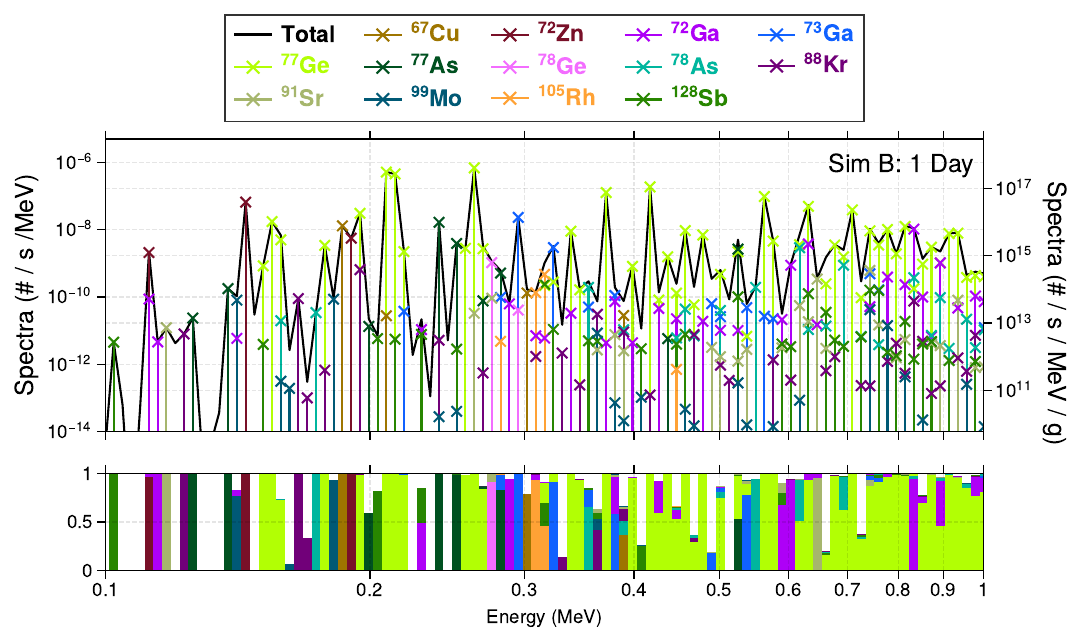}
\includegraphics{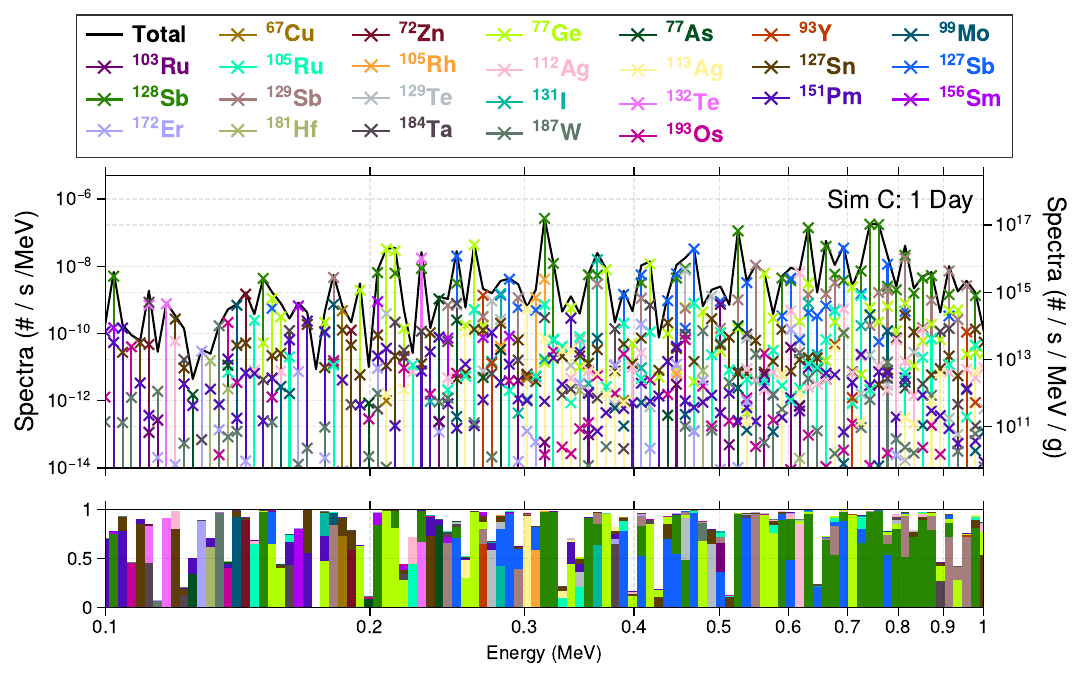}
\includegraphics{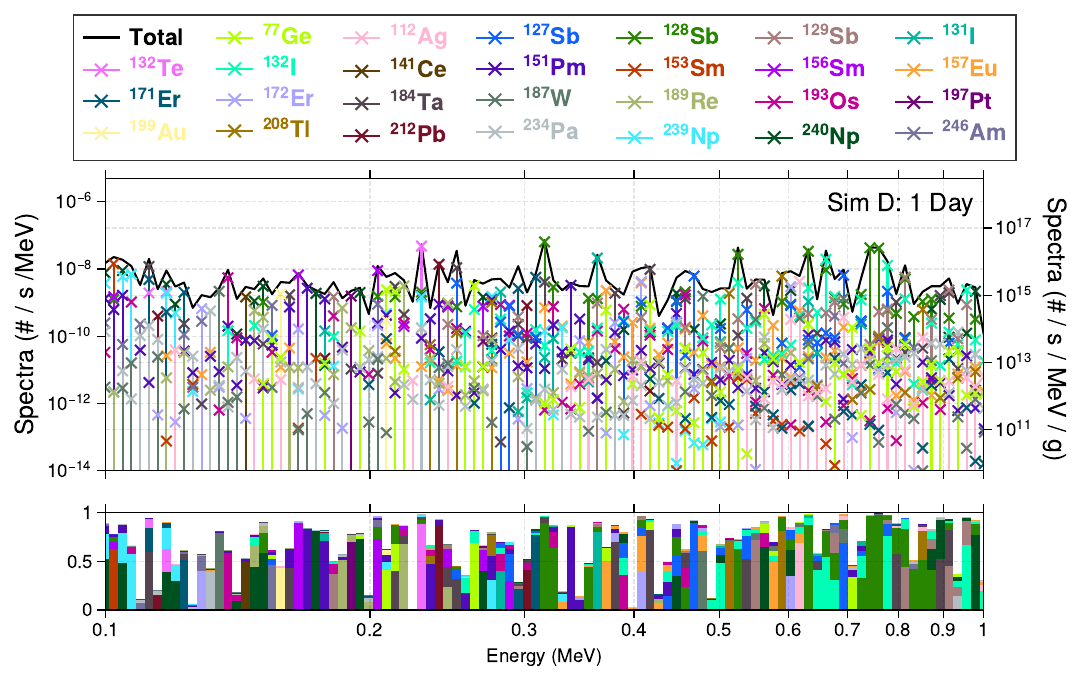}

\includegraphics{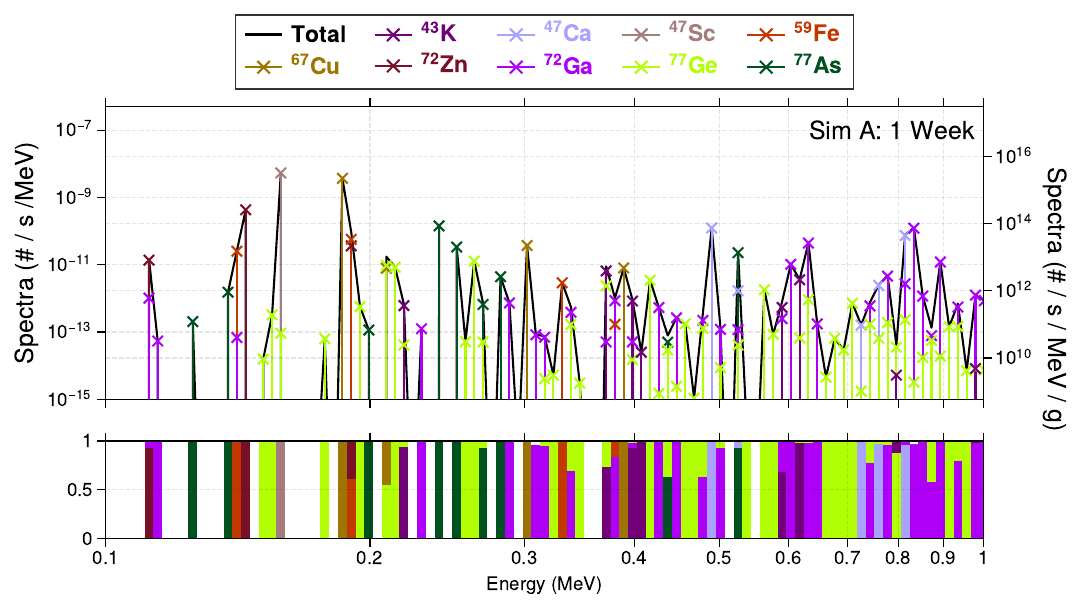}
\includegraphics{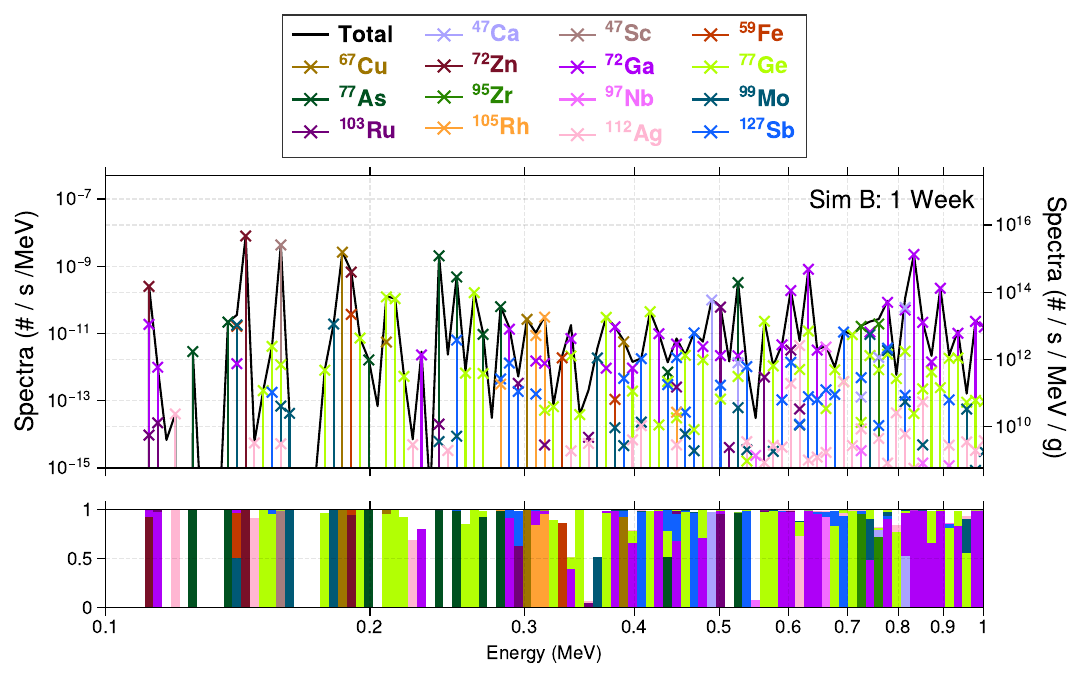}
\includegraphics{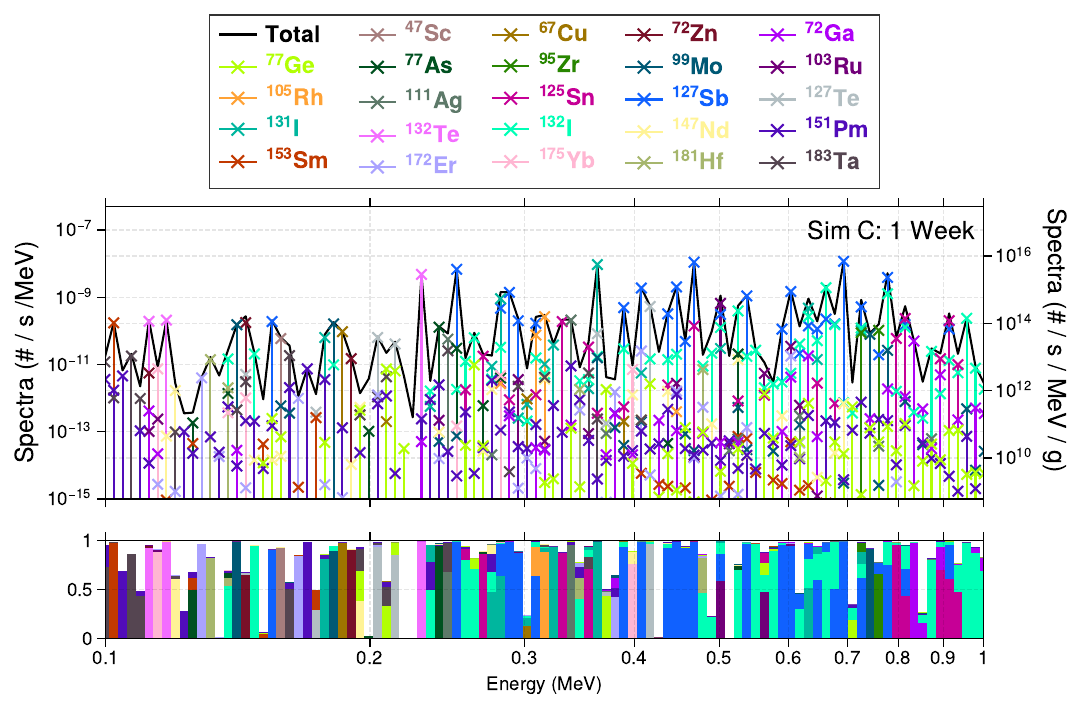}
\includegraphics{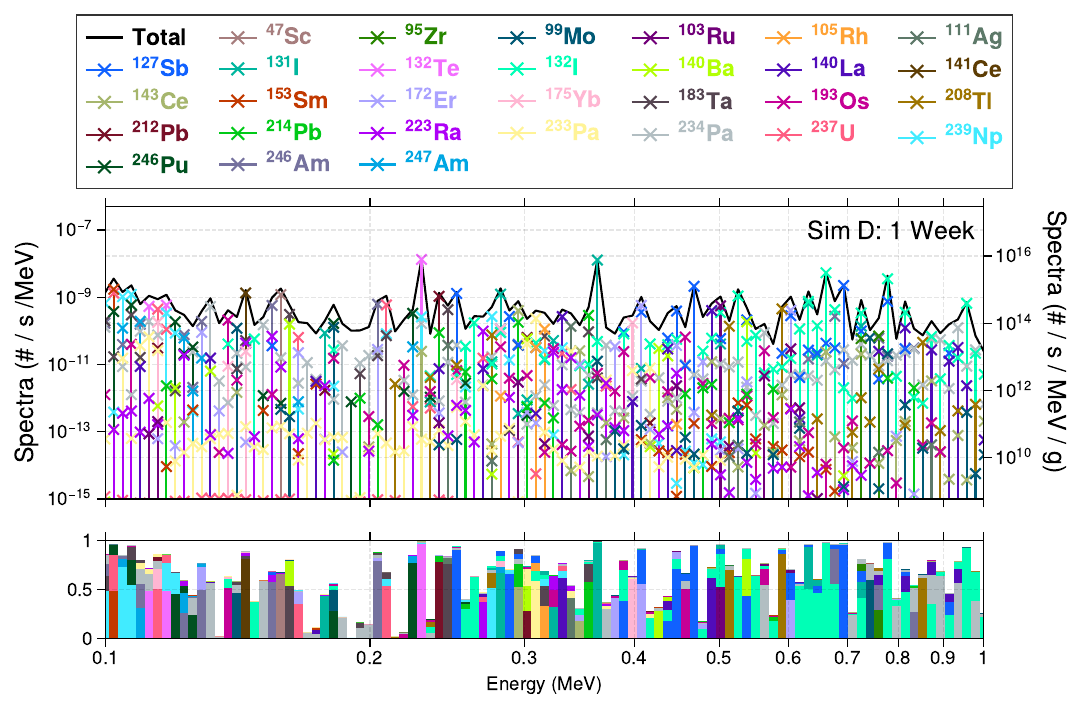}

\includegraphics{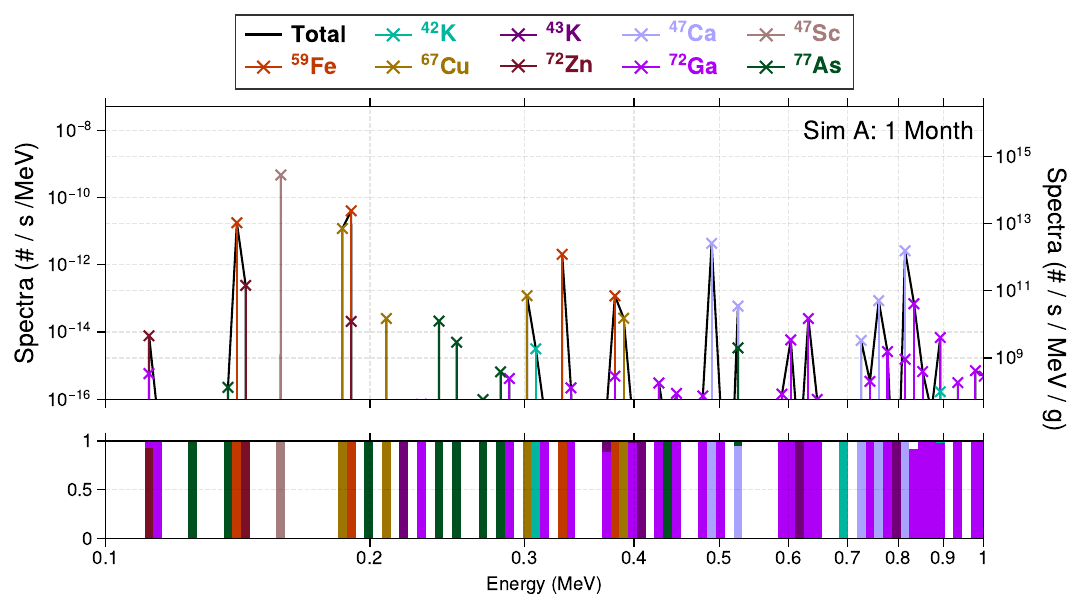}
\includegraphics{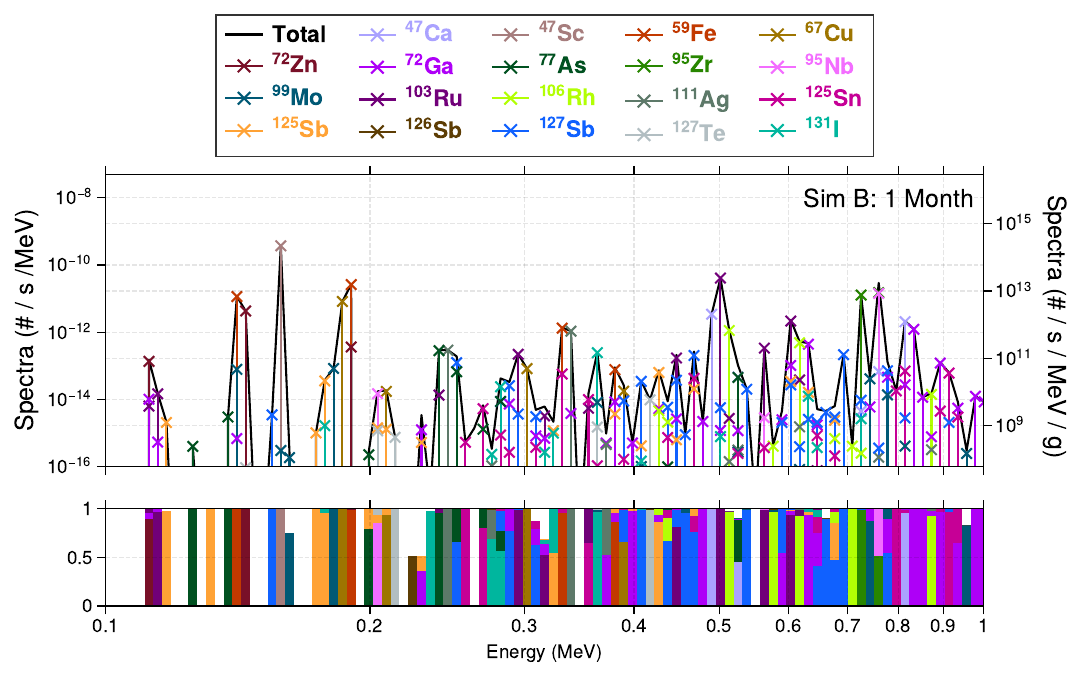}
\includegraphics{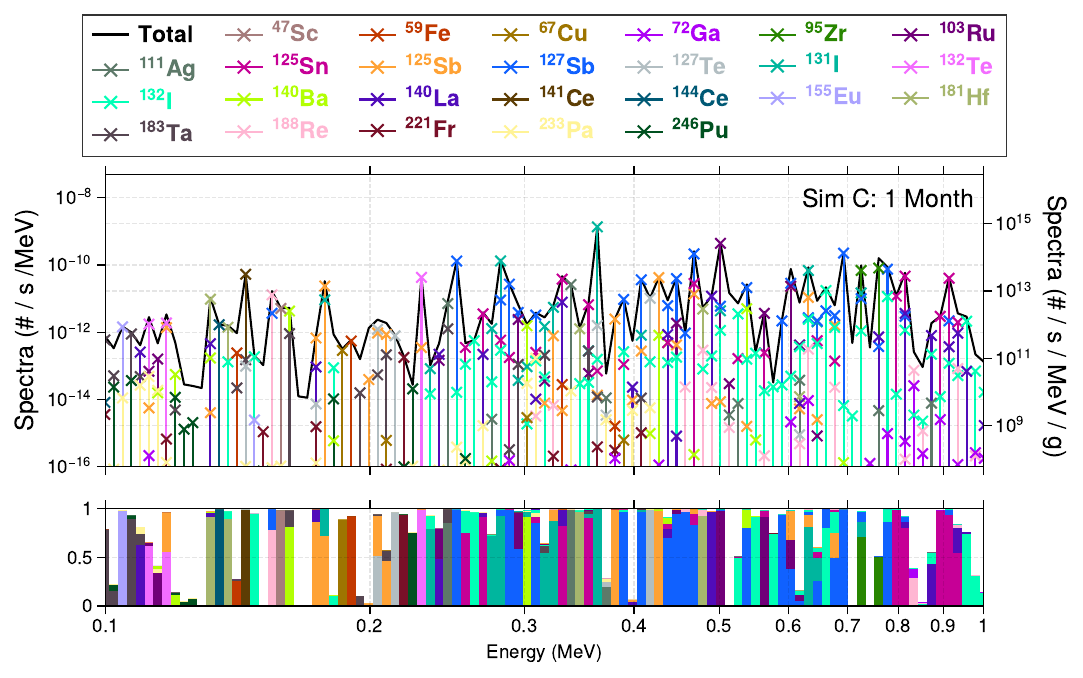}
\includegraphics{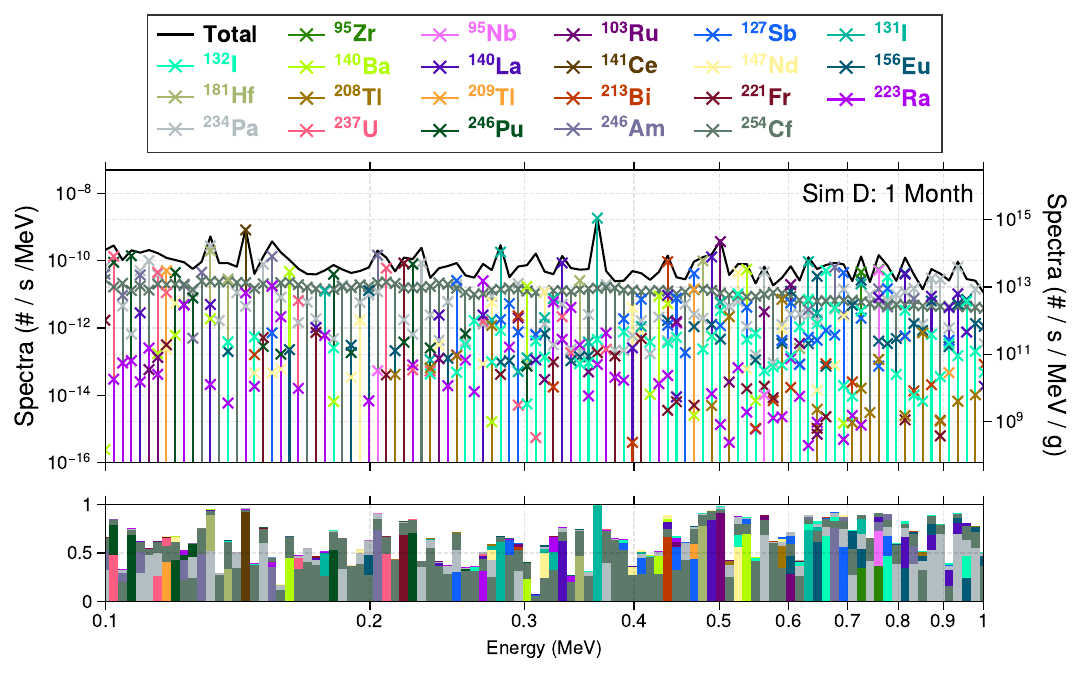}

\includegraphics{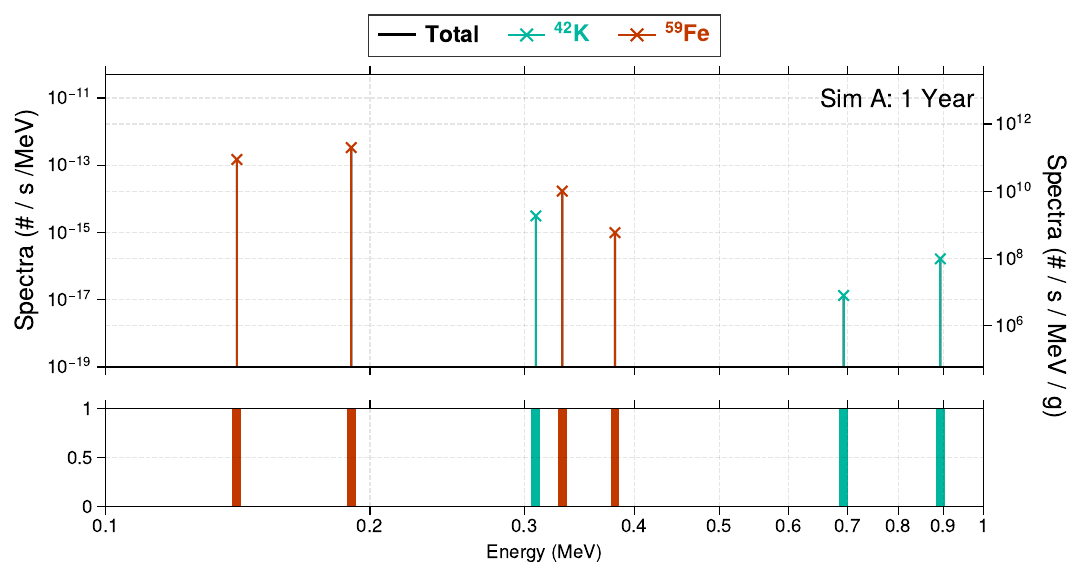}
\includegraphics{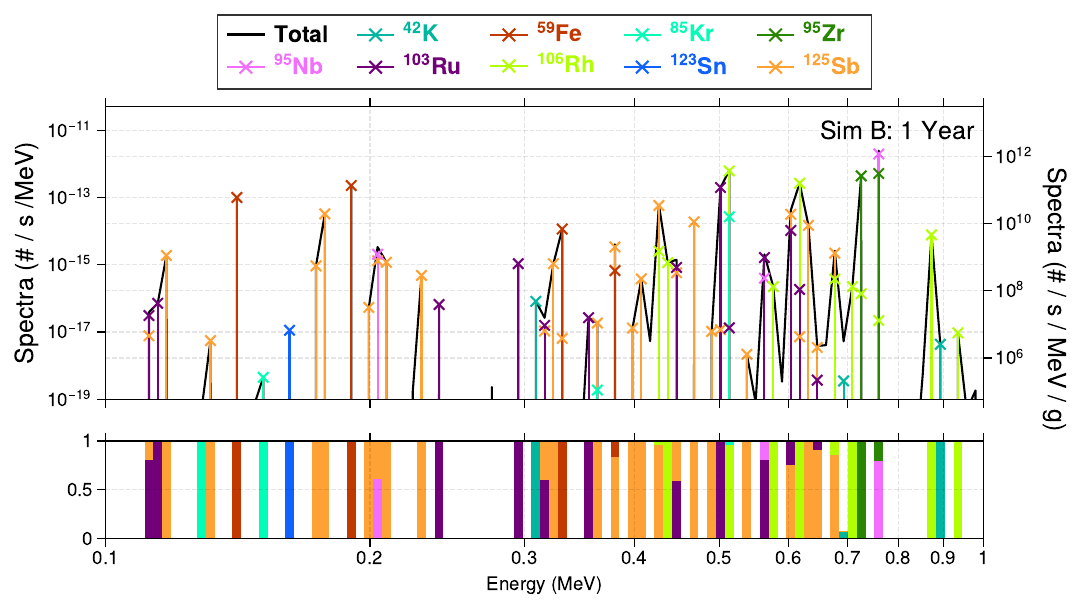}
\includegraphics{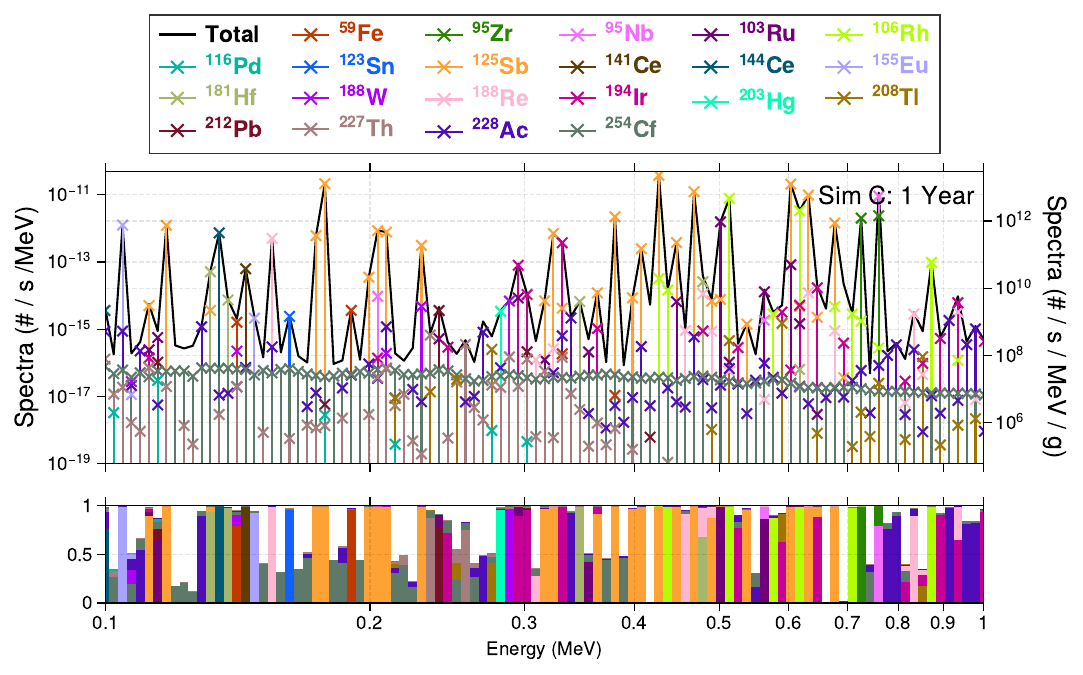}
\includegraphics{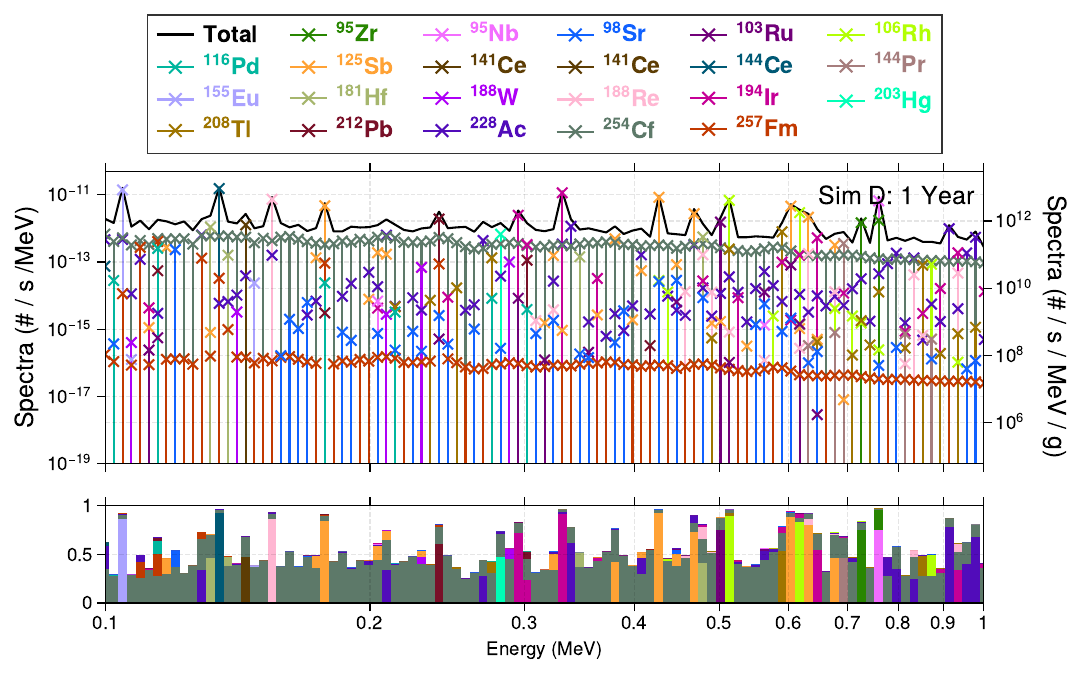}

\includegraphics{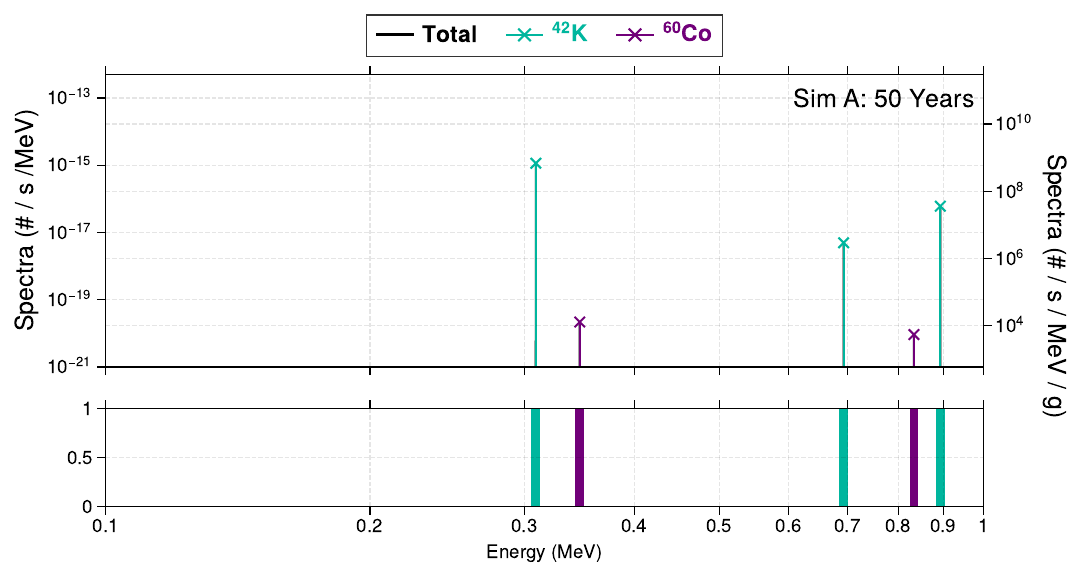}
\includegraphics{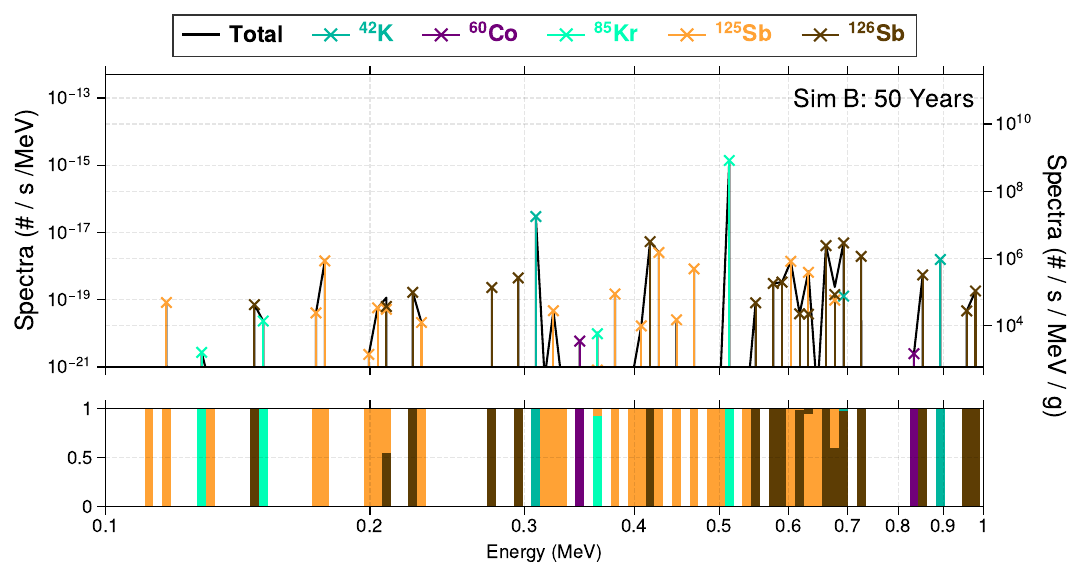}
\includegraphics{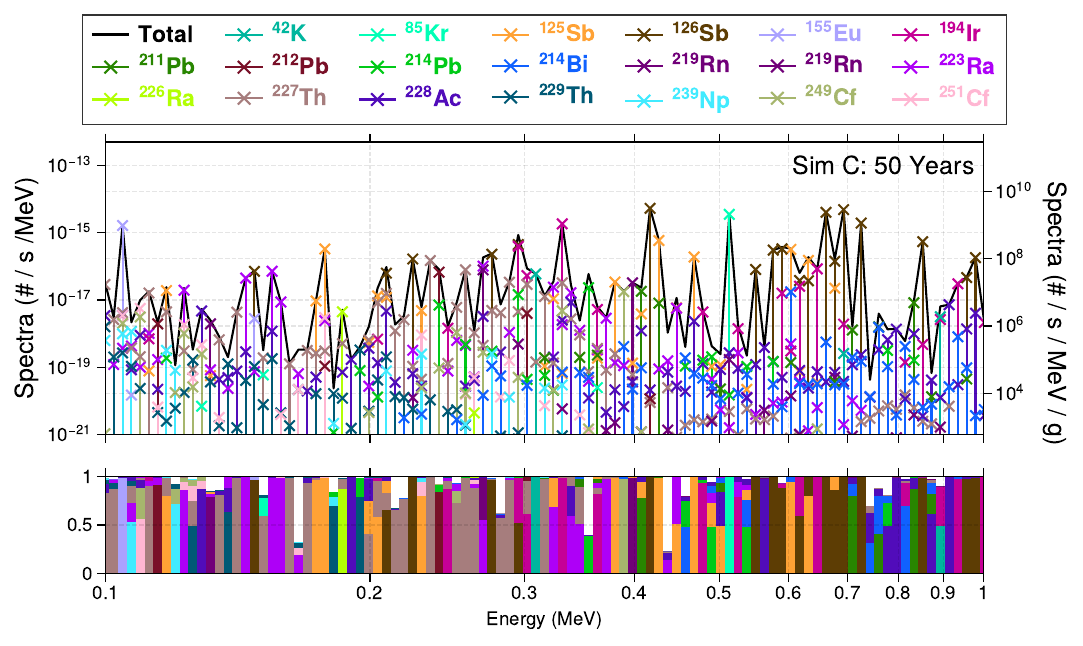}
\includegraphics{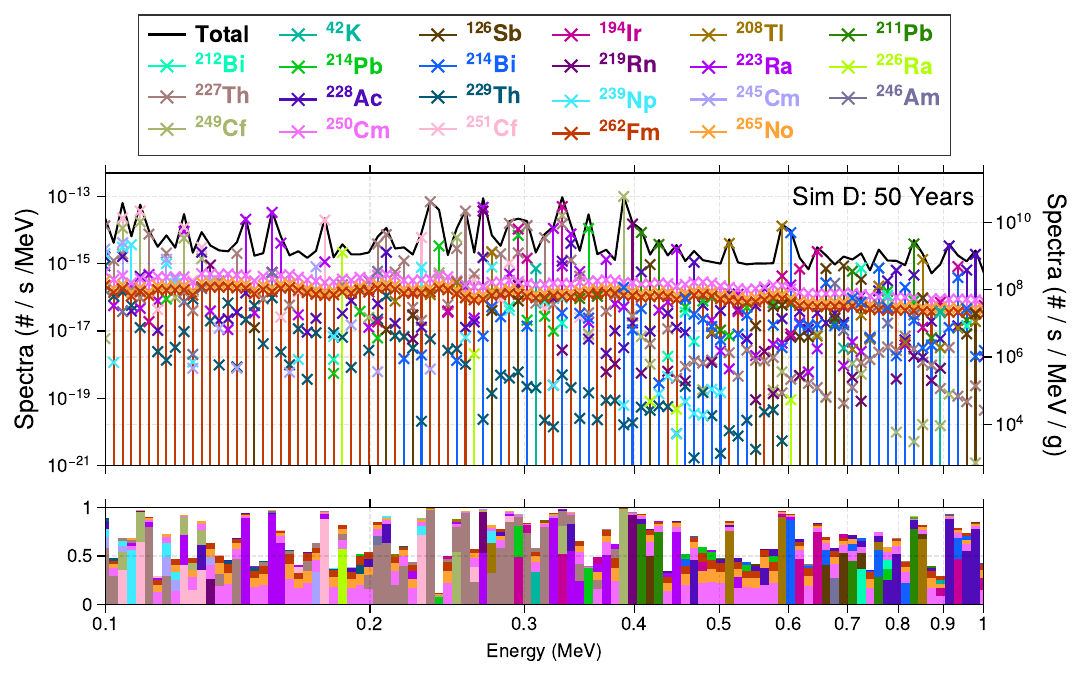}

\includegraphics{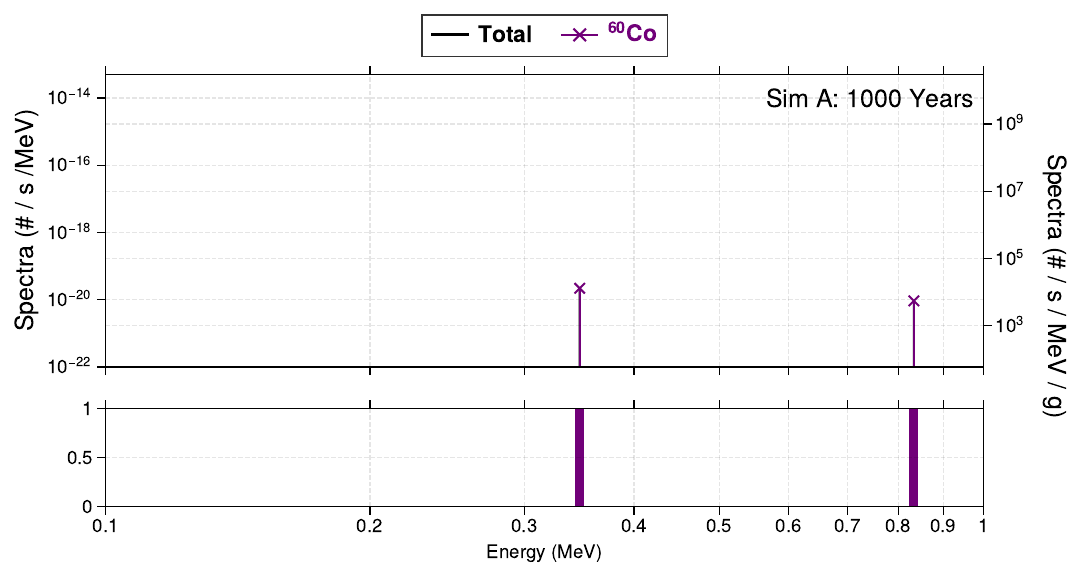}
\includegraphics{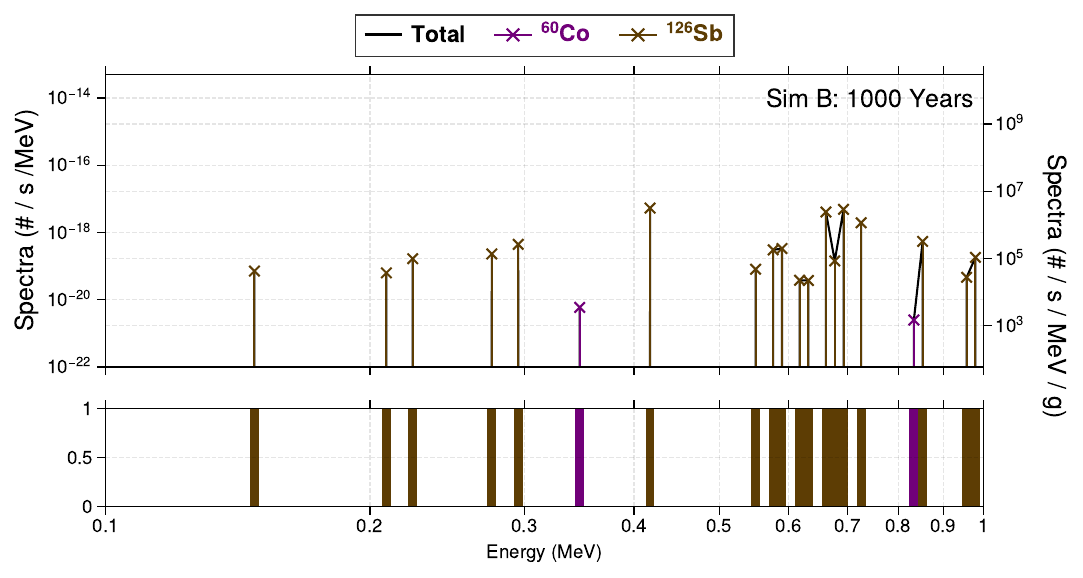}
\includegraphics{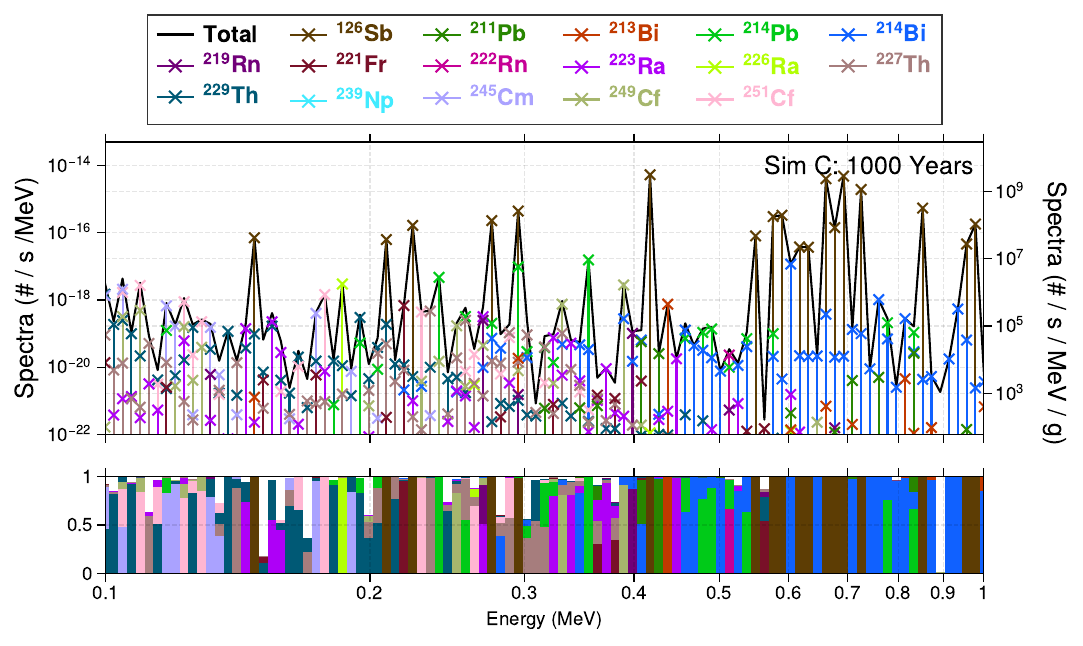}
\includegraphics{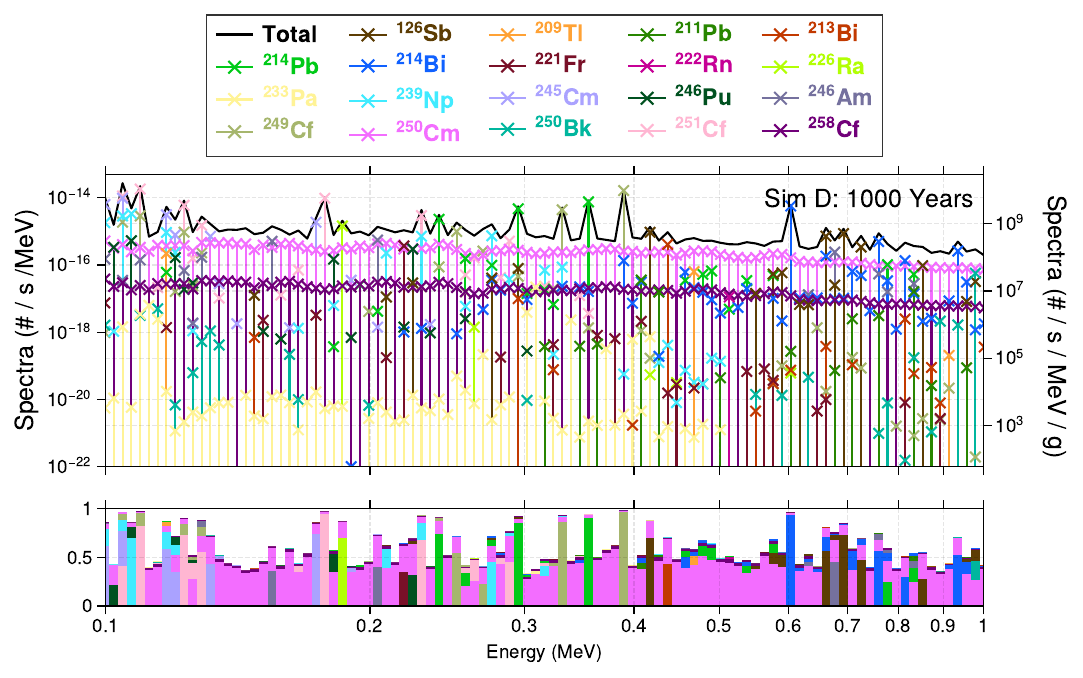}

\includegraphics{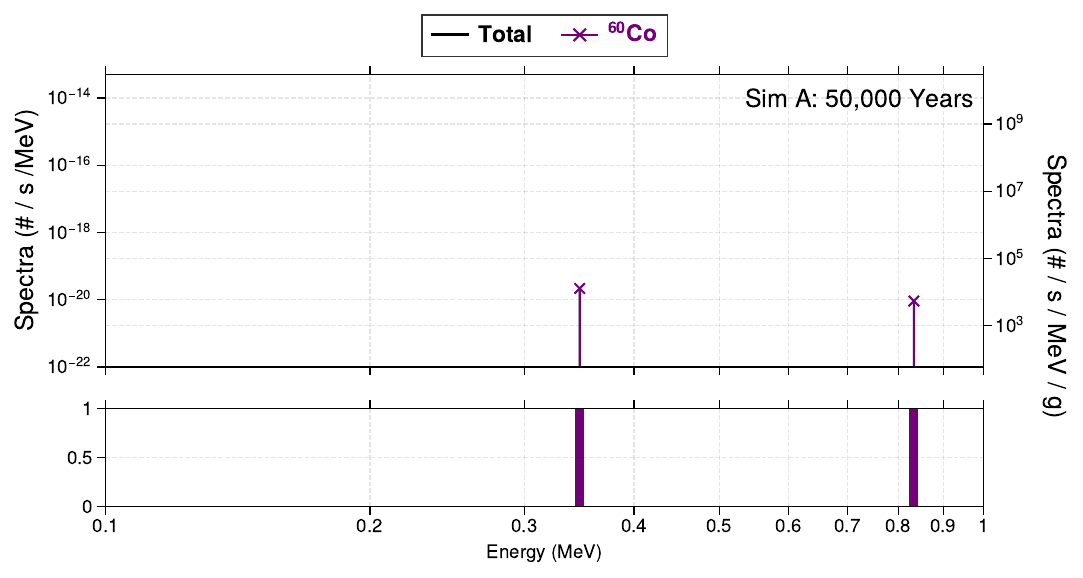}
\includegraphics{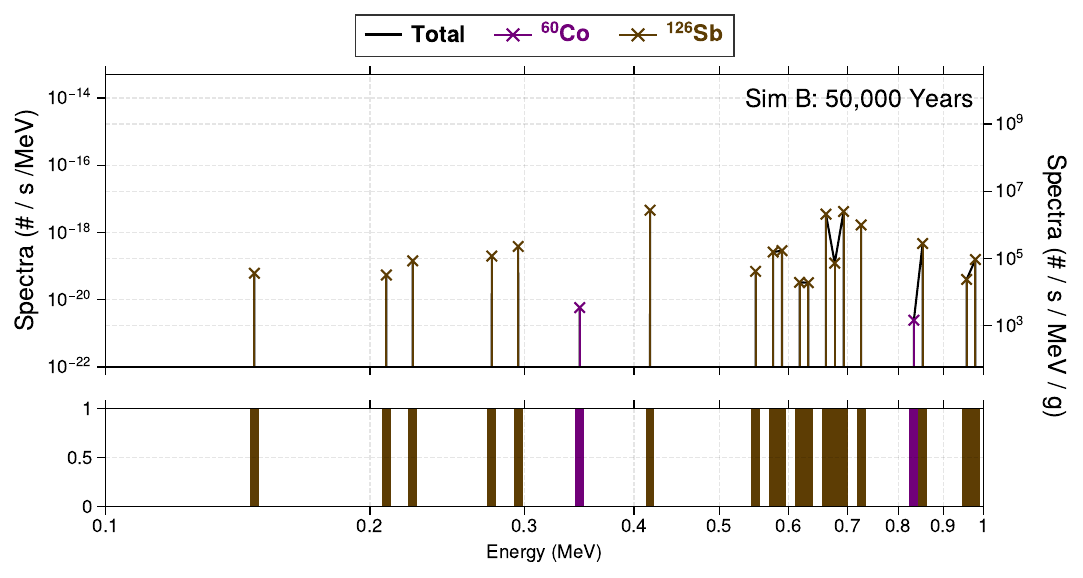}
\includegraphics{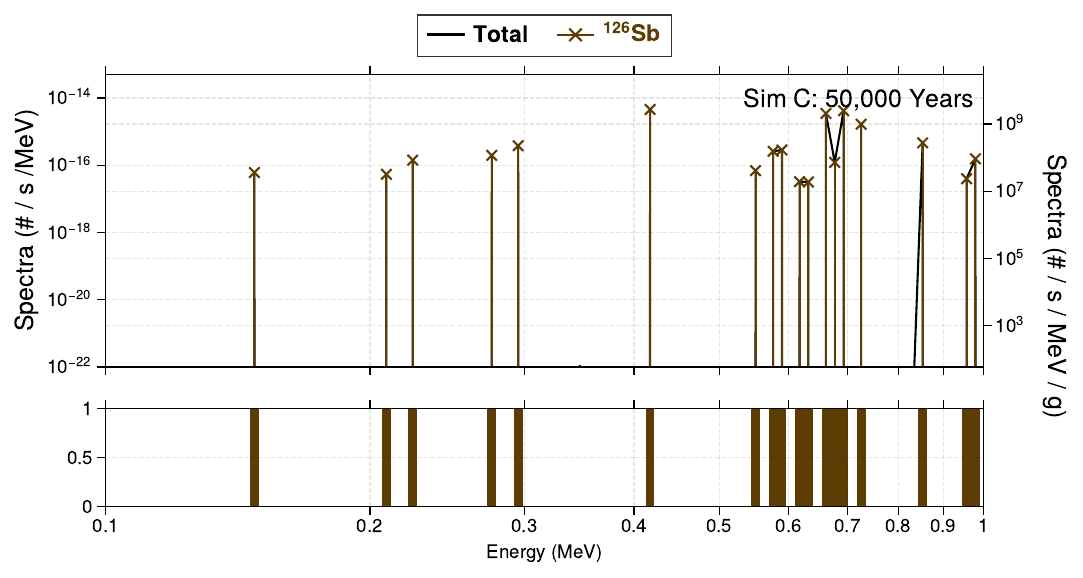}
\includegraphics{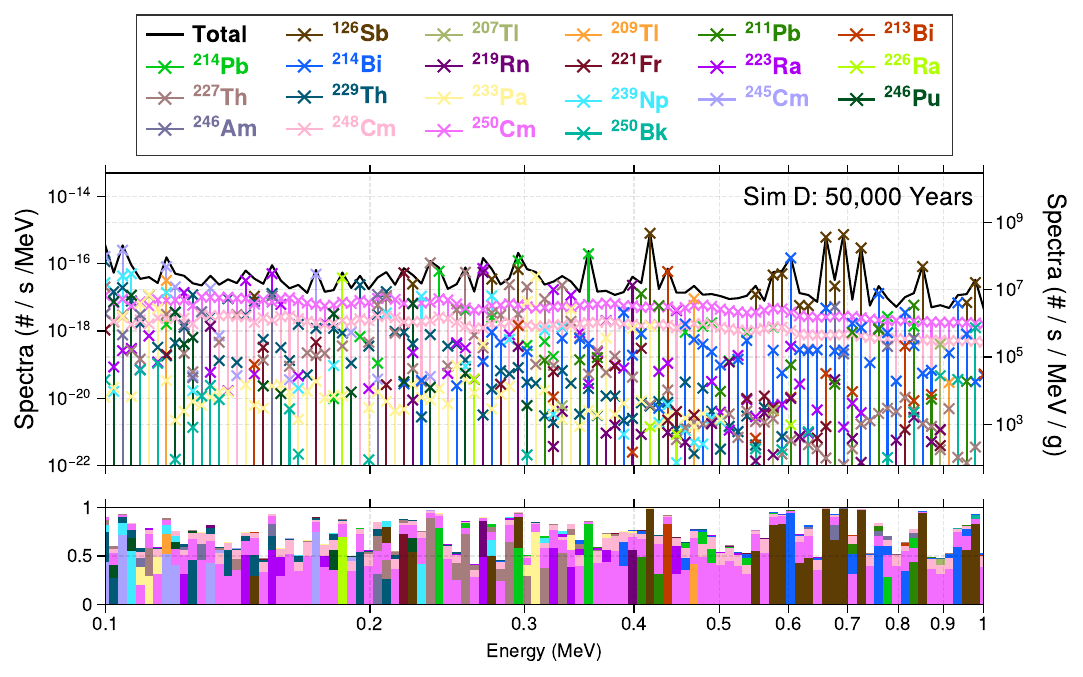}

\includegraphics{HighETime1SimA.pdf}
\includegraphics{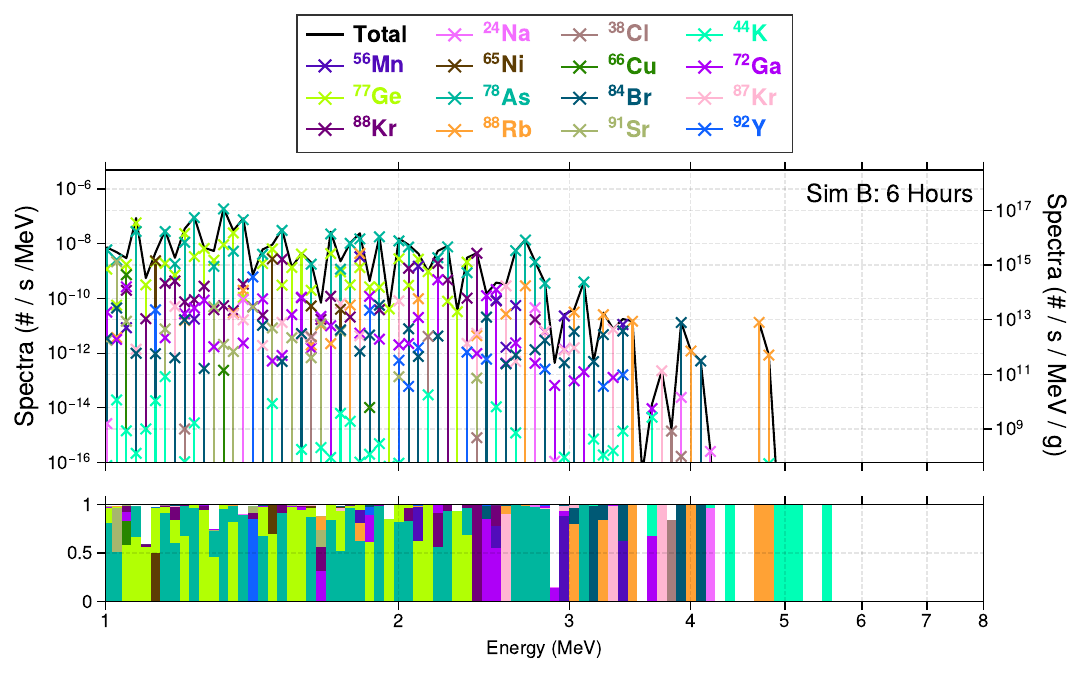}
\includegraphics{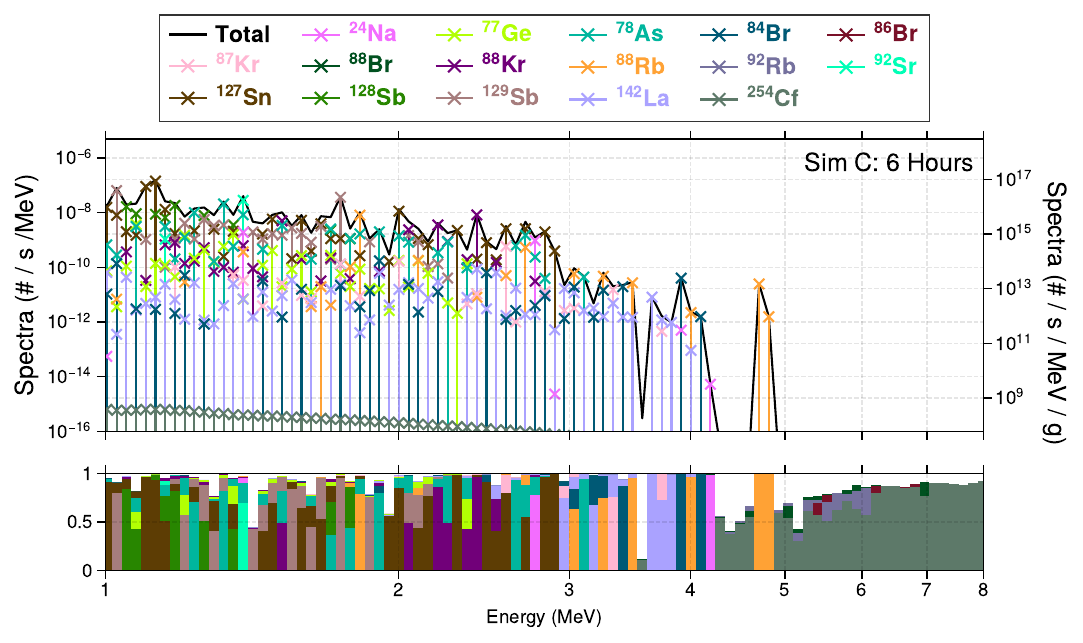}
\includegraphics{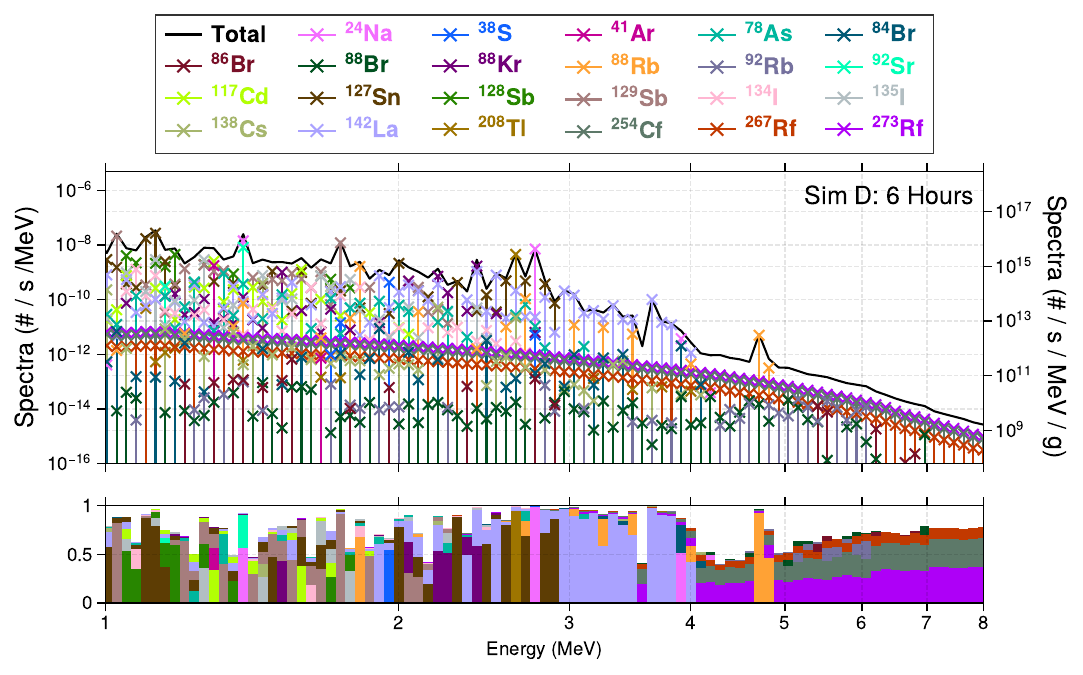}

\includegraphics{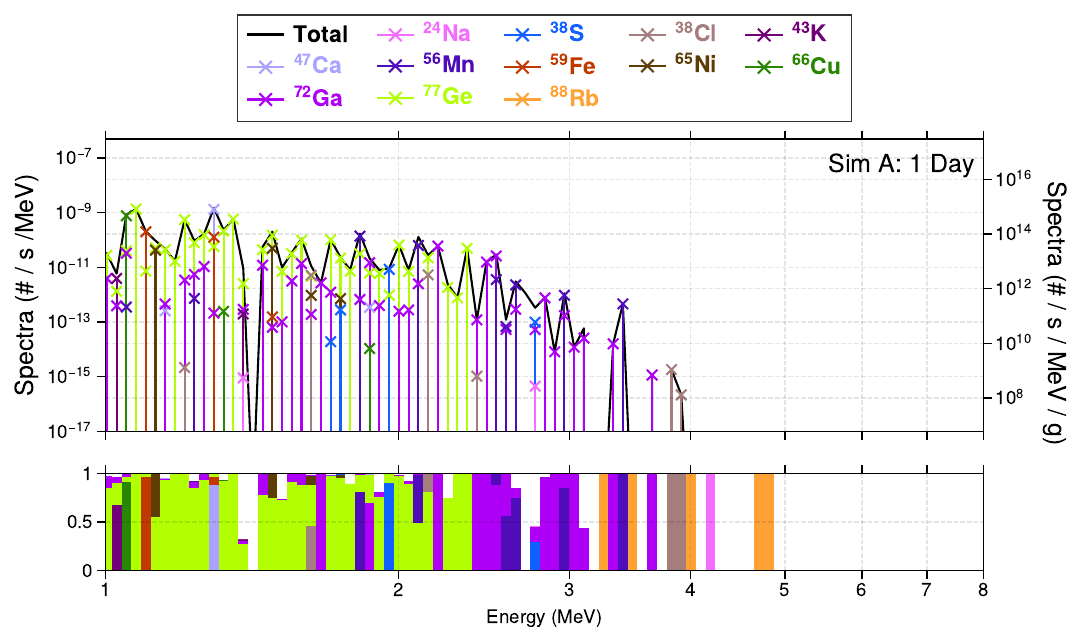}
\includegraphics{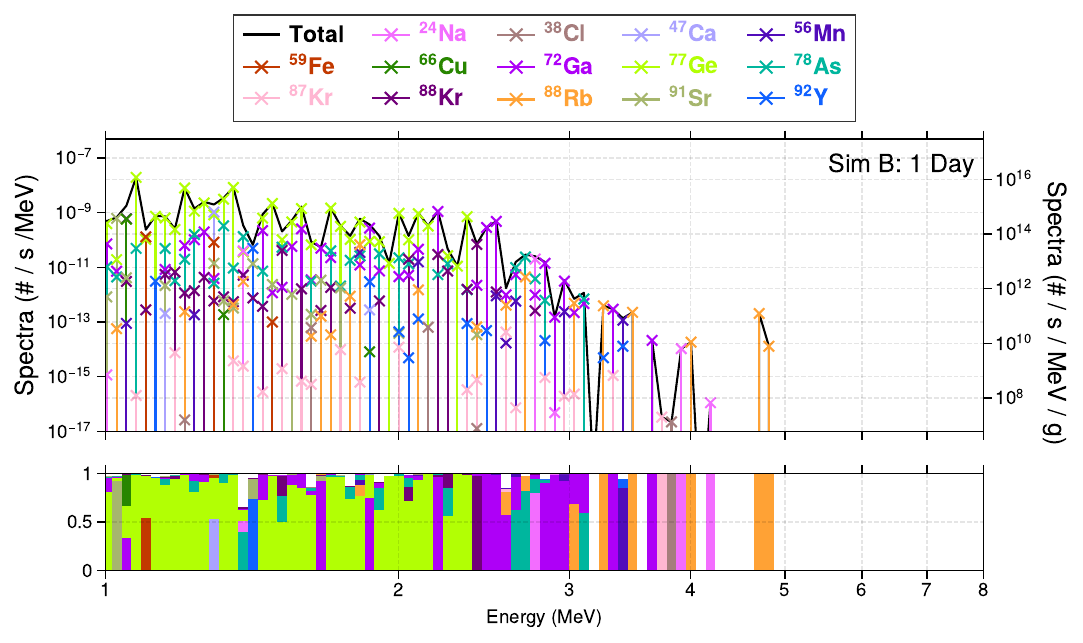}
\includegraphics{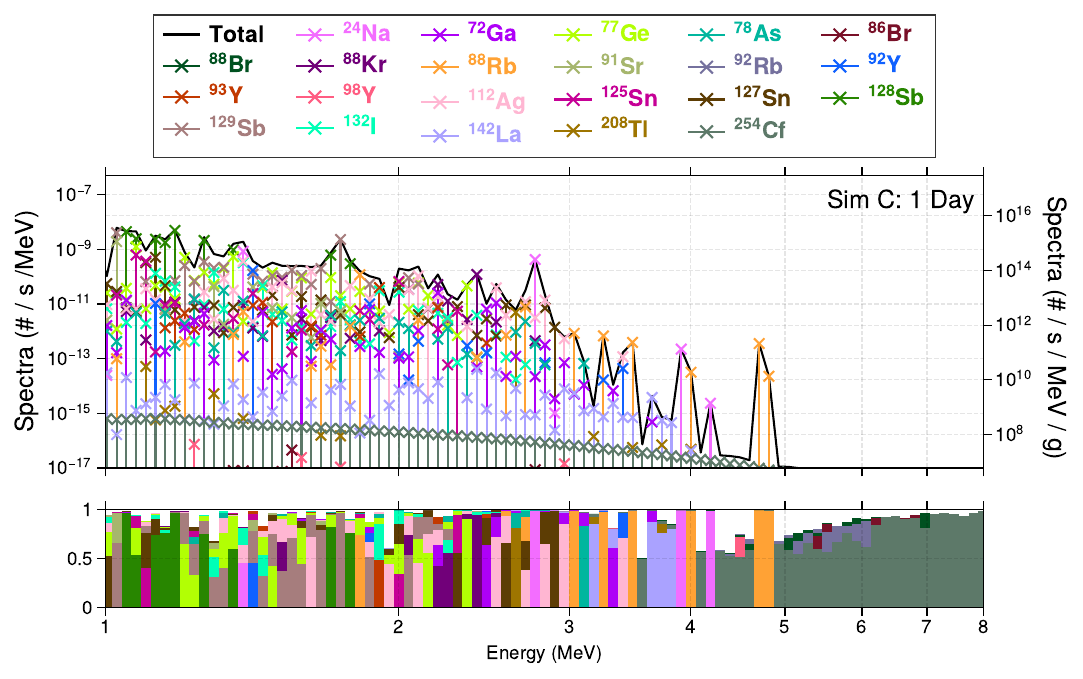}
\includegraphics{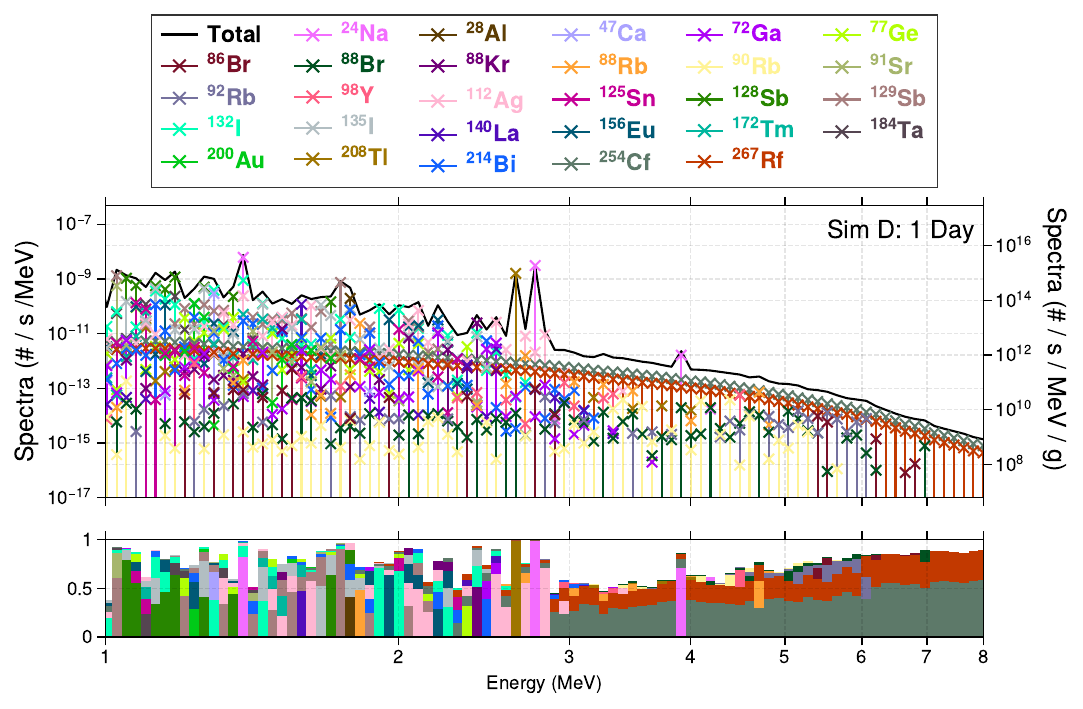}

\includegraphics{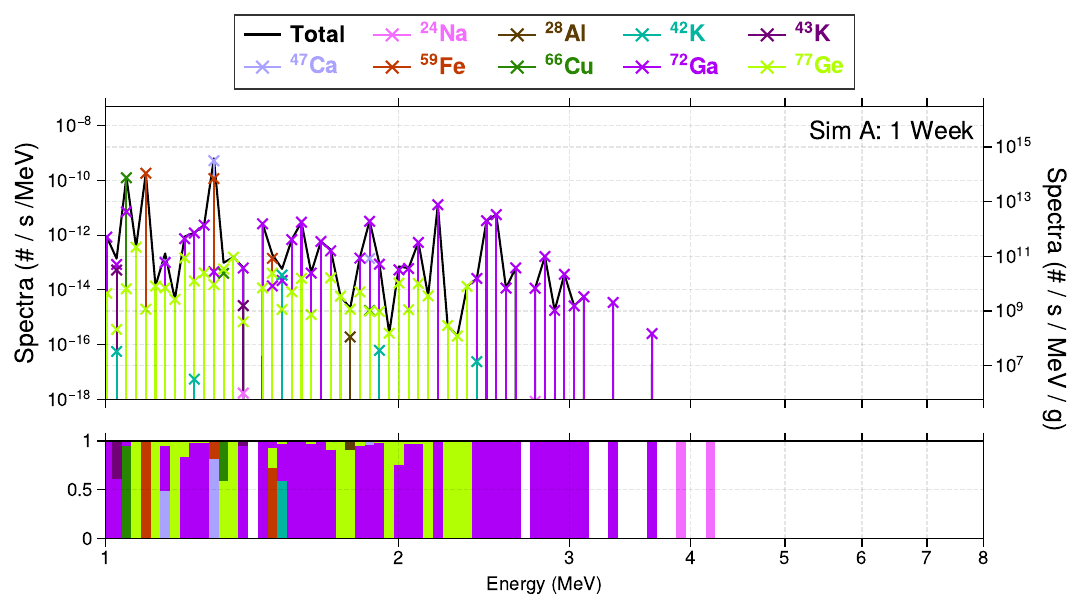}
\includegraphics{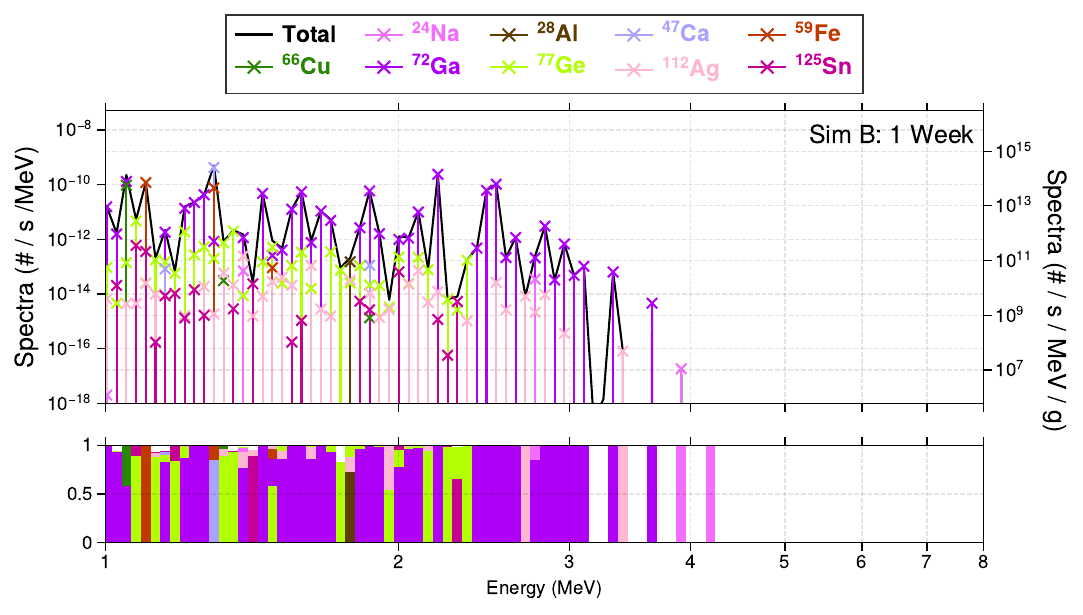}
\includegraphics{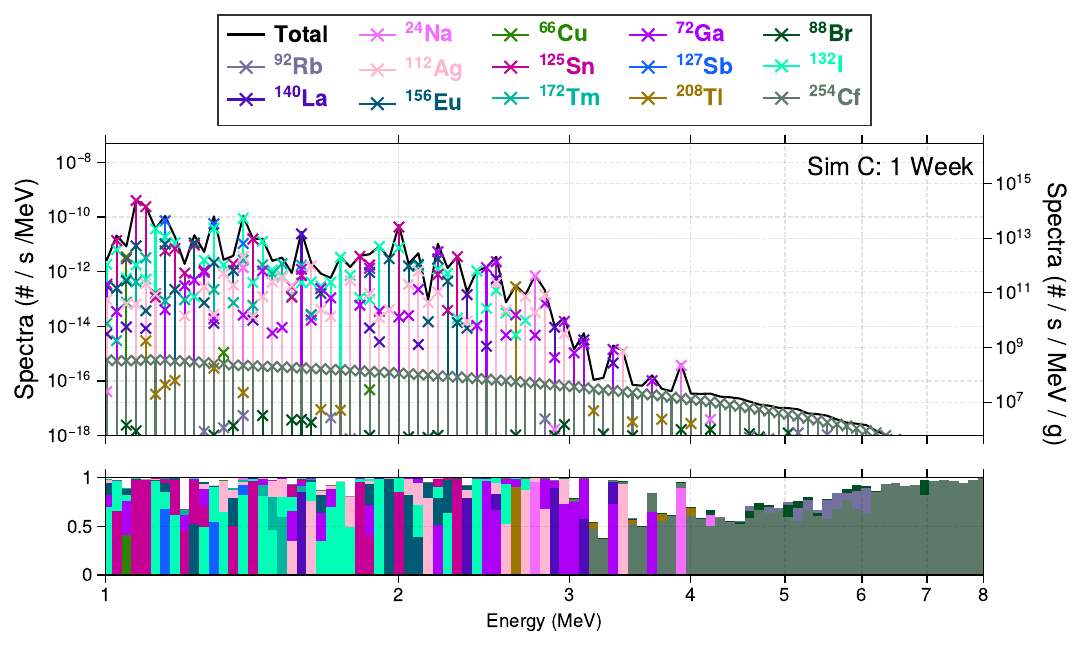}
\includegraphics{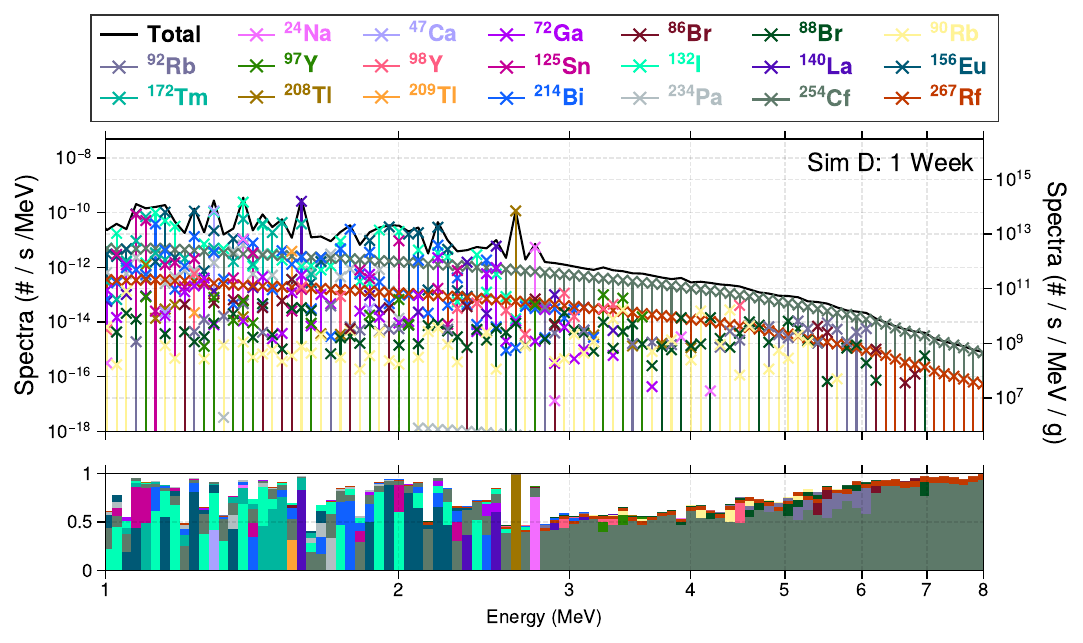}

\includegraphics{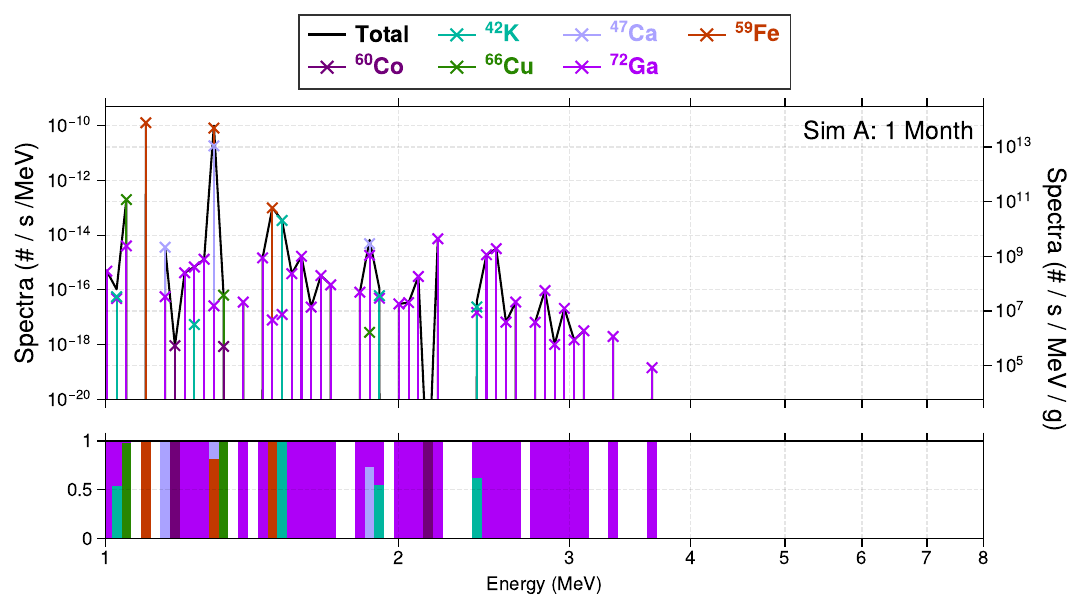}
\includegraphics{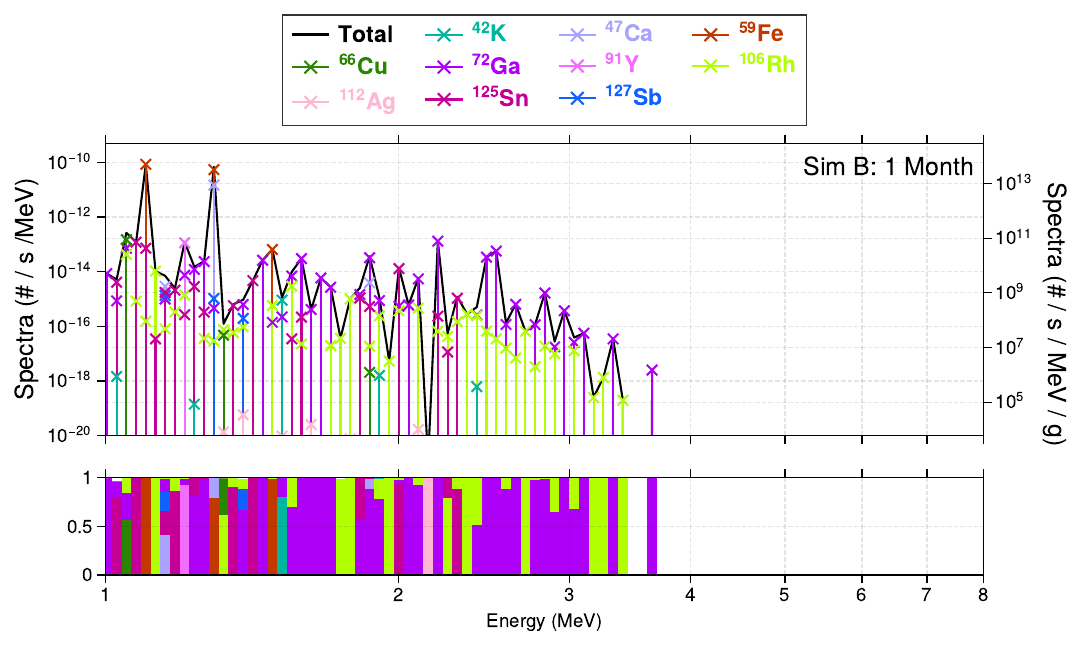}
\includegraphics{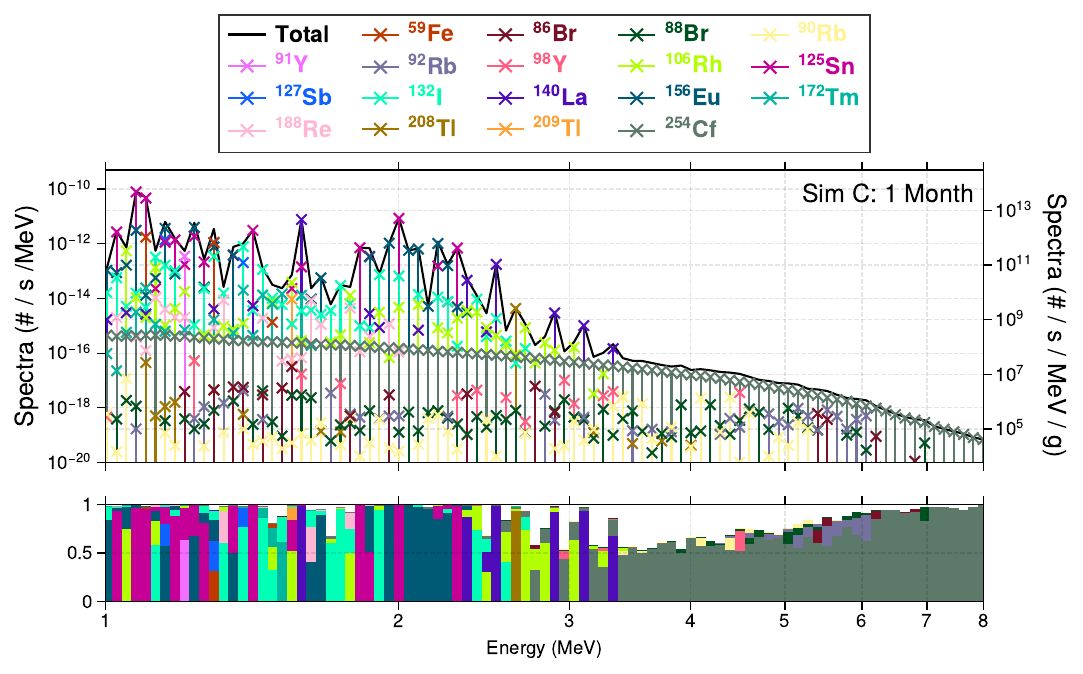}
\includegraphics{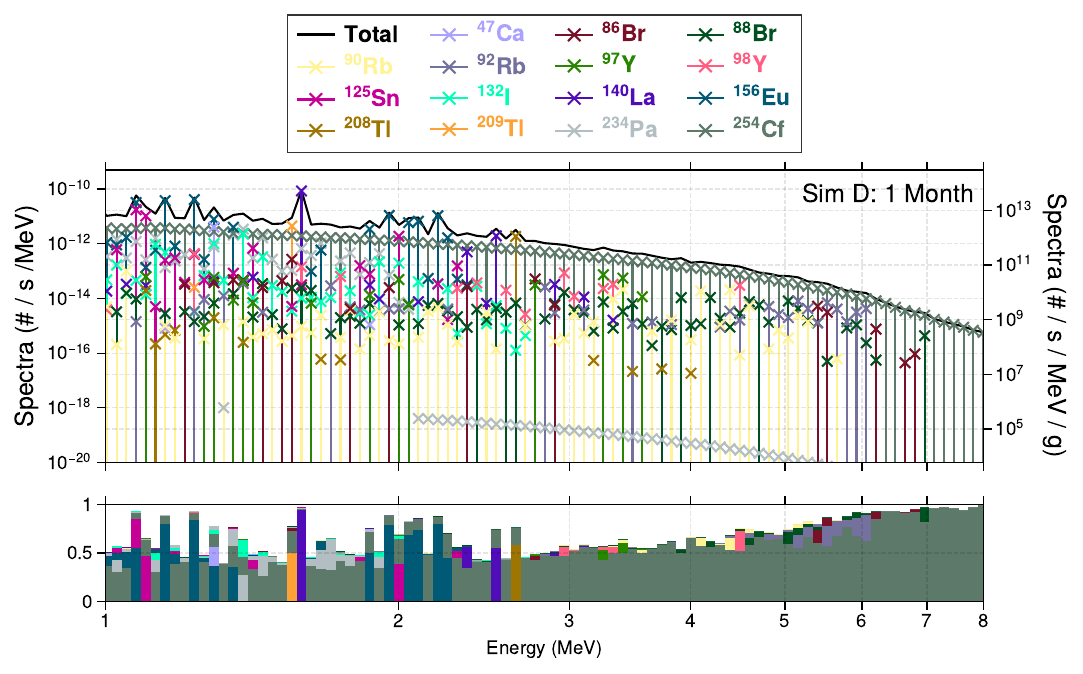}

\includegraphics{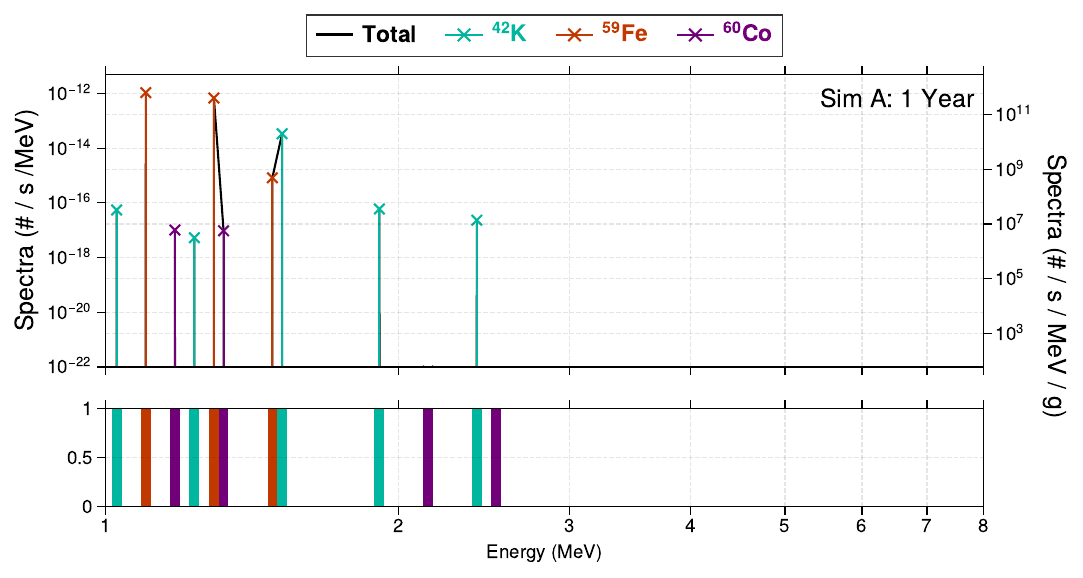}
\includegraphics{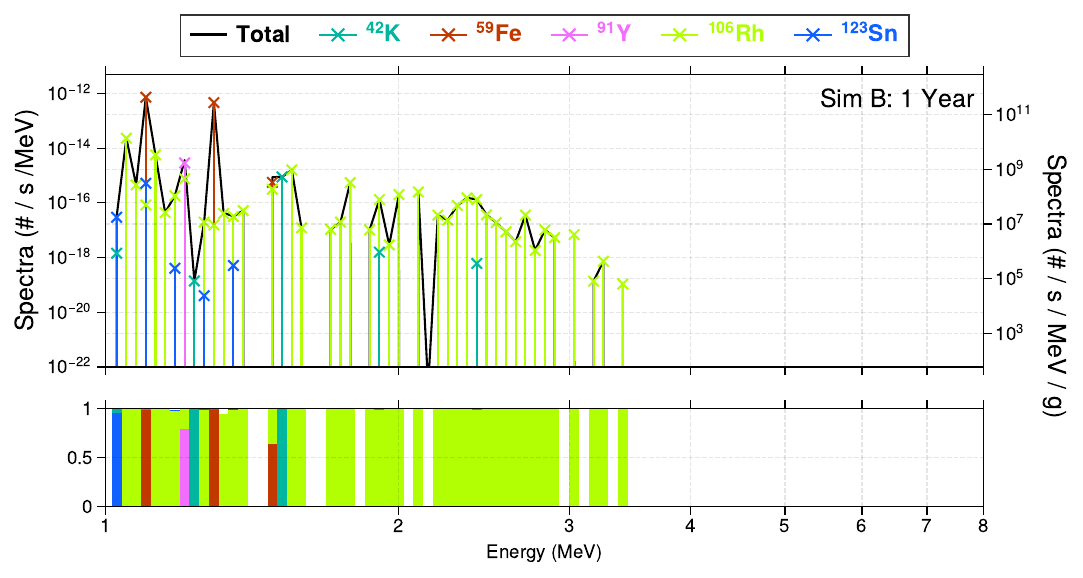}
\includegraphics{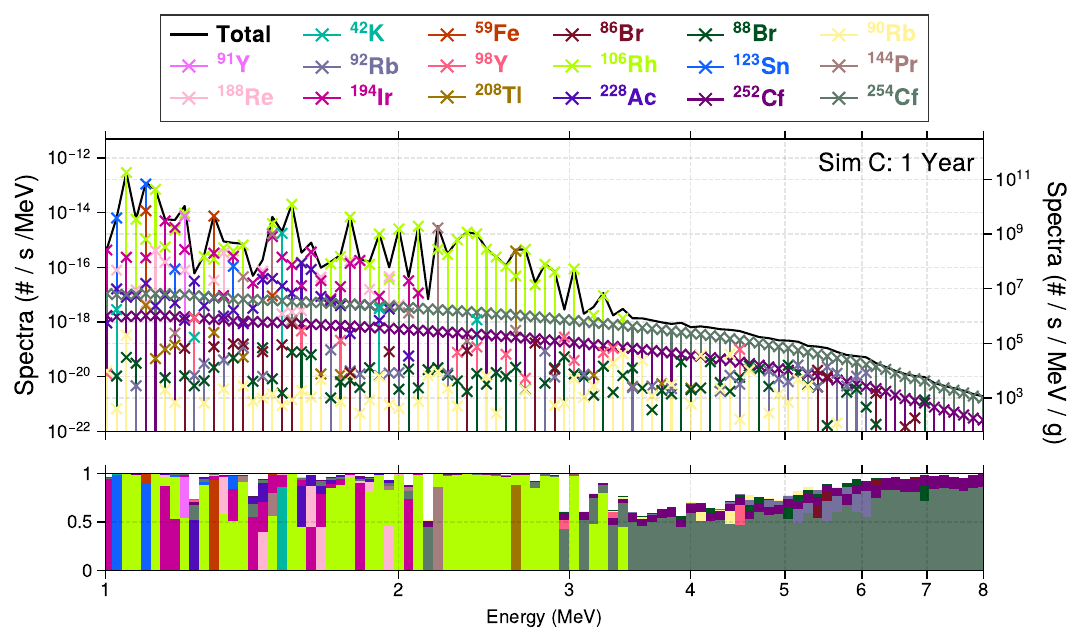}
\includegraphics{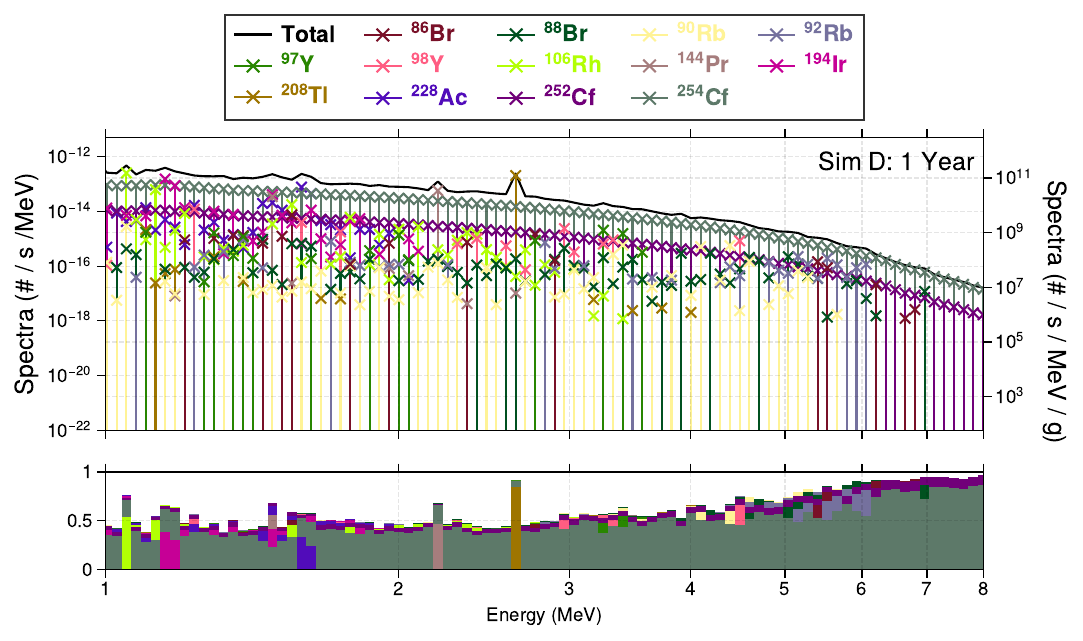}

\includegraphics{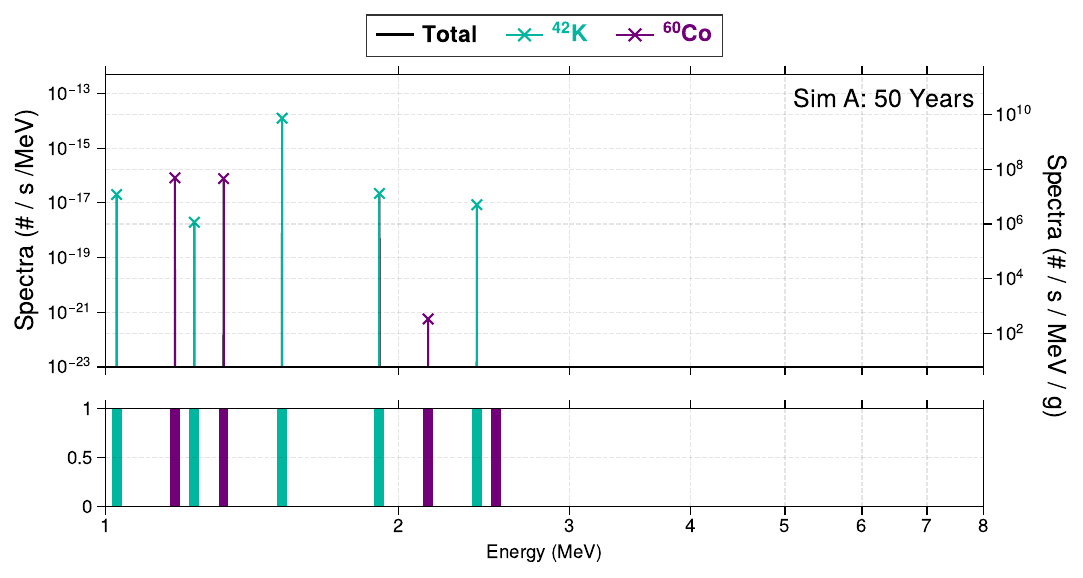}
\includegraphics{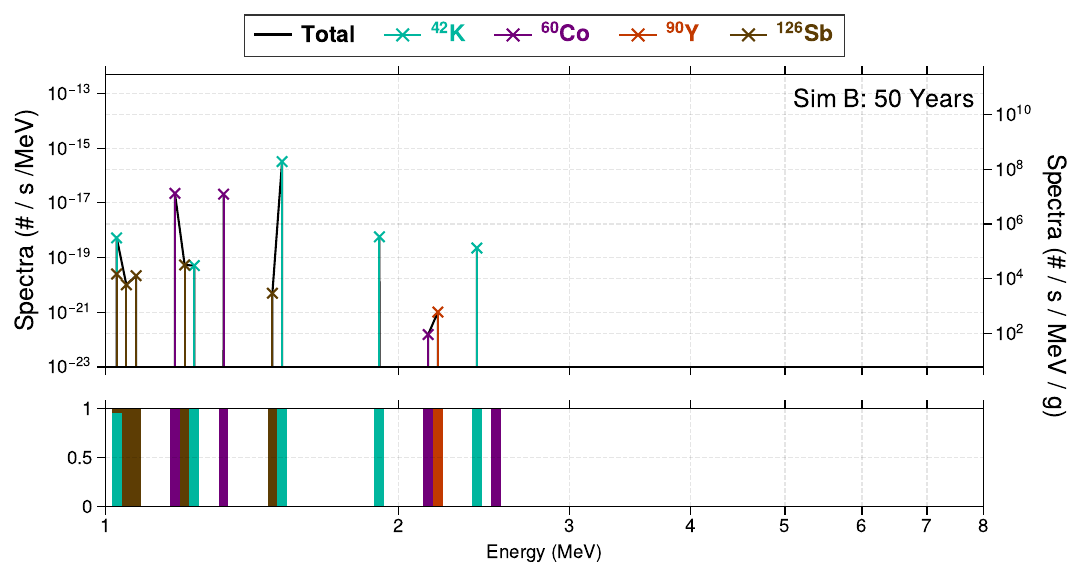}
\includegraphics{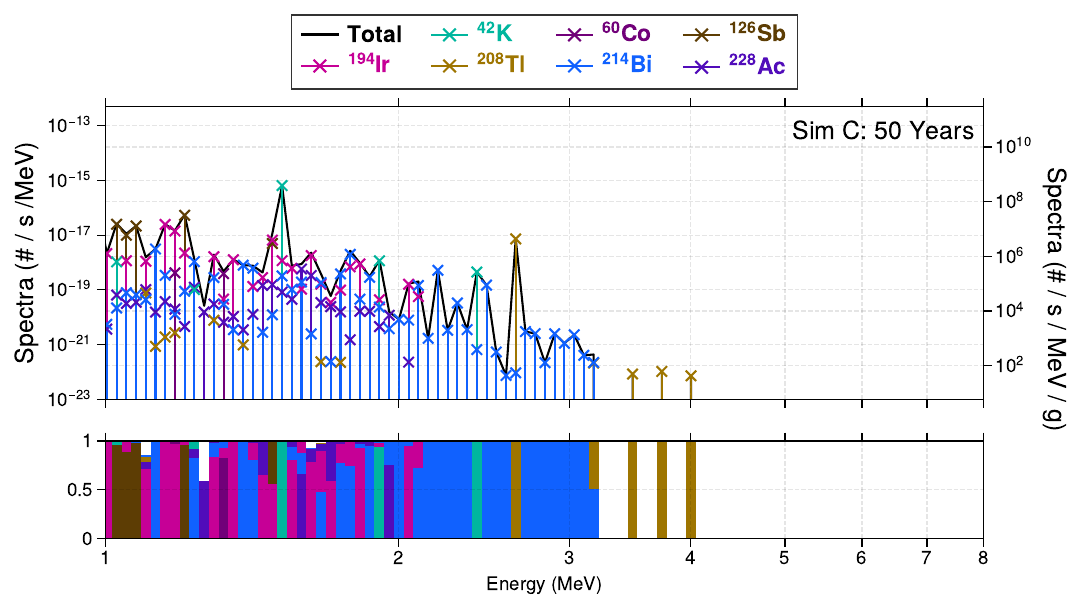}
\includegraphics{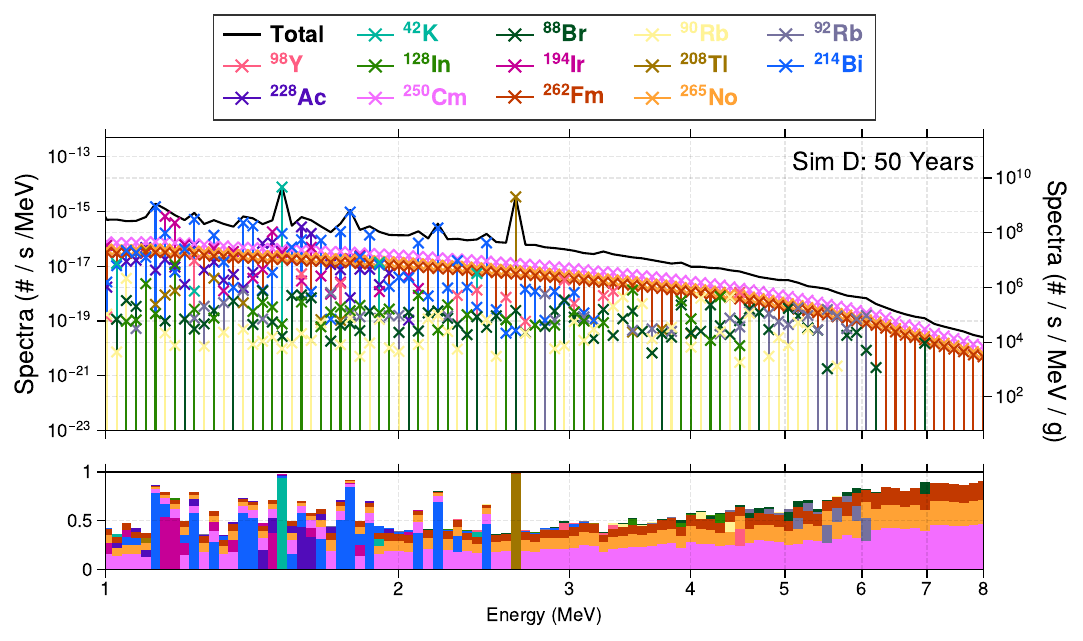}

\includegraphics{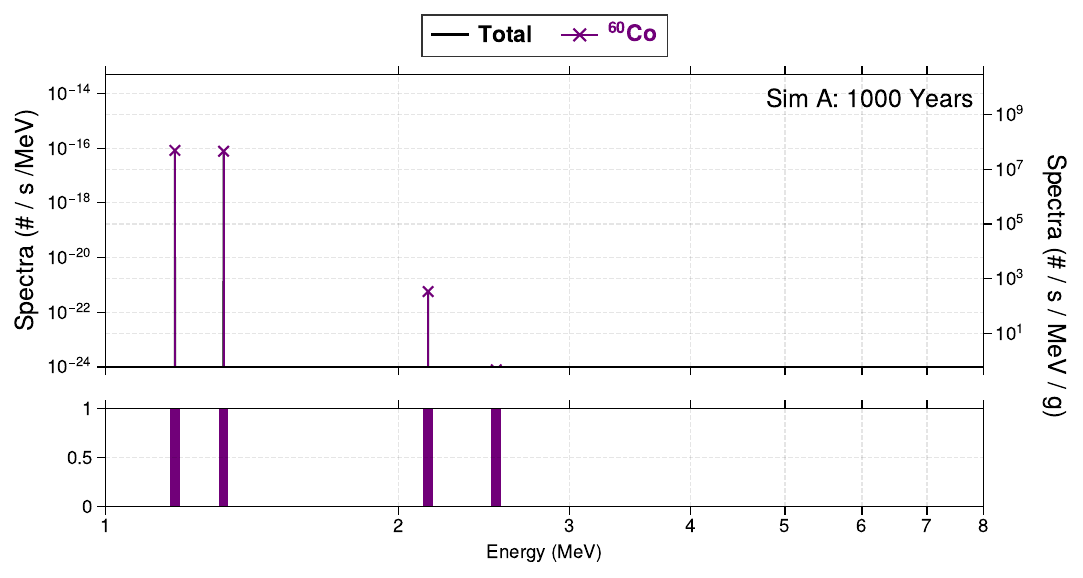}
\includegraphics{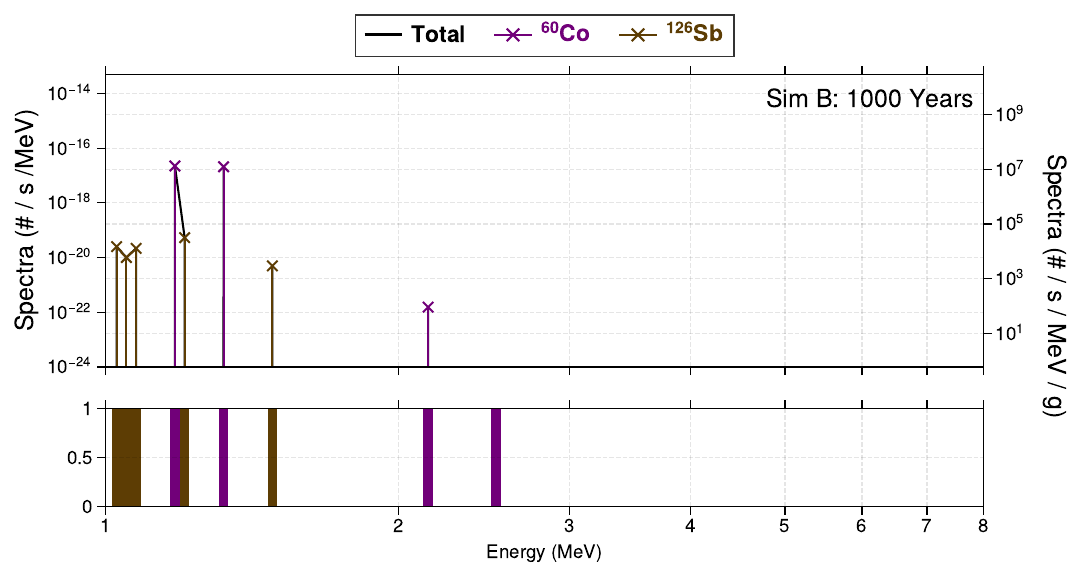}
\includegraphics{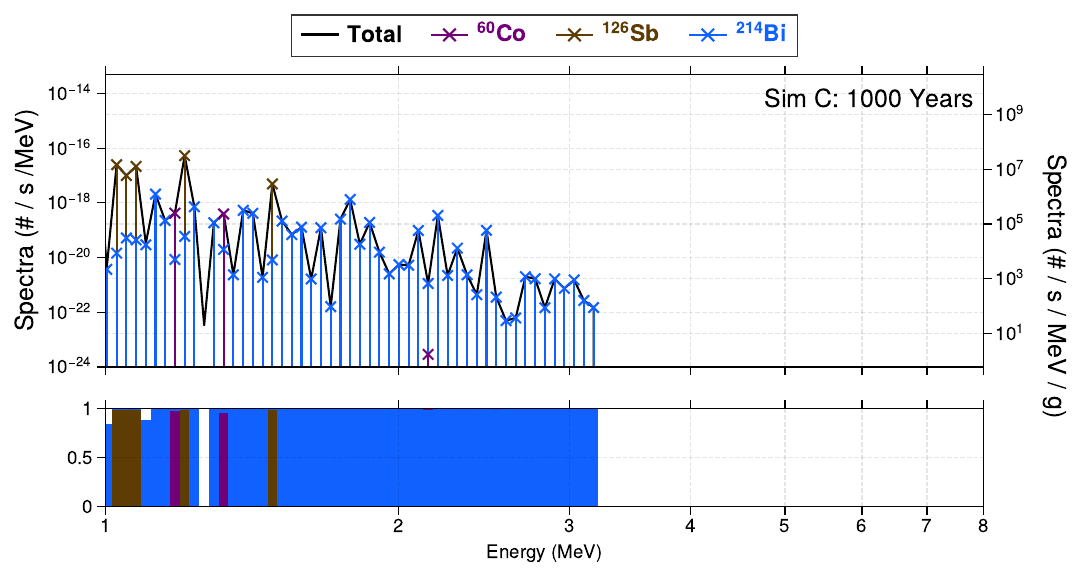}
\includegraphics{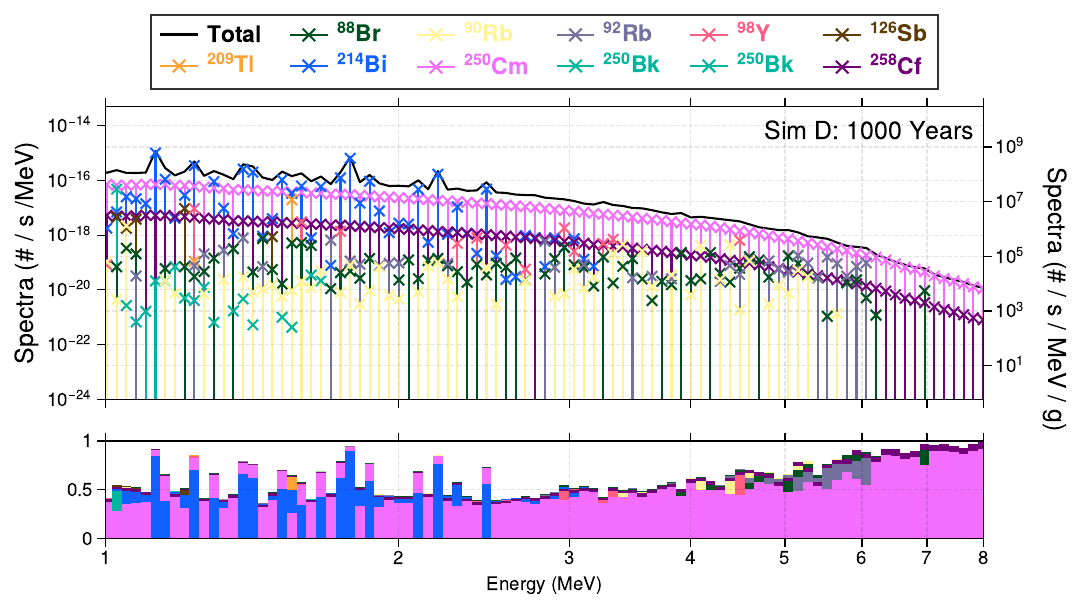}

\includegraphics{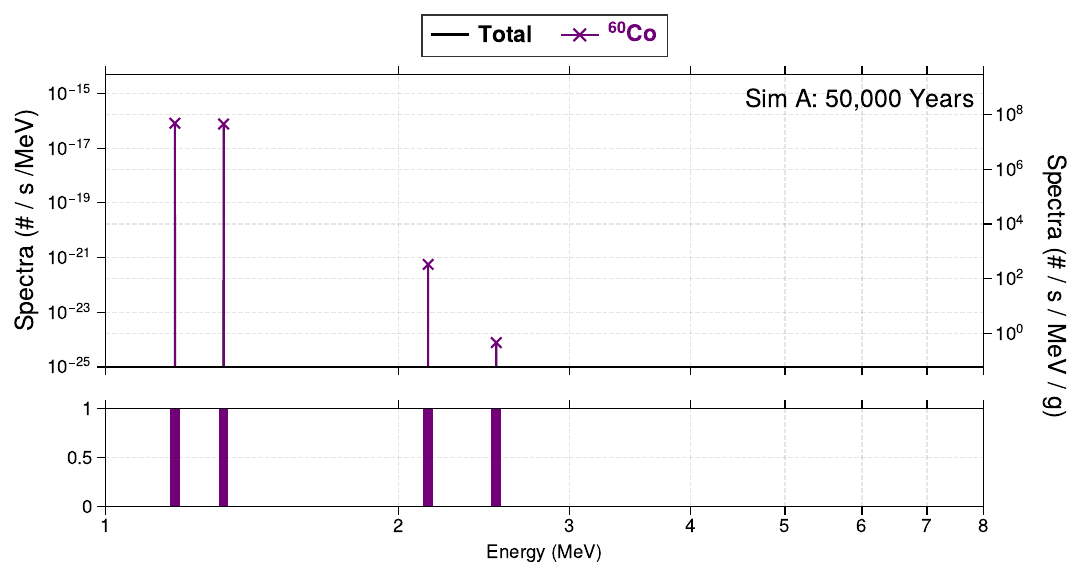}
\includegraphics{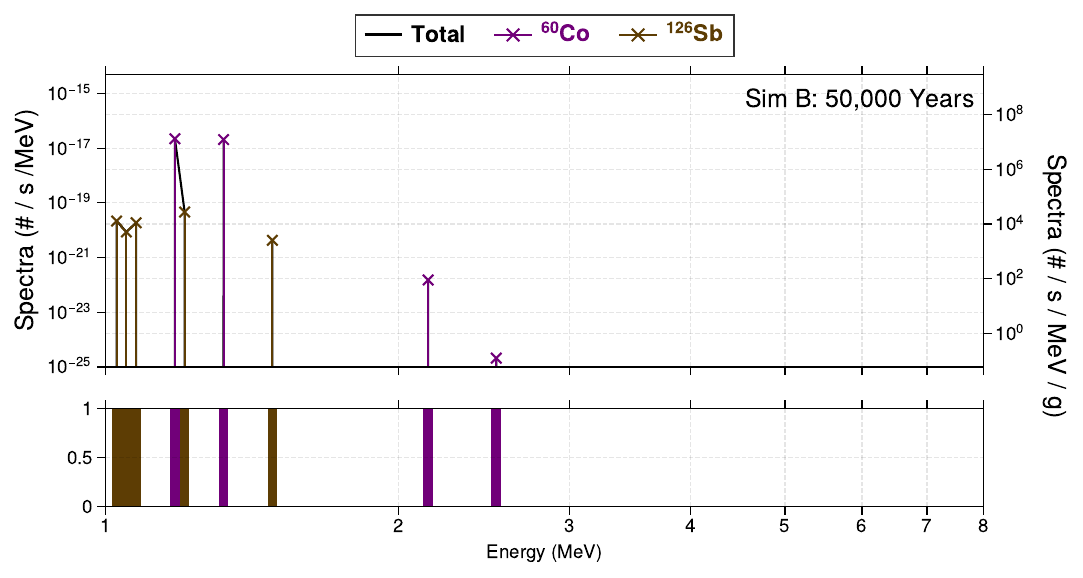}
\includegraphics{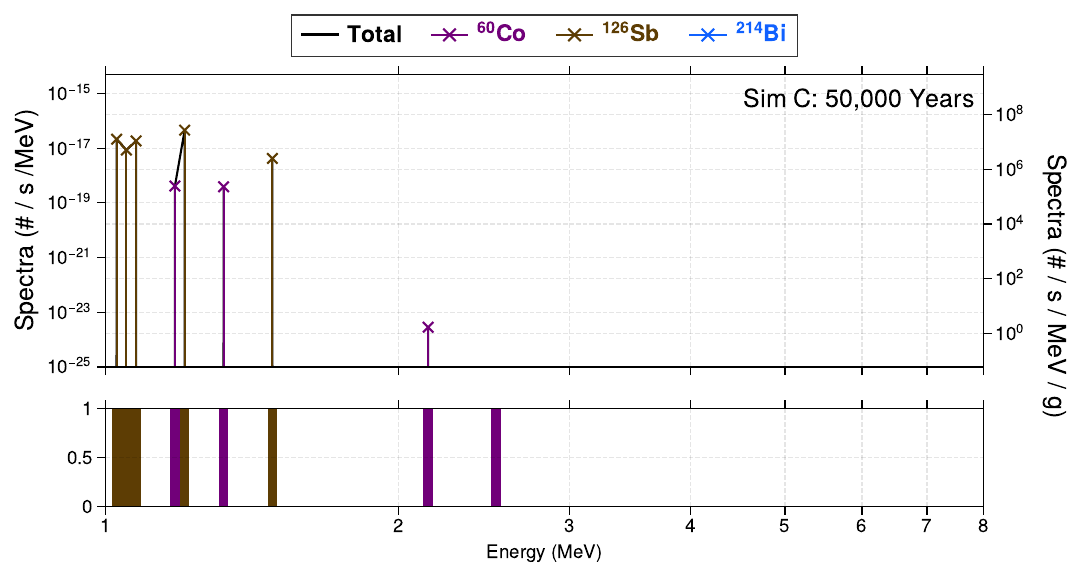}
\includegraphics{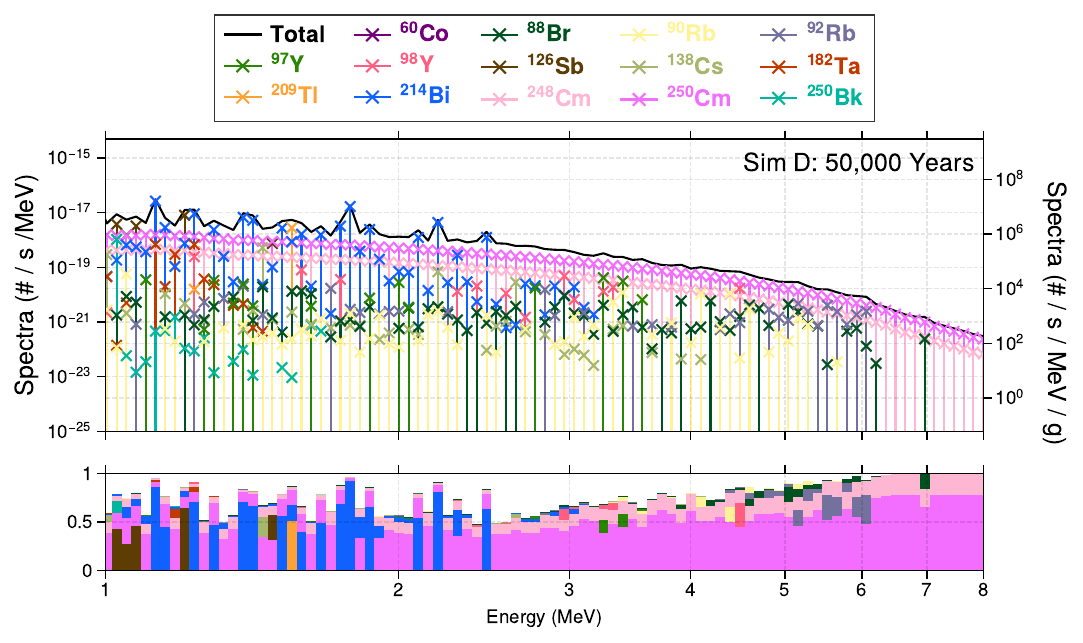}

\end{document}